\documentclass[a4paper,11pt]{article}
\pdfoutput=1
\usepackage{jheppub}

\usepackage[T1]{fontenc}
\usepackage{amsfonts}
\usepackage{amsmath}
\usepackage{amssymb}
\usepackage{multirow}
\usepackage{graphicx}
\usepackage{subfig}

\usepackage{scrextend}

\usepackage{tikz-cd}

\usepackage{mathtools}
\usepackage{array,ragged2e}

\usepackage{multicol}
\usepackage{blindtext}
\usepackage{floatrow}
\newfloatcommand{capbtabbox}{table}[][\FBwidth]

\usepackage{xfrac} 

\DeclareMathOperator*{\RE}{Re}
\DeclareMathOperator*{\IM}{Im}

\DeclareMathOperator{\diag}{diag}

\newcommand{\F}{{\mathbb{F}}}
\newcommand{\Z}{{\mathbb{Z}}}

\newcommand{\C}{{\mathbb{C}}}
\newcommand{\1}{{\mathbb I}}  
\newcommand{\vek}[1]{{\boldsymbol{#1}}}
\newcommand{\thetaUD}[2]{{\renewcommand{\arraystretch}{1}\theta\begin{bmatrix}#1 \\ #2 \end{bmatrix}}}

\newcommand{\ncu}[2]{#2\text{\scriptsize{$(\times#1)$}}}

\numberwithin{equation}{section}
\usepackage{graphicx}

\begin{document}

\title{Averaging over codes and an $SU(2)$ modular bootstrap}

\author[a]{Johan Henriksson,}
\emailAdd{johan.henriksson@df.unipi.it}
\affiliation[a]{Department of Physics, University of Pisa and INFN, \\Largo Pontecorvo 3, I-56127 Pisa, Italy}
\author[a]{Brian McPeak}
\emailAdd{brian.mcpeak@df.unipi.it}

\abstract{Error-correcting codes are known to define chiral 2d lattice CFTs where all the $U(1)$ symmetries are enhanced to $SU(2)$. In this paper, we extend this construction to a broader class of length-$n$ codes which define full (non-chiral) CFTs with $SU(2)^n$ symmetry, where $n=c+\bar c$. We show that codes give a natural discrete ensemble of 2d theories in which one can compute averaged observables. The partition functions obtained from averaging over all codes weighted equally is found to be given by the sum over modular images of the vacuum character of the full extended symmetry group, and in this case the number of modular images is finite. This averaged partition function has a large gap, scaling linearly with $n$, in primaries of the full $SU(2)^n$ symmetry group. Using the sum over modular images, we conjecture the form of the genus-2 partition function. This exhibits the connected contributions to disconnected boundaries characteristic of wormhole solutions in a bulk dual.}

\maketitle

\section{Introduction}

Error-correcting codes\footnote{In this paper, we shall only be interested in \emph{binary linear codes}, which will always be what is meant by ``codes'' unless otherwise specified. See section~\ref{sec:Symmetry} for a review of some coding theory terminology.} and lattices have enjoyed a long and deep relationship. At the heart of this bond is Construction A \cite{LeechSloane1971}, which associates a dimension-$n$ lattice to every length-$n$ code, and which was originally used to find a number of the densest known sphere packings. The result is that any statement about codes can be directly translated to a statement about lattices, and many results in the theory of lattices correspond to a result in the theory of codes.

Lattices are known to define CFTs if they are even and self-dual \cite{Narain:1985jj}. The resulting theories of Narain type are free, representing bosons moving in a compact space. Doubly-even self-dual codes are related by Construction A to even self-dual Euclidean lattices, so they define chiral CFTs. This fact was first used in the classification of $c = 24$ chiral CFTs \cite{Dolan:1989kf, Dolan:1994st}. More recently, the construction has been extended to fermionic CFTs arising from codes over $\F_3$ \cite{Gaiotto:2018ypj}, and more general sets of (Narain) CFTs arising from codes over $\F_p \times \F_q$ \cite{, Dymarsky:2020qom, Dymarsky:2020bps, Dymarsky:2021xfc, Yahagi:2022idq, Angelinos:2022umf}.

One obvious question is: to what degree do error-correcting codes define an interesting ensemble of 2d CFTs? The importance of ensembles of theories was demonstrated by the striking example of random matrix theory \cite{Cotler:2016fpe, Saad:2018bqo, Saad:2019lba}, and was recently highlighted by the observation that the ensemble average of all Narain CFTs has a number of holographic properties at large $c$ \cite{Maloney:2020nni, Afkhami-Jeddi:2020ezh}. These properties include a large gap in the spectrum of $U(1)^c \times U(1)^c$ primaries, and a partition function which arises from a sum over modular images just like the Poincar\'e series of \cite{Dijkgraaf:2000fq,Maloney:2007ud, Keller:2014xba}. The result is the conjecture that such an ensemble average is dual to a theory of $c$ Chern--Simons gauge fields in the bulk. 

It is easy to define the ensemble of codes, which are simply collections of binary vectors (typically satisfying other requirements as well, such as self-duality in the present case). However, ensembles of code theories \cite{Dymarsky:2020qom, Dymarsky:2020pzc} have resisted a clean holographic interpretation, primarily because the construction that defines lattices from codes cannot yield a large gap in $U(1)^c \times U(1)^c$ primaries. The reason is that lattices defined by codes through Construction A have vectors with length-squared equals $2$, so a code theory always has light primaries in the spectrum of $U(1)^c \times U(1)^c$ primaries, even in the large $c$ limit.\footnote{Though see \cite{Angelinos:2022umf} for an interesting construction of an ensemble of code theories where a large gap is obtained by scaling the size of the code's base field $\F_p \times \F_q$ at the same time as $c$.}

In this paper, we give a new perspective on the ensemble average of classical error-correcting codes by considering the \emph{full} symmetry group of the code theories, which turns out to be (at least) $SU(2)^n$.\footnote{We use $SU(2)^n$ as a shorthand for $SU(2)^c\times SU(2)^{\bar c}$ and always have $n=c+\bar c$. Moreover, such CFTs only exist for $c-\bar c$ divisible by $8$. We shall refer to all such theories as ``$SU(2)$ theories.''} We shall find that a large gap in $SU(2)^n$ primaries \emph{does} appear in the average over all codes, and that this average can be obtained by a sum over all unique modular images of the $SU(2)^n$ vacuum character. In fact, this enhanced symmetry was already known in \cite{Dolan:1994st}, where the original construction of chiral CFTs from type II codes was given. In this paper, we shall generalize the construction by associating Lorentzian lattices to a broader class of codes -- type I -- which define full (non-chiral) CFTs.

The resulting code theories have the advantage of being extremely tractable, even more so than the lattice theories of which they are a subset. This is primarily because their partition functions satisfy a very restrictive ansatz. As a result, for a given code of length $n$ there are a finite number of possible partition functions which are consistent with both modular invariance and positive $SU(2)$ character degeneracies. Imposing higher-genus modular invariance \cite{Henriksson:2021qkt, Henriksson:2022dnu} leads to an even smaller set of possible partition functions. In section~\ref{sec:Symmetry}, we apply this approach to our new construction of non-chiral $SU(2)^n$ theories, and compare to the number of theories obtained directly from the list of codes \cite{Harada2015}. We find that for length $n \leqslant 10$, genus 1 modular invariance fully determines the set of possible $SU(2)^n$ theories to be equal to the number code theories. With the additional constraints of genus 2 modular invariance, the number of $SU(2)$ theories matches the number of code theories for $n \leqslant 14$. In practice we cannot exceed genus 2 before the number of variables becomes too large to handle, but in principle we can extend this algorithm to arbitrary genus to completely fix the set of code theories of a given length $n$. Based on this, we make the following two conjectures about $SU(2)$ theories:

\begin{quote}
 \emph{(a) the set of $SU(2)$ theories is the same as the set of code theories,}
\end{quote}

\begin{quote}
 \emph{(b) the set of $SU(2)$ theories is fixed by higher-genus modular invariance.}
\end{quote}
If our conjectures are true, then the ensemble of length-$n$ code theories is identical to the ensemble of theories with $SU(2)^n$ symmetry, where $n = c + \bar c$. From here on out, we will assume that this is the case, calling these theories ``code theories'' and ``$SU(2)$ theories'' interchangeably.

The connection we discuss between codes and CFTs shares some similarities with the relation between sphere-packings and CFTs \cite{Hartman:2019pcd} -- in fact, it is in some sense directly analogous. The problem of finding 2d CFTs with the biggest possible gap in $U(1)^c \times U(1)^c$ primaries is exactly equivalent to the mathematical problem of finding optimal lattices -- those with the longest possible shortest vector. Lattices define (a subset of all possible) sphere packings, which allowed the authors of \cite{Hartman:2019pcd} to relate the functionals used in the modular bootstrap to the functionals used to derive optimal sphere packings \cite{Viazovska2017,Cohn2017}. In the present case, our results immediately imply that the problem of finding optimal error-correcting codes -- those with the largest possible Hamming distance -- is exactly equivalent to finding the theory with the largest possible gap in $SU(2)^n$ primaries. As a consequence, all bounds on the Hamming distance of self-dual codes can be directly translated into a statement about 2d CFTs. We have summarized the analogy between lattice CFTs and code CFTs in table~\ref{tab:introtable}.

\begin{table}
\centering
{\renewcommand{\arraystretch}{1.75}
\begin{tabular}{l|lll}
Theory & all 2d CFTs & lattice theories & code theories \\ \hline \hline
symmetry  & Virasoro  &  $U(1)^n$ & $SU(2)^n$
\\\hline
characters $\chi_h(q)$ & $ \frac{q^{h - \frac{n - 1}{24}}}{\eta(\tau)}$ & $\frac{q^h}{\eta(\tau)^n}$ & $\left(\frac{\theta_3(q^2)}{\eta(\tau)}\right)^{n-4h} \left(\frac{\theta_2(q^2)}{\eta(\tau)}\right)^{4h} $ \  \\\hline
vacuum character $\chi_0(q)$ & $\chi_{h = 0} - \chi_{h =1}$ & $\chi_{h = 0}$ & $\left(\frac{\theta_3(q^2)}{\eta(\tau)}\right)^n $ \  \\\hline
optimization problem & $\Delta_{\mathrm{gap}}$ & sphere-packing & Hamming distance
\\ \hline
optimal $n = 24$ example & Monster theory & Leech lattice & Golay code \\\hline
\end{tabular}
}
\caption{Summary of the relation between all theories, lattice theories, and code theories. The Leech lattice may be obtained by a $\Z_2$ twist of the Golay code's Construction A lattice, and the Monster CFT may be obtained by a $\Z_2$ orbifold of the Leech lattice CFT. }
\label{tab:introtable}
\end{table}

A central object in this paper is the \emph{enumerator polynomial}, which is a degree-$n$ homogeneous polynomial which roughly counts the weights (number of 1s) in each codeword. Enumerator polynomials can be directly mapped to the CFT's torus partition function by simply plugging in the $SU(2)^n$ characters:
\begin{align}
    x \mapsto \frac{\theta_3(q^2)}{\eta(\tau)} \, , \qquad y \mapsto \frac{\theta_2(q^2)}{\eta(\tau)} \, , \qquad \bar{x} \mapsto \frac{\overline{\theta_3(q^2)}}{ \overline{\eta(\tau)}} \, , \qquad \bar{y} \mapsto \frac{\overline{\theta_2(q^2)}}{\overline{\eta( \tau)}} \, . 
\end{align}
An average enumerator polynomial for type II codes has been known for a long time \cite{Pless1975},
\begin{align}
    W^{\text{II}}_c = \frac{1}{2^{c/2}+ 4} \left[ 2^{c/2} (x^c + y^c) + (x + y)^c + (x-y)^c + (x + iy)^c + (x-iy)^c \right].
    \label{eq:typeIIaverNewIntro}
\end{align}
One of the main results of this paper is the new construction of 2d CFTs from type I codes, given in section~\ref{sec:Symmetry}, and a formula for the average enumerator polynomial of such codes, 
\begin{align}
\nonumber
\overline{W}^{\hat{\text I}}_{c,\bar c}\ &= \ \frac1{2^{(c+\bar c)/2} + 4}\bigg[ 2^{(c+\bar c)/2} (x^c \bar x^{\bar c} + y^c \bar y^{\bar c})  + (x+y)^c(\bar x + \bar y)^{\bar c} +(x-y)^c(\bar x - \bar y)^{\bar c}
    \\
    & \quad \qquad\qquad\qquad+  (x+ i y)^c(\bar x - i \bar y)^{\bar c}+ (x- i y)^c(\bar x +i \bar y)^{\bar c}
    \bigg].
        \label{eq:typeIaverNewIntro}
\end{align}
Here $\hat{\text{I}}$ refers to a certain class of CFTs constructed from type~I codes, to be defined in the main text. 
By ``average,'' we mean the sum over all codes weighted by the number of different ways of writing that code.\footnote{
In this formulation, a code is an equivalence class, where the equivalence relation is given by the permutation group $\mathrm S_n$ that reorders the entries of the codewords. Later we will denote the individual members of the equivalence class as ``ordered codes.'' 
In the average formulas, every ordered code contributes with equal weight. Some elements of $\mathrm S_n$ map an ordered code to itself by permuting the codewords, and is called an automorphism. Taking these automorphisms into account is the reason for the appearance of factors $|\mathrm{Aut(C)}|^{-1}$ in the average formulas in section~\ref{sec:Symmetry}.}
For the type II codes, the average is known \cite{Pless1975} to be given by~\eqref{eq:typeIIaverNewIntro}, and we find that it is equal to the sum over modular images of the vacuum $SU(2)^c$ character. The formula~\eqref{eq:typeIaverNewIntro} was computed using the sum over modular of the vacuum character of $SU(2)^c \times SU(2)^{\bar c}$ symmetry. We conjecture based on the type II case, but have not proven, that it is equal to the average over all codes, weighted by the number of ways of writing that code \emph{that maintain the property of being a type $\hat{{I}}$ code}. We have confirmed this conjecture explicitly by enumerating all codes for $n = c + \bar c$ up to $16$.

Section~\ref{sec:Holography} will be devoted to studying the properties of the ensemble average proposed in~\eqref{eq:typeIaverNewIntro}. Our results can be summarized by the following:
\begin{itemize}
    \item By expanding \eqref{eq:typeIaverNewIntro} in $SU(2)$ characters, we see that there is a gap that for large $n = c + \bar c = 2 c$ grows linearly with $n$. It is given by 
    \begin{align}
        \Delta_{\text{gap}} \ = \ h^{-1}\left(\tfrac{1}{2} \right) \frac{n}{4 } \ \simeq \ \frac{c}{18.177} \, ,
        \label{eq:gap_intro}
    \end{align} 
    where $h(x)$ is the Shannon entropy of the random variable $x$.\footnote{Compare to the gap in $U(1)^c \times U(1)^c$ primaries in the Narain ensemble \cite{Maloney:2020nni, Afkhami-Jeddi:2020ezh}, $\Delta_{\text{gap}} = \tfrac{c}{2 \pi e} \simeq \frac{c}{17.079}$.}
    \item By expanding the formula \eqref{eq:typeIaverNewIntro} in Virasoro characters, we find that the average number of currents approaches the value $3(c+\bar c)$ as $c\to\infty$, implying that theories with further symmetry enhancement becomes increasingly rare, see \eqref{eq:averagecodeexamples}.
    \item Bounds on the Hamming distance imply bounds on the gap in $SU(2)^n$ primaries. 
    \begin{itemize}
        \item The gap in the average partition function~\eqref{eq:gap_intro} implies a \emph{lower} bound on the maximal gap of at least $c / 18.177$.
        \item  The strongest \emph{upper} bound known for asymptotically large self-dual codes is \cite{Krasikov2000, Rains2003}
    \begin{align}
        \Delta_{\text{gap}} \ \lesssim \ \frac{c}{12.075} \, .
        \label{eq:bound_intro}
    \end{align} 
    \end{itemize}
\end{itemize}

Beyond genus 1, we conjecture a formula for the average genus-2 partition function \eqref{eq:genus2average}, again using the sum over modular images starting from the (genus-2) character of the identity multiplet, recently discussed in \cite{Henriksson:2021qkt, Henriksson:2022dnu}. This allows us to compute the expectation value of the product of the partition function on two disconnected tori. We find that in the ensemble average, this product does \emph{not} factorize, which has been interpreted as a signal that bulk wormholes contribute to this path integral \cite{Maldacena:2004rf, Witten:1999xp}. The (normalized) difference between the expectation value of the product and the product of the expectation value behaves like
\begin{align}
    \nu_{\text{wh}}(\tau = i, \tilde \tau = i) \ \sim \ \frac{1}{2} (\sqrt 8 - 2)^{-2c}
\end{align}
for large $c$, indicating that the wormhole contributions are suppressed at large $c$.

In section~\ref{sec:Other} we discuss some open directions, including more general code CFT constructions and higher-level $SU(2)$ symmetry. We leave a number of technical and numerical results for the appendices, including some explicit calculations for chiral theories defined on Euclidean lattices, as well as a discussion on the denominator of the higher-genus partition functions.

\section{Codes and $SU(2)$ theories}
\label{sec:Symmetry}

Error-correcting codes are mathematical objects which use redundancy to protect information against corruption.\footnote{The relation between codes and CFTs has recently been reviewed in \cite{Dymarsky:2020qom, Henriksson:2021qkt}.} A commonly studied class are binary codes, which take the form of maps from binary $k$-vectors to binary $n$-vectors. The classic example is the triple repetition code $i_3$, $(0) \mapsto (000)$, $(1) \mapsto (111)$, for which $k = 1$ and $n = 3$. Error-correction is performed by projecting the received message to the nearest codeword. Therefore the error-correcting capacity of a code is proportional to the distance between its codewords. For example, the code $i_3$ can correct a single error -- if the message $(111)$ is corrupted to $(101)$ for instance, then the receiver will still correctly interpret the message because $(111)$ is the nearest codeword. If there are two errors at different points, the message will be incorrectly interpreted. In this paper, the double repetition code $i_2$, defined by $(0) \mapsto (00)$, $(1) \mapsto (11)$, shall be quite important. It is less commonly considered in the code literature, since this code cannot correct any errors. 

For our purposes, we may think of a code $\mathcal{C}$ as a collection of $n$-vectors -- in the case of $i_3$, it would be $\{(000)^T, \, (111)^T\}$. In this paper, we shall be concerned with \emph{linear} codes, which have the property that the sum (mod 2) of two codewords is always a codeword. Such codes form a linear subspace of the vector space $\F_2^n$, where $\F_2$ is the field with two elements. Now we will introduce some of the main notions needed in the rest of the paper.

\paragraph{Enumerator polynomial} An important quantity characterizing codes, and one that will be central for this paper, is the enumerator polynomial, 
\begin{align}
    W_\mathcal{C} = \sum_{c \in \mathcal {C}} x^{n - w(c)} y^{w(c)} \, .
\end{align}
The enumerator polynomial is defined by the sum over all codewords $c \in \mathcal{C}$. Each codeword has an associated \emph{Hamming weight} $w(c)$, defined as the number of 1s in the codeword, \emph{e.g.} $w(111) = 3$. See appendix~\ref{app:Hamming} for an explicit calculation using the Hamming code as an example.

The enumerator polynomial is a coarse-grained measure of the code. It is clear from its definition that it is a homogeneous polynomial with positive integer coefficients, and that the sum of all coefficients is equal to the number of codewords. 

We will also be interested in studying code CFTs at higher genus, which will involve the weight-$g$ enumerator polynomial, defined as a sum over all $g$-tuples of codewords,
\begin{equation}
    W^{(g)}=\sum_{\mathbf M\in \mathcal C^g}\prod_{i=1}^nx_{\mathrm{row}_i(\mathbf M)}\,,
\end{equation}
where the matrix $\mathbf M$ has the $g$ codewords as columns. It is a polynomial in $2^g$ variables $x_{0\cdots00}$, $x_{0\cdots01}$, \ldots, $x_{1\cdots11}$. The ordinary (weight 1) formula follows by noting that $\prod_{i=1}^nx_{c_i}=x_0^{n-w(c)}x_1^{w(c)}$ and renaming $(x_0,x_1)$ as $(x,y)$. The weight-2 enumerator polynomial involves the sum over pairs of codewords and is given by
\begin{align}
    W^{(2)}_{\mathcal{C}}(x_0,x_1,x_2,x_3) = \sum_{c_1, \, c_2 \, \in \mathcal C} x_0^{n + c_1 \cdot c_2 - w(c_1) - w(c_2)} x_1^{w(c_2) - c_1 \cdot c_2} x_2^{w(c_1) - c_1 \cdot c_2}  x_3^{c_1\cdot c_2} \, ,
    \label{eq:g2_EP}
\end{align}
where $x_{00}=x_0$, $x_{01}=x_1$, $x_{10}=x_2$, $x_{11}=x_3$.

\paragraph{Type I and Type II codes} 

A classical code is called \emph{even} if the Hamming weight of all codewords are divisible by 2, and \emph{doubly-even} if all codewords are divisible by 4. A code is called \emph{self-dual} if the code is equal to its dual, defined by 
\begin{align}
    \mathcal{C}^* = \left\{ c' \in \F_2^n \ | \ c' \cdot c = 0 \ (\mathrm{mod} \, 2) \ \mathrm{for \ all} \ c \in \mathcal{C} \right\}.
\end{align}
Importantly, both of these properties can be checked from the enumerator polynomial alone. Even and self-dual codes have enumerator polynomials which are invariant under the transformations
\begin{align}
\begin{split}
    \text{even:}& \qquad y\ \mapsto\ -y \,,\\
    \text{self-dual:}& \qquad x \ \mapsto \ \frac{x + y}{\sqrt{2}} \, , \quad y\ \mapsto\ \frac{x - y}{\sqrt{2}}
    \,.
    \label{eq:evenSD} 
\end{split}
\end{align}
Codes which are even and self-dual are said to be type I. 

Doubly-even and self-dual codes have enumerator polynomials which are invariant under the transformations
\begin{align}
\begin{split}
    \text{doubly-even:}& \qquad y\ \mapsto\ i y ,\\
    \text{self-dual:}& \qquad x\ \mapsto\ \frac{x + y}{\sqrt{2}} \, , \quad y \ \mapsto\ \frac{x - y}{\sqrt{2}}
    .
    \label{eq:dubevenSD} 
\end{split}
\end{align}
Such codes are said to be type II. Conventions differ, but in this work, we think of type II as a subset of type I.

\paragraph{Construction~A} Each codeword may be thought of as a vector in $\F_2^n\cong \mathbb{Z}_2^n$, which we then embed into $\mathbb{Z}^n$. Because codewords are binary vectors, they sit at the vertices of the unit cube. This leads to a simple way to identify lattices from codes: you simply think of the codewords as the unit cell, and the lattice is formed from all possible integer sums of codewords plus the unit vectors $2 e_i$, where $e_1 = (1,0,0,\ldots)$, etc. It is also necessary to scale by $\sqrt{2}$ so that the lattice is self-dual for self-dual codes. The result is the formula
\begin{align}
\Lambda(\mathcal{C}) = \left\{ 
v / \sqrt{2} \ \middle| \ v \in \mathbb{Z}^n, \ v \equiv c \, \text{ (mod 2)}, \ c \in \mathcal{C} \ \right\}
\label{eq:defConAlattice}
\end{align}
This is the Construction~A of Leech and Sloane \cite{LeechSloane1971}, which underpins the relationship between sphere-packing and error-correcting codes. It allows for a direct relationship between many quantities arising from lattices and quantities arising from codes. A central example is the lattice theta function,
\begin{align}
    \Theta_{\Lambda}(q) = \sum_{v \in \Lambda} q^{v^2 / 2} \, ,
\end{align}
which can be computed directly from the code enumerator polynomial by\footnote{
Recall the definition of the Jacobi theta functions, 
\begin{align*}
    \theta_2(q) = \sum_{m = - \infty}^\infty q^\frac{(m+1/2)^2}{2} \, , \qquad \theta_3(q) = \sum_{m = - \infty}^\infty q^\frac{m^2}{2} \, .
\end{align*}}
\begin{align}
    \Theta_{\Lambda(\mathcal{C})}(q) = W_\mathcal{C} \left( \theta_3(q^2), \theta_2(q^2) \right) \, .
    \label{eq:TH-EPrelation}
\end{align}
The proof of \eqref{eq:TH-EPrelation} is reviewed in \cite{Henriksson:2021qkt}. It tells us that the lattice theta function, a sum over lattice vectors, may be computed from the enumerator polynomial, a sum over codewords. 

Another important result is that Construction~A maps self-dual codes to self-dual lattices. Doubly-even codes get mapped to even lattices (thanks to the $\sqrt{2}$). These are crucial properties because theories defined by compactifying free bosons on a lattice only form CFTs if the lattice is even and self-dual. General lattices will not lead to modular invariant partition functions.

\subsection{Chiral $SU(2)$ theories from Type II codes}

Throughout this paper, we shall use the term ``$SU(2)$ theories'' to refer to 2d CFTs with enhanced $SU(2)^c$ global symmetry, or $SU(2)^c \times SU(2)^{\bar{c}}$ in the non-chiral case. The purpose of this section will be to review the construction of chiral 2d CFTs from Type II error-correcting codes, and then to show that all such theories have $SU(2)^c$ symmetry. Therefore all chiral code theories are chiral $SU(2)$ theories.\footnote{It would be interesting to prove that \emph{all} $SU(2)$ theories arise from codes, but we leave that to future work.} Both of these facts were demonstrated in the original papers on code CFTs \cite{Dolan:1989kf, Dolan:1994st}. It is useful to review them here in preparation for the following subsection, where we shall give a new construction of non-chiral $SU(2)$ theories arising from type I codes. 

\subsubsection{CFTs from codes}

The key to constructing CFTs from error-correcting codes is Construction~A, which associates a lattice $\Lambda(\mathcal{C})$ to a code $\mathcal{C}$ using the formula~\eqref{eq:defConAlattice}. For a type II code of length $n$, this leads to a Euclidean lattice $\Lambda(\mathcal{C})$. Such a lattice defines a 2d chiral CFT whose momenta take values in the lattice \cite{Dolan:1989vr, Dolan:1994st}.

For an $n$-dimensional  Euclidean lattice $\Lambda$, we consider the theory whose states include the momentum states $|\lambda \rangle$ for $\lambda \in \Lambda$, as well as the descendant states from acting with oscillators $a^i_m$, $m \in \Z$, $1 \leqslant i \leqslant n$. This can be constructed from the chiral half of a boson on a Narain lattice at the points in moduli space where the theory holomorphically factorizes into a chiral and anti-chiral part.
%
%
Such a theory possesses a $ U(1)^n$ symmetry which acts by shifting $\varphi^i\to \varphi^i+\ell^i$. The primaries with respect to this symmetry are the vertex operators $V_\lambda$,  $\lambda \in \Lambda$, defined by 
\begin{align}
    V_\lambda(z) = e^{i \lambda_i \varphi_i(z)}.
\end{align}
The zero vector $\vec 0\in\Lambda$ corresponds to the $ U(1)^n$ vacuum multiplet, which contains the generators $J^i = i \partial \varphi_i $, in addition to the Virasoro vacuum.

The enhancement to $SU(2)^n$ symmetry arises because Construction~A lattices, by the definition in~\eqref{eq:defConAlattice}, include vectors $e_1 = \sqrt{2} (1, 0, 0,\ldots)$, $e_2 = \sqrt{2} (0, 1, 0,\ldots)$, etc. These are exactly the vectors which are equivalent (mod 2) to the 0-vector, so they are included in every Construction~A lattice. Denote the vertex operators arising from such lattice vectors by $E_i^\pm$, so that
\begin{align}
    E_i^\pm = e^{\pm i e_i \cdot \varphi_i} = e^{\pm i \sqrt 2 \varphi_i}.
\end{align}
Then we can check the OPEs
\begin{align}
    E_i^+(z) E_j^-(w) \sim  (z - w)^{-e_i \cdot e_j} e^{i \sqrt 2 (\varphi_i(z)- \varphi_j(w))} \ &= \ \frac{\delta_{ij}}{(z - w)^2} +  \frac{ \sqrt 2\delta_{ij}}{z - w} J_j(w) + \ldots, \\
    J_i(z) E_j^\pm(w) \sim \frac{\pm  \sqrt{2} \delta_{ij} }{z - w }  E_j^\pm(w) .
\end{align}
This is the OPE of $n$ independent level-1 $SU(2)$ Kac--Moody symmetries, meaning that the symmetry algebra contains $n$ independent copies of $\mathfrak{su}(2)$. This is precisely the conclusion of Lemma 6.6 of \cite{Dolan:1994st}. Note that in many cases we find additional enhanced symmetry, such as for the Hamming code, which gives rise to a theory with symmetry algebra $E_8$ (see appendix~\ref{app:Hamming}).

\subsubsection{$SU(2)$ characters}

The partition function for an $n$-dimensional chiral lattice theory takes the form
\begin{align}
    Z_\Lambda(\tau) = \frac{\Theta_\Lambda(q)}{\eta(\tau)^n} \, ,
\end{align}
where the Dedekind $\eta$ function is defined by
\begin{equation}
    \label{eq:DedekindEta}
    \eta(\tau)=q^{\frac1{24}}\prod_{m=1}^\infty(1-q^m), \qquad q=e^{2\pi i \tau}.
\end{equation}
Therefore the relationship between the code enumerator polynomial and the lattice theta function~\eqref{eq:TH-EPrelation} immediately implies
\begin{align}
    Z(q) &= W_\mathcal{C}(x, y),
\end{align}
where $x$ and $y$ are defined by
\begin{align}
    x = \frac{\theta_3(q^2)}{\eta(\tau)}\, , \qquad \qquad  y = \frac{\theta_2(q^2)}{\eta(\tau)} \, .
\end{align}
In fact, $x$ and $y$ are nothing other than the characters of the extended $SU(2)$ symmetry. This can be demonstrated exactly by the formula for the $\mathfrak{su}(2)$ characters (see for instance  \cite{DiFrancesco:1997nk})
\begin{align}
    \chi_{\lambda}^{\mathfrak{su}(2)}(\tau) = q^{(\lambda + 1)^2/4(k+2)-1/8} \frac{  \sum_{n \in \mathbb{Z}} (\lambda+1+2n(k+2))q^{n(\lambda+1+n(k+2))} }{\sum_{n \in \mathbb{Z}}(1 + 4n )q^{n(1+2n)}} \, ,
    \label{eq:su2chars1}
\end{align}
where $k$ is the level of the symmetry and $\lambda$ is the weight. Choosing $k = 1$ \footnote{At present, we do not know of any code interpretation of the higher-level theories 
-- see section \ref{sec:Other} for a brief discussion.} yields
\begin{align}
    \chi_0^{\mathfrak{su}(2)}(\tau) =   \frac{\theta_3(q^2)}{\eta(q)}\, , \qquad \chi_1^{\mathfrak{su}(2)}(\tau)  = \frac{\theta_2(q^2)}{\eta(q)} \, .
    \label{eq:su2chars2}
\end{align}
The modular transformations for these characters are exactly the invariances of the type II codes:
\begin{align}
    S:& \qquad x \to \frac{x + y}{\sqrt{2}}\, , & &y \to \frac{x - y}{\sqrt{2}}\, ,\\
    T:& \qquad x \to e^{-\frac{2 \pi i}{24}} x\, ,   &  &y \to i e^{-\frac{2 \pi i}{24}} y \,;
    \label{eq:genus1modular}
\end{align}
compare with \eqref{eq:dubevenSD}.

The $24^{\text{th}}$ roots of unity that appear essentially arise from the $\eta$ functions in the denominator. They are a manifestation of the fact that in the chiral case, modular invariant genus-1 partition functions are only possible when $c$ is divisible by 24. The characters \eqref{eq:su2chars2} are the natural building blocks for $ SU(2)^c \times SU(2)^{\bar{c}} $ theories. Partition functions for such theories will be homogeneous polynomials of degree $c$ in the holomorphic characters $x$ and $y$, and degree $\bar{c}$ in the antiholomorphic ones, which we will call $\bar x$ and $\bar y$.

\paragraph{Higher-genus characters} Higher-genus partition functions depend on the period matrix $\Omega$, which generalizes the modular parameter $\tau$ and is a symmetric $g\times g$ matrix with positive-definite imaginary part.
The characters at higher-genus are created from a generalization of the Jacobi theta functions to higher genus. These are defined by 
\begin{equation}
\label{eq:highergenustheta}
    \theta\begin{bmatrix}\vek a\\\vek b\end{bmatrix}(\vek z,\Omega)=\sum_{\vek m\in\Z^g}\exp\left( \pi i(\vek m+\vek a)\cdot\Omega\cdot(\vek m+\vek a)+2\pi i(\vek m+\vek a)\cdot(\vek z+\vek b)\right).
\end{equation}
Here $\vek a$ and $\vek b$ are length-$g$ vectors whose entries are either $0$ or $1/2$. The $SU(2)$ characters are given by theta-functions where $\vek b = 0$, and evaluated at argument $2 \Omega$. We define these by
\begin{align}
\label{eq:thetaconstantsdefn}
    \vartheta_{\vek a}(\Omega) := \thetaUD{\vek a/2}{\vek 0}(0,2 \Omega) \, ,
\end{align}
they are sometimes known as theta constants of second-order characteristics.
These characters were shown to form the higher-genus partition functions of code CFTs in \cite{Henriksson:2021qkt}, and they were discussed in the context of the $c =1$ $SU(2)$ theory in \cite{Dijkgraaf:1987vp}.

The higher-genus lattice theta function is obtained directly from the higher weight enumerator polynomial through the formula
\begin{align}
    \Theta_{\Lambda}(\Omega) = W^{(g)}_{\mathcal{C}}\left( \vartheta_{0\ldots00}(\Omega), \,   \vartheta_{0\ldots01}(\Omega) \, , \ldots ,  \vartheta_{1\ldots11}(\Omega) \right).
\end{align}
Therefore we see that these characters comprise the numerator of the genus-$g$ partition function via
\begin{align}
    Z^{(g)}_\Lambda(\Omega) = \frac{ \Theta_{\Lambda}(\Omega)}{\Phi_g^c} \, ,
\end{align}
where $\Phi_g$ is a universal term taking into account oscillator modes. It has the appropriate weight to cancel the modular transformations of the numerator, up to phases, and is not relevant for the rest of this work. For completeness, we give some more details on $\Phi_g$ in appendix~\ref{app:denominator}.

\subsection{$SU(2)$ theories from codes: general construction}

The purpose of this section is to construct $SU(2)$-invariant theories from error-correcting codes of type I. In general this will give rise to theories with\footnote{This requirement follows from evaluating the genus-1 characters at $\tau=i$ and demanding invariance under the (trivial) modular transformation $(ST)^3$}
\begin{equation}
    c-\bar c \,\equiv\, 0\ \text{(mod $8$)}.
\end{equation}
In this construction, which we call the \emph{Lorentzian Construction~A},
it is important to think of the ordering of codes as collections of \emph{ordered vectors}. This is in contrast with previous constructions of CFTs from codes, where CFTs can be unambiguously assigned to equivalence classes of ordered codes, under the $\mathrm S_n$ permutation group of the entries of a length $n$ vector.\footnote{For instance, there are 30 different ordered codes in the equivalence class of the Hamming code $H$. These 30 ordered codes are perturbed by elements of the quotient $\mathrm S_8/\mathrm{Aut}(H)$, where $\mathrm{Aut}(H)$ are the permutations that leave a given ordered code invariant by permuting the ordered codewords, $|\mathrm{Aut}(H)|=1344$. Under the usual Construction A, the 30 ordered codes in the Hamming class all give rise to the same CFT ($E_8$), while, as will follow from the definitions below, only 6 of these 30 ordered codes suffice to build CFTs with $c=\bar c=4$ (table~\ref{tab:typeIaver}).} Denoting an ordered code by $\hat{\mathcal C}$, the common notion of a code is then an equivalence class $\mathcal C$,
\begin{equation}
    \mathcal C=[\hat{\mathcal C}].
\end{equation}
To give a concrete example, consider the equivalence class $\mathcal C=i_2^2$. It contains three ordered codes, with generator matrices\footnote{For each ordered code, the set of $2^2$ codewords is given by multiplying $G$ with all the length-two binary vectors $\left(\begin{smallmatrix}0\\0\end{smallmatrix}\right)$, $\left(\begin{smallmatrix}0\\1\end{smallmatrix}\right)$, $\left(\begin{smallmatrix}1\\0\end{smallmatrix}\right)$, and $\left(\begin{smallmatrix}1\\1\end{smallmatrix}\right)$.}
\begin{equation}
    G^T=\begin{pmatrix}
        1&1&0&0
        \\
        0&0&1&1
    \end{pmatrix}
    ,\qquad 
        G^T=\begin{pmatrix}
        1&0&1&0
        \\
        0&1&0&1
    \end{pmatrix}
    ,\qquad 
        G^T=\begin{pmatrix}
        1&0&0&1
        \\
    0&1&1&0
    \end{pmatrix}\,.
    \label{eq:threegeneratormatrices}
\end{equation}
The first is the standard expression found in for instance in the database of by Harada and Munemasa \cite{Harada2015}, and the other two are given by swapping entries $2\longleftrightarrow3$ and $2\longleftrightarrow4$ respectively.

The strategy of the Lorentzian Construction A is to use an ordered code to define a signature $(n,\bar n)=(c,\bar c)$ Lorentzian lattice. This lattice will be the momentum lattice of the corresponding $SU(2)$ theory. Considering the metric $\mathrm{diag}(+,+,\ldots,+|-,-,\ldots,-)$, write each codeword as $c=(c_+|c_-)$ and define the weights
\begin{align}
\label{eq:wpwmDef}
\begin{split}
    w_+(c)\ &=\ w(c_+)\ =\ c\cdot(1,1,\ldots,1|0,0,\ldots,0), \\
    w_-(c)\ &=\ w(c_-)\ = \ c\cdot(0,0,\ldots,0|1,1,\ldots,1) \, .
    \end{split}\end{align}
Then we define the class of (ordered) codes, 
\begin{equation}
\text{type $\text I^{n,\bar n}$ }= \left\{\hat{\mathcal C}\ \middle|\ \text{$\hat{\mathcal C}$ is type I of length $n+\bar n$};\ \forall c\in\hat{\mathcal C}, w_+(c)-w_-(c)\equiv 0\ (\mathrm{mod}\ 4) \right\}\,.
\label{eq:defIhat}
\end{equation}
For instance, out of the three ordered codes in \eqref{eq:threegeneratormatrices}, only the last two satisfy the constraint to be type $\text I^{2,2}$ codes.

Our new type $\text I^{n,\bar n}$ codes can be characterized by the enumerator polynomial,
\begin{equation}
    \hat W(x,\bar x,y,\bar y)=\sum_{c\in \hat{\mathcal C}}x^{n-w_+(c)}\bar x^{\bar n-w_-(c)}y^{w_+(c)}\bar y^{w_-(c)}\,.
\end{equation}
Note that the usual (Hamming) enumerator polynomial is recovered upon setting $\bar x=x$ and $\bar y=y$. We shall also commonly use type $\hat{\text{I}}$ to refer to ordered codes where $n = \bar n$. 

\paragraph{Lorentzian Construction A}
We now describe the Lorentzian Construction A that gives rise to CFTs with central charges $(c,\bar c)$ and $SU(2)^c\times SU(2)^{\bar c}$ symmetry. The construction is a direct generalization of the original Construction A, where the primary operators (under the $U(1)^c\times U(1)^{\bar c}$ subgroup) are given by points on the Lorentzian lattice defined by the code,
\begin{equation}
    \Lambda = \left\{
    \frac v{\sqrt2}\middle| v\equiv c \ (\text{mod $2$}), \, c\in \hat{\mathcal C}
    \right\}
\end{equation}
The code lattice is the momentum lattice of the conformal field theory. 
The condition that $w_+(c)-w_-(c)\equiv 0\ (\text{mod $4$})$ implies that the Lorentzian lattice is even. It is important to note that we cannot uniquely assign a CFT to a given type~I code, since there are different orderings of the entries of the codewords.\footnote{Ordered codes related by separate permutations of the $c$ ``plus'' entries and the $\bar c$ ``minus'' entries give rise to the same CFT.}

The genus-1 partition function is given by the substitutions
\begin{equation}
    x\mapsto \theta_3(q^2), \quad
    \bar x\mapsto \overline{\theta_3(q^2)}, \quad
    y\mapsto \theta_2(q^2), \quad
    \bar y\mapsto \overline{\theta_2(q^2)}, 
\end{equation}
and then dividing with the suitable denominator,
\begin{equation}
    Z=\frac{W
    \left(\theta_3(q^2), \overline{\theta_3(q^2)},  \theta_2(q^2), \overline{\theta_2(q^2)} \right)}{\eta(\tau)^c\overline{\eta(\tau)}^{\bar c}}\,.
\end{equation}
Equivalently, since $W$ is a homogeneous degree $(c,\bar c)$ polynomial we may write
\begin{equation}
    Z=W\left(\frac{\theta_3(q^2)}{\eta(\tau)}, \frac{\overline{\theta_3(q^2)}}{\overline{\eta(\tau)}},\frac{  \theta_2(q^2)}{\eta(\tau)},\frac{ \overline{\theta_2(q^2)}}{\overline{\eta(\tau)}}\right)\,.
\end{equation}

For $c=\bar c$, the resulting theories are Narain CFTs \cite{Narain:1985jj,Narain:1986am} -- specifically they lie at special points in Narain moduli space with symmetry enhancement to $SU(2)^c\times SU(2)^{ c}$. 
This can be seen from the fact that the momentum lattice $\Lambda$ is even and self-dual. 

The genus-$g$ partition functions for $SU(2)$ theories are related to weight-$g$ enumerator polynomials (see \cite{Henriksson:2021qkt} for a description in the case of chiral theories). The weight-$g$ enumerator polynomial is given by
\begin{equation}
    \hat W^{(g)}(x_i,\bar x_i)=\sum_{\mathbf M\in\hat{\mathcal C}^g}\prod_{i=1}^cx_{\mathrm{row}_i(\mathbf M)}\prod_{i=c+1}^{c+\bar c}\bar x_{\mathrm{row}_i(\mathbf M)}
    \,.
\end{equation}
The higher-genus partition function will be given by
\begin{equation}
    Z^{(g)}(\Omega)=\frac{W^{(g)}(\vartheta_i(2\Omega), \overline{\vartheta _i(2\Omega)})}{\Phi_g^c\overline{\Phi}_g^{\bar c}}
    \label{eq:genusgPF}
      \,.
\end{equation}
Like above, the denominator appearing in~\eqref{eq:genusgPF} is analogous to the powers of $\eta(\tau)$ appearing in the genus 1 denominator -- see appendix~\ref{app:denominator} for more details.

\subsection{Higher genus bootstrap for $SU(2)$ theories}

The theories described in the previous subsection have two key properties: they have a finite number of characters (of the $SU(2)^c \times SU(2)^{\bar c}$ symmetry) and these characters transform in a very simple way under the modular group \eqref{eq:genus1modular}. This will allow us to find the most general possible partition functions which are compatible with this symmetry and modular invariance. Once we require that all character degeneracies are positive integers, we shall find a finite number of solutions, and furthermore we shall find that imposing higher-genus modular invariance reduces this number of possible solutions.\footnote{For simplicity, in this section we consider theories with $c=\bar c$. However, see appendix~\ref{app:cc12theory} for an example where $c \neq \bar c$.} The idea of bootstrapping code theories by considering invariant polynomials was introduced in \cite{Dymarsky:2020qom} and higher-genus constraints were incorporated in \cite{Henriksson:2021qkt, Henriksson:2022dnu}.

\subsubsection{Invariant polynomials}

The characters for $SU(2)$ symmetry are simply $x$ and $y$. Since all the $SU(2)$ factors commute, and thus act independently, this means that the characters for $SU(2)^n$ are $n^\text{th}$ order monomials of $x$ and $y$. 
Here we will be concerned with the non-chiral case, where we have both left-handed characters $x$, $y$ as above, and right-handed characters $\bar x = \overline{ \chi_{0} ( \tau)} =\overline{ \theta_{ 3}(q^2)} /\overline{ \eta( \tau)}$ and  $\bar y = \overline{\chi_{1} (\tau)} = \overline{\theta_{2}(q^2)} /\overline{ \eta( \tau)}$, which transform with an opposite sign under the $T$ modular transformation. The result is that 
\begin{align}
    S:& \qquad x \mapsto \frac{x + y}{\sqrt{2}},   & &y \mapsto \frac{x - y}{\sqrt{2}}\, , & &\bar x \mapsto \frac{ \bar x + \bar y}{\sqrt{2}},   & &\bar y \mapsto \frac{\bar x - \bar y}{\sqrt{2}}\, ,
    \label{eq:g1Strans}
    \\
    T:& \qquad x \mapsto x,    & &y \mapsto i y\, , & &\bar x \mapsto \bar x,    & &\bar y \mapsto -i \bar y . 
    \label{eq:g1Ttrans}
\end{align}

Now we would like to ``bootstrap'' these theories -- that is, we will find all possible partition functions which are consistent with modular invariance. First of all, each partition function must be a homogeneous polynomial where each term is degree-$c$ in $x$s and $y$s, as well as degree-$\bar c$ in $\bar x$s and $\bar y$s. The vacuum is $x^{c} \bar x^{ c}$, so its degeneracy must be one. The other terms will, for now, have undetermined coefficients (though in a moment we will impose that they are positive integers). So the most general polynomial is given by
\begin{align}
    P^c_{\text{gen}} = x^{c} \bar x^{c} + a_{c-1, c} x^{c-1} y \bar x^{c} + a_{c, c-1} x^{c}  \bar x^{c-1} \bar y + \ldots \,.
\end{align}
After writing the most general polynomial, we impose modular invariance, which results in relations that reduce the number of linearly independent coefficients $a_{i,j}$. This leads to the most general modular invariant polynomial, which we call $P^c_{\text{inv}}$ Finally, we impose that all $a_{i, j}$ are positive integers, and after this step we find a finite number of solutions. 

Let us illustrate this with a few examples. The simplest is $c = 1$, where we have
\begin{align}
    P^1_{\text{gen}} = x \bar x + a_{1,0} x \bar y + a_{0,1} y \bar x + a_{0,0} y \bar y \, .
\end{align}
Solving invariance under~\eqref{eq:g1Strans} and~\eqref{eq:g1Ttrans} over non-negative integers gives the unique solution
\begin{align}
    P^2_{\text{inv}} = x \bar x + y \bar y \, .
\end{align}
This is consistent with the fact that there is a single length-$2$ type I code -- the repetition code $i_2$, and its enumerator polynomial is $x^2 + y^2$. Integer powers of $P^1_{\text{inv}}$ will always satisfy our constraints and provide a sequence of allowed partition functions. These correspond to the theories that arise from direct sums of the repetition code $i_2$. For $c = 2$ and $c = 3$, such theories are the only solutions that exist. 

For $c = \bar c= 4$ the situation is a little more interesting. By writing $ P^4_{\text{gen}}$ and demanding modular invariance, we find
\begin{align}
\begin{split}
    P^4_{\text{inv}} &= x^4 \bar x^4 + (4 - 4 a_{0,4}) x^3 y \bar x^3 \bar y + (4 - 4 a_{0,4}) x y^3 \bar x \bar y^3 + a_{0,4} x^4 \bar y^4 \\
    & \qquad + a_{0,4} \bar x^4 y^4 + (6 + 6 a_{0,4}) x^2 \bar x^2 y^2 \bar y^2 + y^4 \bar y^4 \, . 
    \end{split}
\end{align}
It is apparent that $ P^4_{\text{inv}} $ will only have positive integer coefficients in two cases: $a_{0,4} = 0$, which leads to 
\begin{align}
    P^4_{\text{inv}} |_{a_{0,4} = 0} = \left( P^1_{\text{inv}}\right)^4,
\end{align}
and $a_{0,4} = 1$, in which case
\begin{align}
    P^4_{\text{inv}} |_{a_{0,4} = 1} = x^4 \bar x^4 + x^4 \bar y^4 + 12 x^2 \bar x^2 y^2 \bar y^2 +  \bar x^4 y^4  + y^4 \bar y^4 
\end{align}
These two options arise from the two length-$8$ type I codes: the fourth power of the repetition code, and the Hamming code. In fact, the theory that arises from the Hamming code using the Lorentzian Construction A with $c=\bar c=4$ is the unique CFT with these central charges and $SO(8)$ symmetry. More details of this particular example are given in appendix~\ref{app:Hamming}.

\subsubsection{Higher genus constraints}

As we push our results to larger central charge, we eventually find that the number of theories that arise from Lorentzian Construction A is fewer than the number of invariant polynomials with positive integer coefficients. The lowest $c$ where this happens is $c = 6$, where Lorentzian Construction A gives only 4 theories but we find 16 modular invariant partition functions (see table \ref{tab:results}). The existence of such ``fake polynomials,'' which appear to be perfectly valid partition functions (for instance, they have positive integer degeneracies of Virasoro primaries as well) was pointed out in \cite{Dymarsky:2020qom}. 

It is not surprising that modular invariance alone might not entirely fix the set of theories, as it only encodes information about the spectrum of the theory. More information, such as averaged OPE coefficients, appears in the higher-genus modular invariance. For the $SU(2)$ theories of interest, higher-genus modular invariance will reduce the number of fake polynomials. The procedure to do this, for example at genus 2, is
\begin{enumerate}
    \item Write the most general possible polynomial $Q^c_{\text{gen}}$ which is homogeneous in the genus 2 characters given in~\eqref{eq:thetaconstantsdefn}.
    \item Impose modular invariance to find $Q^c_{\text{inv}}$.
    \item List all solutions with positive integer coefficients.
    \item Remove all solutions which do not factorize into two genus 1 partition functions in the limit where the genus 2 surface degenerates into two genus 1 surfaces.
    \item Consider the list factorizing genus 1 partition functions that arise this way. Any ``valid'' genus 1 partition function not on this list is in fact ``fake.''
\end{enumerate}

The logic and technical details behind these steps were developed recently for the case of chiral $SU(2)^n$ theories in \cite{Henriksson:2021qkt}, so we will not review them here. The result of this procedure is a shorter list of genus 1 partition functions. For $c \geqslant 6$, there will be some partition functions which are valid for genus 1 but do not arise via factorization from genus 2 partition functions. It is theoretically possible that such ``fake partition functions'' do actually correspond to actual CFTs whose genus 1 partition functions happen to arrange into $SU(2)$ characters, though this seems unlikely. What we know for sure is the following: such partition functions cannot arise from $SU(2)$ theories because these must have higher-genus partition functions with positive integer numbers of $SU(2)$ characters and which have the correct factorization properties.

\subsubsection{Results}

In table~\ref{tab:results} we list the number of solutions compatible with genus 1 constraints, and the subset of those solutions compatible with genus 2 constraints, along with the number of type I codes and theories from our new Lorentzian Construction~A. For convenience, we also give the number of currents (Virasoro primaries with $\Delta = J = 1$) of each of the theories.

\begin{table}
    \centering
    \label{tab:cn-bootstrap-results}
    \begin{tabular}{|c|c|c|c|c|>{\raggedright\arraybackslash}p{40mm}|}
    \hline
      $c$   & codes &   $g=1$ & $g=2$ & theories & $N_\text{currents}$ \\\hline
      $1$ & $1$ & $1$ & $1$ &  $1$ & $3$
      \\
      $2$ & $1$ & $1$ & $1$ &  $1$ & $6$
      \\
      $3$ & $1$ & $1$ & $1$ &  $1$ & $9$
      \\
      $4$ & $2$ & $2$ & $2$ &  $2$ & $12$, $28$
      \\
      $5$ & $2$ & $2$  & $2$  &  $2$ & $15$, $31$
      \\
      $6$ & $3$ & $16$ & $4$ &  $4$ & $\ncu2{18}$, $34$, $66$
      \\
     $7$ & $4$ & $16$ & $6$ &  $6$ & $\ncu2{21}$, $\ncu2{ 37}$, $69$, $131$
      \\
      $8$ & $7$ & $58$ & $24$ &  $13$ & $\ncu4{24}$, $\ncu3{40}$, $56$, $\ncu2{72}$, $120$, $136$, $248$
      \\
      $9$ & $9$ & $1030$  & ?  &  $22$ & $\ncu6{27}$, $\ncu5{ 43}$, $\ncu2{59}$, $\ncu4{75}$, $\ncu2{123}$, $\ncu2{ 139}$, $251$
      \\
      $10$ & $16$ & $21441$ &?  &  $56$ & $\ncu{13}{30}$, $\ncu{14}{46}$, $\ncu5{62}$, $\ncu9{78}$, $\ncu{3}{94}$, $\ncu4{126}$, $\ncu4{142}$, $\ncu2{190}$, $\ncu2{254}$
      \\\hline
    \end{tabular}
    \caption{Results from the bootstrap of full theories with $SU(2)^c\times SU(2)^c$ symmetry. The ``?'' are simply beyond the computing power on our laptops. Likewise, we have implemented the genus 3 constraints but the number of variables is too large to use it beyond $c = 3$. }
    \label{tab:results}
\end{table}

We see that genus 1 modular invariance completely specifies the set of code of length up to $c = 5$, beyond which we find ``fake partition functions'' which satisfy modular invariance but do not derive from codes. The genus 2 constraints reduce the number of fake partition functions, and in fact completely fix the set of codes for length $c = 6$ and $c = 7$ as well. For the case of type II codes, it is known that the weight-$(c/2-1)$ enumerator polynomial fully specifies the code \cite{Runge1996}, meaning that this algorithm must terminate and eventually uniquely fix the actual set of codes if performed to high enough genus. However this ``critical genus'' increases with $c$, potentially without bound.  It would be interesting to understand the actual value of the critical genus in general, and if a similar result exists for type I codes.

\subsection{Averaging over codes}
\label{sec:codeaverage}

The doubly-even self-dual codes discussed above have been classified up to $n = 40$ \cite{Betsumiya2012}, where there are known to be 94,343 codes
. Type I codes have been classified up to $n= 38$, where there are 38,682,183 codes \cite{Bouyuklieva2012}. Remarkably there exist simple formulas for computing the average over both type I and type II theories that apply for any $n$. The average code enumerator polynomial is conventionally defined by
\begin{align}
\label{eq:average-gen}
    \sum_{\mathcal C}\frac{1}{|\mathrm{Aut}(\mathcal{C})|}W_{\mathcal{C}}(x,y)= \frac{T_n}{n!} \overline{W}_n(x, y) \, .
\end{align}
Here the average is weighted by the number of automorphisms of the code, denoted $|\mathrm{Aut}(\mathcal{C})|$. The factor $T_n$ is simply the number of ordered codes -- that is, the total number of codes before permutation symmetries reduce them to equivalence classes. These are given by
\begin{align}
    T_n^\text{I} \ = \ \frac{1}{2} \prod_{j = 0}^{n/2-1} (2^j + 1) \, , \qquad T_n^{\text{II}} \ = \ \prod_{j = 0}^{n/2-2} (2^j + 1),
\end{align}
and the averaged enumerator polynomials are given by
\cite{Pless1975}
\begin{align}    \label{eq:typeIaver}
    \overline{W}^{\text{I}}_n \ &= \ \frac{2^{n/2} (x^n + y^n) + (x+y)^n + (x-y)^n }{2^{n/2} + 2}\,, \\
    \overline{W}^{\text{II}}_n \ &= \  \frac{2^{n/2} (x^n + y^n) + (x+y)^n + (x-y)^n + (x+ i y)^n + (x- iy)^n}{2^{n/2} + 4}\,.
    \label{eq:typeIIaver}
\end{align}
When constructing $c,\bar c$ theories from type $\text I^{c,\bar{c}}$ ordered codes, another 
average appears which is not present in the code literature. This is the average over all $SU(2)^c \times SU(2)^{\bar c}$ theories:
\begin{align}
    \overline{W}^{\hat{\text I}}_{c,\bar c} \ &= \ \frac1{2^{(c+\bar c)/2} + 4}\bigg[ 2^{(c+\bar c)/2} (x^c \bar x^{\bar c} + y^c \bar y^{\bar c})  + (x+y)^c(\bar x + \bar y)^{\bar c} +(x-y)^c(\bar x - \bar y)^{\bar c}
    \nonumber\\
    & \quad \qquad\qquad\qquad+  (x+ i y)^c(\bar x - i \bar y)^{\bar c}+ (x- i y)^c(\bar x +i \bar y)^{\bar c}
    \bigg].
        \label{eq:typeIaverNew}
\end{align}
For $\bar c=0$, it reduces to the type II average.

In the rest of this section, we shall explain how these formulas arise, and in particular how the derivation of the standard formulas \eqref{eq:typeIaver}--\eqref{eq:typeIIaver} must be modified to give the new average formula \eqref{eq:typeIaverNew}. Section~\ref{sec:Holography} will largely be devoted to exploring the consequences of these formulae.

\paragraph{Why code averages arise}

A code is a set containing $2^k$ vectors of length $n$. 
Consider the collection of such sets that are linear codes satisfying some additional constraints to be an ordered type $i$ code ($i=\text I$ or $\text{II}$). Denote this collection $X_{n}(\text{type $i$})$, 
    \begin{equation}
        X_{n}(\text{type $i$})=\left\{\hat C 
        \middle| \text{ $\hat C$ is a type $i$ ordered code}\right\}\,.
    \end{equation}
    The average of an observable $\mathcal{O}$ is defined as the average over ordered codes with weight one:
    \begin{equation}
    \label{eq:averagePre}
        \overline{\mathcal{O}} =\frac1{T^i_n}\sum_{\hat C\in X_{n}(\text{type $i$})} \mathcal{O}_{\hat C}
    \end{equation}
    where $T^i_n=|X_{n}(\text{type $i$})|$. Many observables are independent of the ordering, and therefore only depend on the equivalence classes $\mathcal C$ of ordered codes. In this case one can reorganize the sum in \eqref{eq:averagePre} into a sum over equivalence classes of codes. The enumerator polynomial is an example of such an observable, so we find
    \begin{equation}
    \label{eq:averagePost}
        \overline W=\frac1{T_n^i}\sum_{\mathcal C\text{ type $i$ code}}\frac{n!}{|\mathrm{Aut}(\mathcal C)|} W_{\mathcal C}\,.
    \end{equation}
    Here we used the fact that a given equivalence class contains $n!/|\mathrm{Aut}(\mathcal C)|$ ordered codes. The $n!$ corresponds to all orderings of the elements of the codewords. Different orderings lead to equivalent ordered code. However, sometimes a reordering leads to the same exact ordered codes -- such reorderings are automorphisms. Therefore the number of equivalent (but distinct) ordered codes is equal to $n! / |\mathrm{Aut}(\mathcal C)|$. Finally, the prefactor is precisely the $T_n$ counting the number of codes, so we find \eqref{eq:average-gen}.

\paragraph{Example: $c = 24$ type II codes} For an explicit example, let us check this with the 9 doubly-even self-dual codes. We find that
\begin{align}
    \frac{T_{24}}{24!} = \frac{453091}{1637898780672} \sim 2.77 \times 10^{-7} \, ,
\end{align}
and 
\begin{align}
\begin{split}
    \overline{W}^{\text{II}}_c(x,y)\  &= \ x^{24} + \frac{10626}{1025} x^{20} y^4 +\frac{735471}{1025} x^{16} y^8 + \frac{2704156}{1025} x^{12} y^{12}  \\
    & \qquad + \frac{735471}{1025}   x^8 y^{16} + \frac{10626}{1025}  x^4 y^{20}  + y^{24} \, .
\end{split}
 \end{align}
This expression can be easily verified to match the left-hand-side of~\eqref{eq:average-gen} using the list of codes and their automorphism groups in table \ref{tab:codes}, given in appendix \ref{app:data}.

\subsubsection{Average of $SU(2)^c\times SU(2)^{\bar c}$ theories}

We will make some further considerations of type $\hat{\text I}$ codes to show how our new average formula \eqref{eq:typeIaverNew} arises. The expected form of such an average formula is an equal-weight sum over all members in the class.
    \begin{equation}
       \frac1{|X_{n}(\text{type $\hat{\text I}$})|}\sum_{\hat C\in X_{n}(\text{type $\hat{\text I}$})} W_{\hat C} =  \overline{W}^{\hat{\text  I}}_n 
       \label{eq:typeIhatpre}.
    \end{equation}
In general, because type $\hat{\text{I}}$ codes satisfy the extra condition $w_+(c)-w_-(c)\equiv 0\ \text{(mod $4$)}$, see \eqref{eq:wpwmDef}, the number of summands will be smaller than the number  $T_n^{\text I}$ of ordered type I codes. 
The conjectured average in the right-hand side is given by \eqref{eq:typeIaverNew} above, and was determined based on the Poincar\'e sum in the next section. We have checked it explicitly for $c = \bar c$ up to $12$. 

\begin{table}[]
    \centering
    \begin{tabular}{|c|cccc|c|}
    \hline
        $n$ & code $\mathcal C$ & $|\mathrm{Aut}|$& $\frac{n!}{|\mathrm{Aut}|}$  &  $\frac{n!}{|\mathrm{Aut}|}r_{\mathcal C,k}$ & Symmetry \\
         \hline
        $2$ & $i_2$ & $2$ & $1$ & $1$ & $SU(2)$ 
        \\\hline
        $4$ & $i_2^2$ & $8$ & $3$ & $2$& $SU(2)^2$ 
        \\\hline
        $6$ & $i_2^3$ & $48$ & $15$ & $6$& $SU(2)^3$ 
        \\\hline
        $8$ & $i_2^4$ & $384$ & $105$ & $24$& $SU(2)^4$ 
        \\
        $8$ & $e_8$ & $1344$  &  $30$  &  $6$ & $SO(8)$ 
        \\\hline
        $10$ & $i_2^5$ & $3840$ & $945$ & $120$ & $SU(2)^5$ 
        \\
    $10$ & $i_{2}\times e_8$ & $2688$  &  $1350$  &  $150$& $SU(2)\times SO(8)$ 
    \\\hline
    $12$ & $i_{2}^6$ & $46080$  &  $10395$  &  $720$ & $SU(2)^6$ 
    \\
    $12$ & $i_2^2\times e_8$ & $10752$ &$44550$  & $2700$ & $SU(2)^2\times SO(8)$ 
    \\
    $12$ & $b_{12}$ & $23040$ &  $20790$  & $\{450,720\}$& $\{SO(12), SU(2)^6\}$ 
    \\\hline
    \end{tabular}
    \caption{Weights for type I averages. $\frac{n!}{|\mathrm{Aut}|}$ denotes the number of ordered codes in each equivalence class $\mathcal C$. Of this number, a share $r_{\mathcal C,k} $ produces a CFT labeled by $\mathcal C $ and $k$. The first case where more than one value of $k$ occurs is for the code $\mathcal C=b_{12}$.}
    \label{tab:typeIaver}
\end{table}

In principle, \eqref{eq:typeIhatpre} is all that is needed. To analyze the situation further and make some checks, we would like to reorganize the left-hand side.
Knowing that each equivalence class of codes $\mathcal C$ can give rise to several distinct $SU(2)$ theories, when organizing the sum into the contribution from different $\mathcal C$, we need to take into account that a given code equivalence class $\mathcal C$ is an equivalence class containing many different ordered codes $\hat{ \mathcal C}$. Only some of them satisfy the conditions to be a type $\hat{\text I}$ ordered code. Moreover, the ordered codes belonging to a given (code equivalence class) $\mathcal C$ may give rise to $N_{\mathcal C}$ different enumerator polynomials, each appearing a fraction $r_{\mathcal C_,k}$ of the cases, with $\sum_kr_{\mathcal C_,k}\leqslant 1$. Taking this into account, we can write \eqref{eq:typeIhatpre} as
        \begin{equation}
      \overline{W}^{\hat{ \text I}}_{c,\bar c} = \frac1{|X_{n}(\text{type $\hat{\text I}$})|}\sum_{\mathcal C\text{ type I}}\sum_{k=1}^{N_{\mathcal C}} r_{\mathcal C,k}\,W_{\hat{\mathcal C}}  
    \end{equation}
    where
    \begin{equation}
        |X_{n}(\text{type $\hat{\text I}$})|=T_n^{\mathrm I}\sum_k r_{\mathcal C,k} \leqslant T_n^\text{I}\,.
    \end{equation}
 For low $n$, the values of the constants $r_{\mathcal C,k}$ can be determined explicitly. We give some values of these constants in table~\ref{tab:typeIaver}. In our example above with the code $\mathcal C=i_2^2$, there is only one resulting CFT, occurring for two out of the three ordered codes \eqref{eq:threegeneratormatrices}. Hence $r_{i_2^2,1}=\frac23$.
 
 The first case exhibiting the most general structure is $n=12$. Here there are three type I codes: $i_{2}^6$, $i_{2}^2\times e_8$ and $b_{12}$ \cite{Pless1972}.\footnote{They can easily be extracted from the database by Harada and Munemasa \cite{Harada2015}, see appendix~\ref{app:data}.} The first two codes give rise to unique CFTs, namely factorizable CFTs with symmetry groups $SU(2)^6$ and $SU(2)^2\times SO(8)$ respectively. The last code, $b_{12}$, on the other hand, gives rise to two different type $\hat{\text I}$ codes. The generating matrix is
\begin{equation}
\label{eq:GTnew}
    G^T=\left(\begin{array}{cccccccccccc}
    1 & 0 & 0 & 0 & 1 & 0 & 0 & 0 & 1 & 1 & 1 & 1 \\
 0 & 1 & 0 & 0 & 1 & 0 & 0 & 1 & 1 & 1 & 1 & 0 \\
 0 & 0 & 1 & 0 & 1 & 0 & 0 & 1 & 0 & 0 & 0 & 1 \\
 0 & 0 & 0 & 1 & 0 & 0 & 0 & 1 & 1 & 0 & 0 & 1 \\
 0 & 0 & 0 & 0 & 0 & 1 & 0 & 1 & 0 & 1 & 0 & 1 \\
 0 & 0 & 0 & 0 & 0 & 0 & 1 & 1 & 0 & 0 & 1 & 1  
    \end{array}\right)
\end{equation}
By swapping columns $6$ and $9$, it satisfies the conditions to make it a type $\hat{\text I}^{6,6}$ code, and we find that the corresponding CFT has $66$ currents, and enumerator polynomial
\begin{equation}
x^6 \bar{x}^6+    3 x^6 \bar{x}^2 \bar{y}^4+3 x^4 y^2 \bar{y}^6+9 x^4 y^2 \bar{x}^4 \bar{y}^2+32 x^3
   y^3 \bar{x}^3 \bar{y}^3+3 x^2 y^4 \bar{x}^6+9 x^2 y^4 \bar{x}^2 \bar{y}^4+3 y^6 \bar{x}^4
   \bar{y}^2+y^6 \bar{y}^6.
   \label{eq:SO12}
\end{equation}
There are in total $20$ possible swaps that give rise to the same enumerator polynomial, so the corresponding number of ordered codes is $20\cdot 20790\big / \left(\begin{smallmatrix}12\\6\end{smallmatrix}\right)=450$, where $20790=12!/|\mathrm{Aut}(b_{12})|$ is the number of ordered codes in the $b_{12}$ equivalence class. Thus $r_{b_{12},1}=20 \left(\begin{smallmatrix}12\\6\end{smallmatrix}\right)=\frac{5}{231}$.
On the other hand, by swapping columns (2 and 5) with columns (7 and 8), we find a CFT with 18 currents and enumerator polynomial
\begin{equation}
\label{eq:new12Q}
    x^6 \bar{x}^6+6 x^5 y \bar{x} \bar{y}^5+15 x^4 y^2 \bar{x}^4 \bar{y}^2+20 x^3 y^3 \bar{x}^3
   \bar{y}^3+15 x^2 y^4 \bar{x}^2 \bar{y}^4+6 x y^5 \bar{x}^5 \bar{y}+y^6 \bar{y}^6.
\end{equation}
We find that there are $32$ such swaps, which corresponds to $32\cdot 20790\big / \left(\begin{smallmatrix}12\\6\end{smallmatrix}\right)=720$ ordered codes and $r_{b_{12},2}=\frac8{231}$. 

The $c = \bar{c} = 6$ CFTs with $su(2)_L^6 \times su(2)_R^6$ symmetry algebra were considered in \cite{Gaberdiel:2013psa} (see also \cite{Harvey:2020jvu} for an interesting relation between these theories and quantum error correcting codes).\footnote{We thank the anonymous reviewer for bringing these references to our attention.} In particular, the $b_{12}$ theory with $SO(12)$ symmetry has the same partition function as the one given in equations~(C.1) plus (C.2) of \cite{Gaberdiel:2013psa}. Our $(i_2)^6$ theory has a partition function given by~(C.5) of \cite{Gaberdiel:2013psa}, while the $b_{12}$ theory with $SU(2)^6$ symmetry requires that we take~(C.5) and subtract~(C.6), which corresponds to the RR sector. Based on the coincidence of the partition functions, we believe that the theories we have identified in this paper are the same as those theories.

\section{Holography for $SU(2)$ theories}
\label{sec:Holography}

A number of interesting recent developments have pointed to the idea that ensemble average of CFTs may produce certain features of quantum gravity \cite{Marolf:2020xie, Maloney:2020nni, Afkhami-Jeddi:2020ezh, Cotler:2020ugk, Chandra:2022bqq, Schlenker:2022dyo}. The most general conditions under which this occurs is still somewhat unclear, but it appears that it holds in cases where the full set of CFTs with a particular symmetry group is known and can be averaged over. In section~\ref{sec:Symmetry}, we have discussed how Construction~A and Lorentzian Construction~A define chiral and non-chiral $SU(2)^n$ theories, and we have seen that such theories can be averaged over. It is then natural to ask if this ensemble can be interpreted holographically. In this section, we will explore this question, investigating the properties that such an average theory, if it exists, should have.

\subsection{Averaged partition functions from the Poincar\'e sum}

The idea that gravity partition functions in negatively curved $3$-space can be computed using a sum over modular images of the vacuum character on the Riemann surface at the asymptotic boundary has been around for some time. It was first discussed in \cite{Dijkgraaf:2000fq}, and has since appeared in a variety of contexts, including the case of where the only boundary symmetry is Virasoro \cite{Maloney:2007ud,Keller:2014xba} -- which leads to a non-sensical partition function with negative degeneracies -- and the case where the boundary theories in the ensemble all have $U(1)^c \times U(1)^c$ symmetry \cite{Maloney:2020nni, Afkhami-Jeddi:2020ezh}. A key formula for such an average theory is the average partition function. The relevant formula is as follows,
\begin{equation}
\label{eq:gravZgen}
    \overline Z(\Omega,\bar \Omega)=\sum_{\gamma\in \Gamma \backslash \mathrm{Sp}(2g,\Z)} \chi_{\mathrm{vac}}(\gamma \Omega,\gamma\bar\Omega) \, .
\end{equation}
Here the sum goes over unique images of the modular group, parametrized by the coset $\Gamma \backslash \mathrm{Sp}(2g,\Z)$. 
The coset represents the modular group $\mathrm{Sp}(2g,\Z)$ modded out by the subgroup $\Gamma$ that leaves the vacuum character $\chi_{\mathrm{vac}}(\gamma \Omega,\gamma\bar\Omega)$ invariant. We think of $\gamma$ as matrices of the form
\begin{equation}
    \gamma=\begin{pmatrix}A&B\\C&D\end{pmatrix} \, .
\end{equation}
%
For genus 1, $A$, $B$, $C$, and $D$ are simply numbers. In \eqref{eq:gravZgen}, the character $\chi_{\mathrm{vac}}(\Omega,\bar \Omega)$ represents the vacuum partition function of a CFT on a Riemann surface with period matrix $\Omega$. Its precise form will depend on the specific considerations.

An example was recently exhibited in \cite{Maloney:2020nni,Afkhami-Jeddi:2020ezh} in a construction relating this Poincar\'e sum to an average over Narain moduli space, i.e.\ an average over theories with $U(1)^c\times U(1)^c$ symmetry. 
Here the coset is with respect to $\Gamma_\infty$ (denoted $P$ in \cite{Maloney:2020nni}), the group of modular transformations with $C=0$. This group is known as the Siegel parabolic subgroup, and is the higher-genus generalization of the subgroup generated by the $T$ transformation $\tau\mapsto\tau+1$. The vacuum character is
\begin{equation}
    \chi_{\mathrm{vac}}( \Omega,\bar\Omega)=\frac1{|\Phi_g|^{2c}},
\end{equation}
and it transforms as the absolute value of a Siegel modular form of degree $-\frac c2$,
 \begin{equation}
    \sum_{\gamma\in \Gamma_\infty \backslash \mathrm{Sp}(2g,\Z)} 
     \chi_{\mathrm{vac}}(\gamma \Omega,\gamma\bar\Omega) =
     \frac{1}{|\det(C\Omega+D)|^{c}|\Phi_g|^{2c}} 
 \end{equation}

To make clearer the connection with \eqref{eq:gravZgen}, we note that $\sqrt{\IM\Omega}|\Phi_g|^2$ is modular invariant. So we can write
\begin{align}
\nonumber
 \sum_{\gamma\in \Gamma_\infty \backslash \mathrm{Sp}(2g,\Z)} 
    \chi_{\mathrm{vac}}(\gamma \Omega,\gamma\bar\Omega)
   & =
    \sum_{\gamma\in \Gamma_\infty \backslash \mathrm{Sp}(2g,\Z)} 
    \frac{1}{|\det(C\Omega+D)|^{c}|\Phi_g|^{2c}}
    \\
&    =\frac1{(\det\IM\Omega)^{c/2}|\Phi_g|^{2c}} \sum_{\gamma\in \Gamma_\infty \backslash \mathrm{Sp}(2g,\Z)}  \frac{(\det\IM\Omega)^{c/2}}{|\det(C\Omega+D)|^{c}}
\end{align}
But the last sum is exactly the definition of the the real-analytic Eisenstein series,
\begin{equation}
\label{eq:non-holoEisenstein}
    \mathcal E_k(\Omega,\bar\Omega)=\sum_{\gamma\in \Gamma_\infty \backslash \mathrm{Sp}(2g,\Z)}  \frac{(\det\IM\Omega)^{k}}{|\det(C\Omega+D)|^{2k}}
\end{equation}
and in conclusion the average partition function can be written as
\begin{equation}
    \overline Z(\Omega,\bar \Omega)=\frac{\mathcal E_{\frac c2}(\Omega,\bar \Omega)}{(\det\IM\Omega)^{c/2}|\Phi_g|^{2c}}.
\end{equation}

 As an aside, an analogous formula exists also for the holomorphic Eisenstein series:
\begin{align}
    E_k(\Omega)=\sum_{\gamma\in \Gamma_\infty\backslash \mathrm{Sp}(2g,\Z)}\frac1{\det(C\Omega+D)^k}
\end{align}
In appendix~\ref{app:lattices} we review and give an example of how the holomorphic Eisenstein series arise from the average over even self-dual lattices. The corresponding CFTs have $U(1)^c$ symmetry.

It is useful to think of the summands in \eqref{eq:gravZgen} as transforming in a certain representation $\mathfrak R_g$ of the modular group $\mathrm{Sp}(2g,\Z)$. In the formulas above producing the Eisenstein series, this is an infinite-dimensional representation. However for our $SU(2)$ invariant theories, $\mathfrak R_g$ will be finite-dimensional. The size of the representation depends on the class of theories -- in particular, it is determined by the subgroup of the modular transformations which leaves the vacuum character invariant.

In table~\ref{tab:poincareSum} we give the number of images in different cases. We contrast the case at hand in this paper with the case of Eisenstein series and the case of CFTs from error-correcting codes over $\F_2\times\F_2$ by the ``new Construction~A'' of \cite{Dymarsky:2020qom}.
In the table, we also added the considerations from \cite{Castro:2011zq} and \cite{Jian:2019ubz}, who discussed how the partition function for the Ising ($\mathcal M_{4,3}$ unitary minimal model) arises as a Poincar\'e sum, however with some over-counting which we have indicated.

\begin{table}[ht]
    \centering
    \begin{tabular}{|c|c|c|c|}
    \hline
    Representation     & $|\mathfrak R_1|$& $|\mathfrak R_2|$& $|\mathfrak R_3|$ \\
         \hline
    Eisenstein & $\infty$ & $\infty$ & $\infty$ 
    \\
      $SU(2)$ (this paper)  & $6$ & $60$ & $1080$
     \\
     $\mathbb F_2\times \mathbb F_2$ & $3$ & $15$  &  ? 
     \\
    Ising & $24=8\cdot 3$ \cite{Castro:2011zq} & $3840=384\cdot 10$ \cite{Jian:2019ubz} & ?
    \\\hline
    \end{tabular}
    \caption{The number of terms in the Poincar\'e sums at different genera, equivalent to the size of the representation given by the coset $|\mathfrak R_g|=|\Gamma\backslash \mathrm{Sp}(2g,\mathbb Z)|$ in the appropriate representation $\mathfrak R_g$.}
    \label{tab:poincareSum}
\end{table}

\subsubsection{Genus 1 average from Poincar\'e sum}

In the case of codes, the sum over modular images is quite simple. Namely, there are 6 unique images of the vacuum character $x^n$ under the modular group. Apart from the original term $x^n$, the five images are \begin{align}
    &S x^n    = \left(\frac{x+y}{\sqrt2}\right)^n, & 
    &TS x^n  =  \left(\frac{x+i y}{\sqrt2}\right)^{n},&
    \nonumber
    &T^2S x^{n}  =  \left(\frac{x-y}{\sqrt2}\right)^{n} ,\\
    &T^3S x^{n}  =  \left(\frac{x-iy}{\sqrt2}\right)^{n}  &
    &ST^2S x^{n} =  y^{n} .
\end{align}
Normalizing for the average to take unit value at $x=1$, $y=0$, we find
\begin{equation}
    \overline W^{\text{II}}=\frac1{1+4\cdot 2^{-n/2}}\left(x^n+y^n+2^{-n/2}\left((x+y)^n+(x-y)^n+(x+iy)^n+(x-iy)^n\right)\right),
\end{equation}
which is precisely the average \eqref{eq:typeIIaver}. 

Also for type $\hat {\text I}$ codes, the sum over modular images is quite simple and we find 6 unique images of the vacuum character $x^c\bar x^{\bar c}$. Apart from the original term $x^c\bar x^{\bar c}$, the five images are \begin{align}
    &S x^c\bar x^{\bar c}    = \left(\frac{x+y}{\sqrt2}\right)^c\left(\frac{\bar x+\bar y}{\sqrt2}\right)^{\bar c}, & 
    &TS x^c\bar x^{\bar c}  =  \left(\frac{x+i y}{\sqrt2}\right)^{c} \left(\frac{\bar x-i \bar y}{\sqrt2}\right)^{\bar c}, \\
    \nonumber
    &T^2S x^c\bar x^{\bar c}  =  \left(\frac{x-y}{\sqrt2}\right)^{c} \left(\frac{\bar x-\bar y}{\sqrt2}\right)^{\bar c} , &
    &T^3S x^c\bar x^{\bar c}  =  \left(\frac{x-iy}{\sqrt2}\right)^{c}  \left(\frac{\bar x+i\bar y}{\sqrt2}\right)^{\bar c} ,\nonumber \\
    &ST^2S x^c\bar x^{\bar c} =  y^{c}\bar y^{\bar c} .
\end{align}
Again, normalizing for the average to take unit value at $x=\bar x=1$, $y=\bar y=0$, we find
\begin{align}
\overline{W}^{\hat{\text{I}}} \ &= \ \frac1{1+4\cdot 2^{-\frac c2-\frac{\bar c}2}}\bigg[  (x^c \bar x^{\bar c} + y^c \bar y^{\bar c})       +2^{-\frac c2-\frac{\bar c}2}\big( (x+y)^c(\bar x + \bar y)^{\bar c} 
\nonumber\\&\qquad\qquad\qquad
+(x-y)^c(\bar x - \bar y)^{\bar c}
+
     (x+ i y)^c(\bar x - i \bar y)^{\bar c}+ (x- i y)^c(\bar x +i \bar y)^{\bar c}
    \big)\bigg]\,,
    \end{align}
which is equal to the average \eqref{eq:typeIaverNew}.

\subsubsection{Higher genus}
 We now turn to the average enumerator polynomial at higher genus. Since we know the vacuum characters for the higher-genus partition functions, we can use the Poincar\'e sum to provide such a formula. In the case of type II codes, the higher-genus average enumerator polynomial was given by Runge \cite{Runge1996} as a sum over admissible monomials. For low genus, we can evaluate that sum and compare with our expressions. We have also tested our formulas for a number of explicit lists of codes making use of the database of error-correcting codes by Harada and Munemasa \cite{Harada2015}. For more details see appendix~\ref{app:data}.

To work out the higher-genus Poincar\'e sum, we proceed with the higher-genus code variables and the modular transformations as reviewed in \cite{Henriksson:2021qkt}.
At genus 2, we find an expression for the average in the chiral case, $n=c$,
\begin{align}
\label{eq:genus2average}
    \overline W^{(g=2)}(x_0,x_1,x_2,x_3)&=\frac1{1+3\cdot 2^{2-n/2}+2^{5-n}}\bigg[\sum_{a=0}^{3}x_a^n+\frac1{2^{n/2}}\sum_{a<b}\sum_{k=0}^3\left(x_a+i^kx_b\right)^n
    \nonumber
    \\
    &\quad+ \frac1{2^n}\sum_{k_1=0}^3\sum_{k_2=0}^3\sum_{k_3=0}^1\left(x_0+i^{k_1}x_1+i^{k_2}x_2+i^{k_1+k_2}(-1)^{k_3}x_3\right)^n\bigg]\,.
\end{align}
It is the Poincar\'e sum of 60 images. Here we renamed $x_{00}=x_0$, $x_{01}=x_1$, $x_{10}=x_2$, $x_{11}=x_3$. The corresponding formula in the non-chiral case is given by replacing $( \ldots )^n$ by $| \ldots |^n$ where $c=\bar c=n/2$, or more generally by adding complex-conjugate factors raised to $\bar c$ next to each factor raised to $c$.

\paragraph{Checks} In the chiral case $\bar c=0$ it is easier to check the average formulas for various $n=c$, since only even self-dual codes (type II) enter, and each such code gives rise to a unique $SU(2)^c$ theory. At $c=24$, there are 9 type II codes, with data given in appendix~\ref{app:typeII24data}. The check at genus 2 is most easily performed by using the parametrization of the genus-2 partition function in terms of the number of currents $N_c$, given in \cite{Keller:2017iql},
\begin{equation}
    Z^{(2)}_{c=24}=\frac1{\Phi^{24}_2}\left(E_4^3+(N_c+984)(N_c-744)\chi_{12}+\frac{N_c-744}{1728}(E_4^3-E_6^2)\right).
    \label{eq:paramintermsofcurrents}
\end{equation}
Using the weights and number of currents for the individual theories, given in table~\ref{tab:codes} of appendix~\ref{app:typeII24data}, we find that the average is
\begin{equation}
    \overline{Z^{(2)}_{c=24}}=\frac1{\Phi_2^{24}}\left(\frac{6523 }{9225}E_4^3+\frac{2702}{9225}E_6^2-\frac{125526016}{205}\chi_{12}\right).
\end{equation}
Then, by expressing the Siegel modular forms $E_4$, $E_6$ and $\chi_{12}$ in code variables, see appendix~B of \cite{Henriksson:2021qkt}, we find perfect agreement with the expression \eqref{eq:genus2average} for the average.

At $c=32$, there are 85 even self-dual codes, described in \cite{Conway1992}. We have confirmed the average formula at genus 1 and 2. Each of the 85 codes have a genus 1 partition function characterized by the number of currents, which ranges from the minimal number 96 (attained by the five codes C81--C85 of \cite{Conway1992}) to 2,016 (attained by $\text{C1}=d_{32}^+$). The genus 2 partition function can be described by the parameterization of \cite{Keller:2017iql}, in terms of an additional parameter $c_{111}$ taking values between 384 and 249,984.

At genus 3, one needs to take the sum of 1080 images. The average formula is given in the ancillary data file of this paper. We have only used this to verify the average of the two codes at $c=16$, which is the first genus where their enumerator polynomials differ.

\subsection{Spectral gap}

An important signature of holographic theories is the appearance of a large gap in states in the large $c$ limit. In the present case, this means the gap in primaries $SU(2)^c$ or $SU(2)^c \times SU(2)^{\bar c}$. We will treat these cases separately, in the latter case limiting to $\bar c =c$.

\paragraph{Chiral theories}

First consider the $q$ expansions of the $SU(2)$ characters:
\begin{align}
 \frac{\theta_3(q^2)}{\eta(q)} \ &= \     q^{-1/24} + 3 q^{23/24} + 4 q^{47/24} + \ldots, \\
 \frac{\theta_2(q^2)}{\eta(q)} \ &= \     2 q^{5/24} + 2 q^{29/24} + 6 q^{77/24} + \ldots.
\end{align}
Hence a term $x^{n-d} y^d$ corresponds to an $SU(2)^n$ primary with scaling dimension $\Delta = d / 4$. A code whose first non-zero coefficient after $x^n$ is $x^{n-d} y^d$ is said to have a \emph{Hamming gap} of $d$, and it has an $SU(2)^c$ primary gap of $d / 4$. We will return to the relation between these quantities below. 

Above we have argued that the average of type II codes defines a bulk theory with chiral $SU(2)^n$ symmetry, so we would like to understand what will be the gap $d$ in the large $n$ limit. This can be computed analytically. First, note that the coefficient $a_d$ of the $x^{n-d} y^d$ term can be computed by
\begin{align}
    a_d = \frac{1}{d!} \left( \frac{\partial}{\partial y} \right)^d W^{\text{II}}_n |_{x = 1, \, y = 0} = \frac{4}{4 + 2^{n/2}} \frac{n!}{d! (n-d)!} \, .
\end{align}
To determine the asymptotic gap, we would like to see at which $d$, in the limit of infinite $n$, this coefficient goes from $a_d \ll 1$ to $a_d \gg 1$. First, we assume that this happens at a fixed fraction of $n$, so we rewrite $d = \delta n$, where we will call $\delta$ the \emph{relative distance}. Then from Stirling's approximation we have
\begin{align}
     \log \left( a_d \right) = \log \left( \frac{4}{2^{n/2}} \frac{n!}{(\delta n)! (n(1- \delta))!}  \right) \simeq  n \log  2  \left( h(\delta) - \frac{1}{2}  \right)  \,.
\end{align}
where $h(\delta)$ is the \emph{entropy function}\footnote{This is the standard term because it is the Shannon entropy of the random variable $x$.}
\begin{align}
    h(x) = x \log_2 \frac{1}{x} + (1-x) \log_2 \frac{1}{1-x} \, .
\end{align}
Therefore we see that, at infinite $n$, the degeneracy of $SU(2)^n$ primaries will be zero until $h(\delta ) = 1/2$. Therefore the gap is given by
\begin{align}
    \delta^* = h^{-1}\left( \frac{1}{2} \right) \approx 0.11002786\,,
\end{align}
leading to a spectral gap for the averaged theory
\begin{align}
    \Delta^{\text{II}}_{\text{gap}} := \frac{\delta^* n }{4 } \simeq \frac{c}{36.354}\,.
\end{align}

\paragraph{Non-chiral theories}

In the case of the non-chiral theories defined from Lorentzian Construction~A, we should consider four possible characters of $SU(2) \times SU(2)$ symmetry:
\begin{align}
 \frac{\theta_3(q^2)}{\eta(q)} \frac{\overline{\theta_3(q^2)}}{\overline{\eta(q)}}  \ &= \     (q \bar q)^{-1/24}  + \ldots  & &(\Delta, \ell) = \left( 0, 0 \right) \,,\\
 \frac{\theta_3(q^2)}{\eta(q)} \frac{\overline{\theta_2(q^2)}}{\overline{\eta(q)}}\ &= \      2 q^{-1/24} \bar{q}^{5/24} + \ldots & &(\Delta, \ell) = \left( \frac{1}{4}, -\frac{1}{4} \right)\,,\\
 \frac{\theta_2(q^2)}{\eta(q)} \frac{\overline{\theta_3(q^2)}}{\overline{\eta(q)}} \ &= \     2 q^{5/24} \bar{q}^{-1/24}  + \ldots & &(\Delta, \ell) = \left(\frac{1}{4}, \frac{1}{4}\right)\,, \\
 \frac{\theta_2(q^2)}{\eta(q)}\frac{\overline{\theta_2(q^2)}}{\overline{\eta(q)}} \ &= \     4 q^{5/24} \bar{q}^{5/24} + \ldots & &(\Delta, \ell) = \left( \frac{1}{2}, 0 \right)\,.
\end{align}
As we see, terms arising from $x \bar{y}$ or $\bar {x} y$ give quarter-integer contributions to the spin, which means that such terms must only appear in powers divisible by four. This is exactly what we expect based on Lorentzian Construction A by the condition $w_+(c)-w_-(c)=0\ \text{(mod $4$)}$. It is also clear from this that the scaling dimension of an $SU(2)^n \times SU(2)^n$ primary is equal to ($\frac14$ times) the number of $y$s and $\bar y$s that appear in it. 

The calculation of the gap proceeds in the same way as the chiral case. We find the same Hamming distance, but for the gap in the spectrum the interpretation changes slightly because now $n=c+\bar c=2c$, and we get
\begin{align}
    \Delta^{\hat{\text{I}}}_{\text{gap}} := \frac{\delta^* n}{4 } \simeq \frac{c}{18.177}\,.
\end{align}

\paragraph{Gilbert--Varshamov bound}

The average formula can be used to derive a simple lower bound on the maximal Hamming distance of any code. To do this, we look not at where the degeneracy $a_d$ becomes order 1, but instead at where the sum of all degeneracies \emph{up to }$d$ becomes order one. However, at asymptotically large $n$, these two things occur at the same place. If the sum of all degeneracies up to a value $d$ is less than one, then there must be a particular code whose gap is at least $d$. Therefore there must be a CFT whose gap satisfies $\Delta_{gap} \geqslant \Delta^{\text{II}} \sim \delta^* $. For a proof of the Gilbert-Varshamov bound (for general codes rather than self-dual codes), see appendix~\ref{app:distance bounds}.

\paragraph{Upper bounds}

The Gilbert--Varshamov bound tells you that there \emph{must} be a code with a relative distance of at least $\delta^*$. It is also possible to prove upper bounds on the maximal Hamming distance. A simple argument for type II codes comes from invariant theory \cite{Mallows1973}. It is known (Gleason's theorem) that the enumerator polynomials of type II codes are elements of the ring generated by $x^8 + 14 x^4 y^4 + y^8$ and $x^4 y^4(x^4-y^4)^4$. The dimension of the degree-$n$ subspace of this ring is $D = \lfloor \frac{n}{24} \rfloor + 1$. This fact can be used to set $D$ of the coefficients in the enumerator polynomial arbitrarily. If we do this to set the leading (in $x$) coefficients in the polynomial to zero, we will find a unique polynomial which is given by $W(x, y) = x^n + \mathcal{O}(y^{4D})$. Therefore $d = 4 (\lfloor \frac{n}{24} \rfloor + 1) $ is the maximal possible Hamming distance of any possible enumerator polynomial, and thus the possible distance of any type~II code. This result was subsequently extended to cover all self-dual codes, leading to the asymptotic bound \begin{align}\delta^* \leqslant \frac{1}{6}. \end{align}

A modest strengthening of the asymptotic bound to 
\begin{align}
    \delta \leqslant \frac{1 - 5^{-1/4}}{2} \simeq 0.1656 \, 
    \label{eq:bestbound}
\end{align}
was obtained for type II codes in \cite{Krasikov2000}, and shown to apply to type I codes as well in \cite{Rains2003}. To the best of our knowledge, this is the best known upper bound for self-dual codes. For non-self-dual codes, the best asymptotic bounds are slightly weaker and come from analyzing the linear programming problem introduced by Delsarte \cite{Delsarte1973}. We mention some of these more general bounds which do not assume self-duality in appendix~\ref{app:distance bounds}.

\subsection{Wormhole contributions to path integral}

\begin{figure}
\begin{floatrow}
\ffigbox[10cm]{%
  \includegraphics[width=10cm]{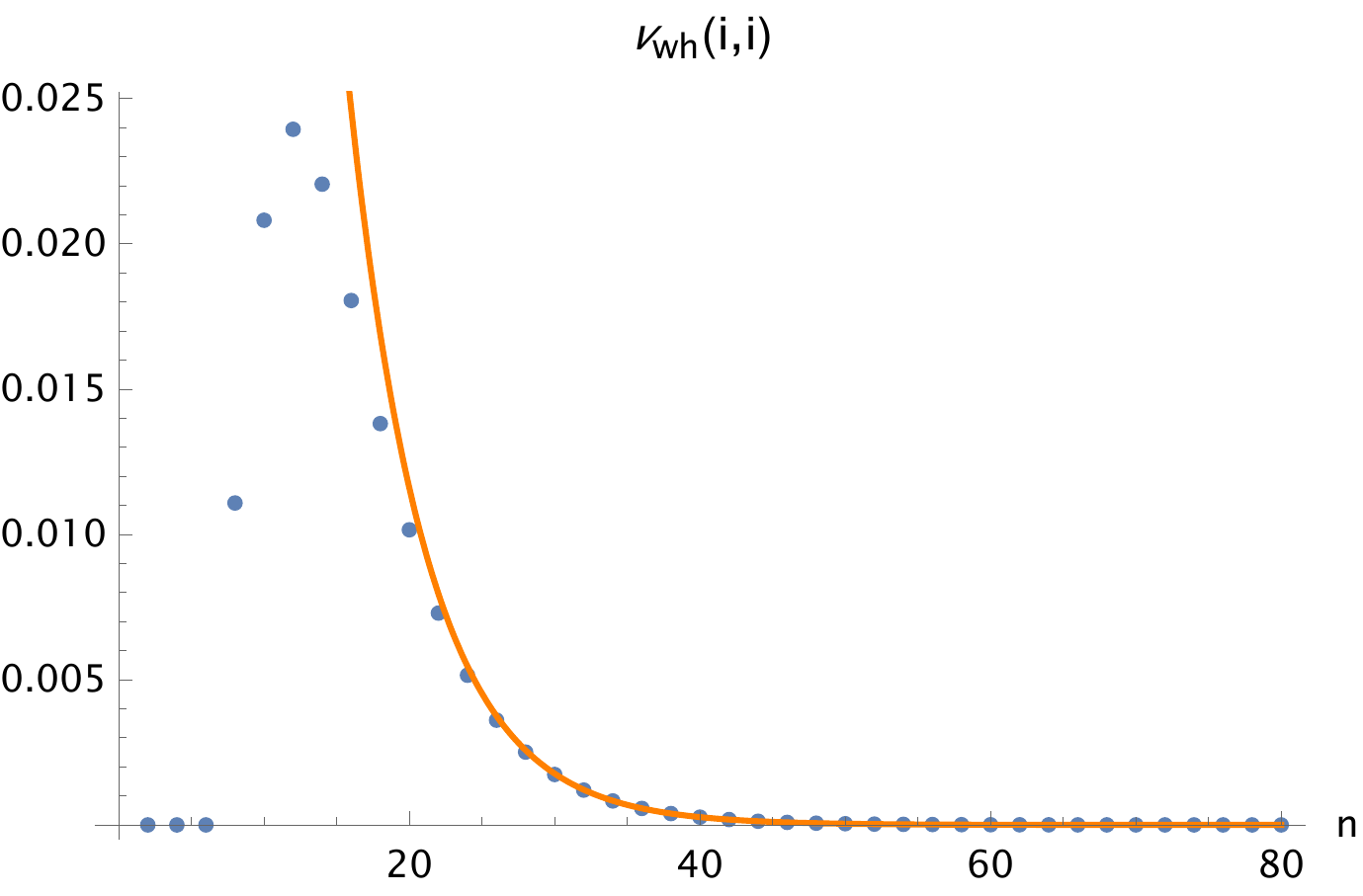}
}{%
  \caption{Blue dots: $\nu_\mathrm{wh}(i,i)$. Orange line:$\frac12(\sqrt8-2)^n$  }%
  \label{fig:whplot}
}
\capbtabbox[5cm]{%
  \begin{tabular}{|cc|}
    \hline
           $n$ &  $\frac{\nu_{\mathrm{wh}}}{\frac12(\sqrt8-2)^n}$
           \\\hline
 $ 24 $&$ 0.94386947 $\\$
 48 $&$ 1.00000714 $\\$
 72 $&$ 1.00003765 $\\$
 96 $&$ 1.00000203 $\\$
 120 $&$ 1.00000008 $
           \\
  $
 144 $&$ 1+3.0401\cdot 10^{-9} $\\$
 168 $&$ 1+1.0562\cdot 10^{-10} $\\$
 192 $&$ 1+3.5541\cdot 10^{-12} $\\$
 216 $&$ 1+1.1724\cdot 10^{-13} $\\$
 240 $&$ 1+3.8180\cdot 10^{-15} $
   \\\hline
    \end{tabular}
}{%
  \caption{numerical values}
    \label{tab:CodeWHnonchiral}
}
\end{floatrow}
\end{figure}

 In the section above we noted that the ensemble average over code CFTs has some holographic features. In such a setup, the average partition function on a Riemann surface $\Sigma$ would be equal to the partition function of a bulk theory with boundary $\Sigma$. 
In general, the bulk path integral would include a sum over all $3$-manifolds $B$ with boundary $\Sigma$,
\begin{equation}
\label{eq:dualitySum}
    \overline Z^{\mathrm{CFT}}_\Sigma(\Omega)=\sum_{\partial B=\Sigma} Z^{\mathrm{bulk}}(B)\,.
\end{equation}
It was pointed out in \cite{Maloney:2007ud} (see also \cite{Dijkgraaf:2000fq, Manschot:2007ha, Castro:2011zq, Keller:2014xba}) that the gravitational path integral will get a contribution from saddle points of different topology. This consideration becomes crucial when considering the case where the boundary is disconnected. For simplicity, consider the case when $\Sigma$ is a pair of disconnected tori,
\begin{equation}
    \Sigma=T\sqcup \tilde{T}, \qquad \Omega=\tau\oplus\tilde{\tau}\,.
\end{equation}
In this case, let us divide the sum in \eqref{eq:dualitySum} into the contributions from disconnected and connected bulk saddles $B$
\begin{equation}
    \overline Z^{\mathrm{CFT}}_{T\sqcup \tilde{T}}(\tau\oplus\tilde{\tau})=\sum_{\text{conn. }B} Z^{\mathrm{bulk}}(B)+\sum_{\partial B=T} Z^{\mathrm{bulk}}(B)\sum_{\partial \tilde{B}=\tilde{T}} Z^{\mathrm{bulk}}(\tilde{B})\,.
    \label{eq:CFTfactor}
\end{equation}
Here we used the fact that the sum over disconnected bulk geometries factorizes. The contribution from connected bulks can be interpreted as a sum over wormhole geometries. At large central charge we expect that such contributions should be suppressed. Equation \eqref{eq:CFTfactor} provides a way to estimate such suppression. Rearranging we find
\begin{equation}
        \sum_{\text{conn. }B} Z^{\mathrm{bulk}}(B)=\overline Z^{\mathrm{CFT}}_{T\sqcup \tilde{T}}(\tau\oplus\tilde{\tau})-\overline Z^{\mathrm{CFT}}_{T}(\tau)\overline Z^{\mathrm{CFT}}_{\tilde{T}}(\tilde{\tau})
\end{equation}
The relative size of the wormhole contribution is then, in schematic notation, given by
\begin{equation}
\label{eq:wormholecontr}
    \nu_{\mathrm{wh}}(\tau,\tilde{\tau})=\frac{\overline W^{(2)}(\tau\oplus\tilde{\tau})-\overline W^{(1)}(\tau)\overline W^{(1)}(\tilde{\tau})}{\overline W^{(1)}(\tau)\overline W^{(1)}(\tilde{\tau})}\,.
\end{equation}
By using our expressions for the average at genus 1~\eqref{eq:typeIaverNew} and genus 2~\eqref{eq:genus2average}, this formula can be readily evaluated. Let us consider the specific values $\tau=\tilde \tau=i$. Then using
\begin{align}
    x_{\tau = i} \ = \ \sqrt{1 + \frac{1}{\sqrt{2}}} \, , \qquad y_{\tau = i} \ = \ \sqrt{1 - \frac{1}{\sqrt{2}}} \, ,
\end{align}
we get an expression for the genus 2 and genus 1 partition functions:
\begin{align}
    Z^{(2)}(i, i)  &=  \frac{2^{n/2+3}(1 + 2^{n/2-1} + 3^{n/2}+ 2^{n-2})+ 4 (2^{n/2} + (2 - \sqrt 2)^{n/2} + (2 + \sqrt 2)^{n/2})^2}{(4 + 2^{n/2})(8 + 2^{n/2})} \nonumber \,,\\
    Z^{(1)}(i) &=  \frac{2 (2^{n/2} + (2 - \sqrt 2)^{n/2} + (2 + \sqrt 2)^{n/2} ) }{ 4 + 2^{n/2}}\,.
\end{align}
From these expressions, we can see that the large-$n$ behavior of $\nu_\mathrm{wh}$ is 
\begin{align}
    \nu_\mathrm{wh}(i,i) \sim \frac{1}{2} (\sqrt 8 - 2)^n \, .
\end{align}
Figure~\ref{fig:whplot} and table~\ref{tab:CodeWHnonchiral} contain the numerical values of $\nu_\mathrm{wh}(i,i) $ for a number of lengths $n$.

\begin{figure}
    \centering
    \includegraphics[width=0.85\textwidth]{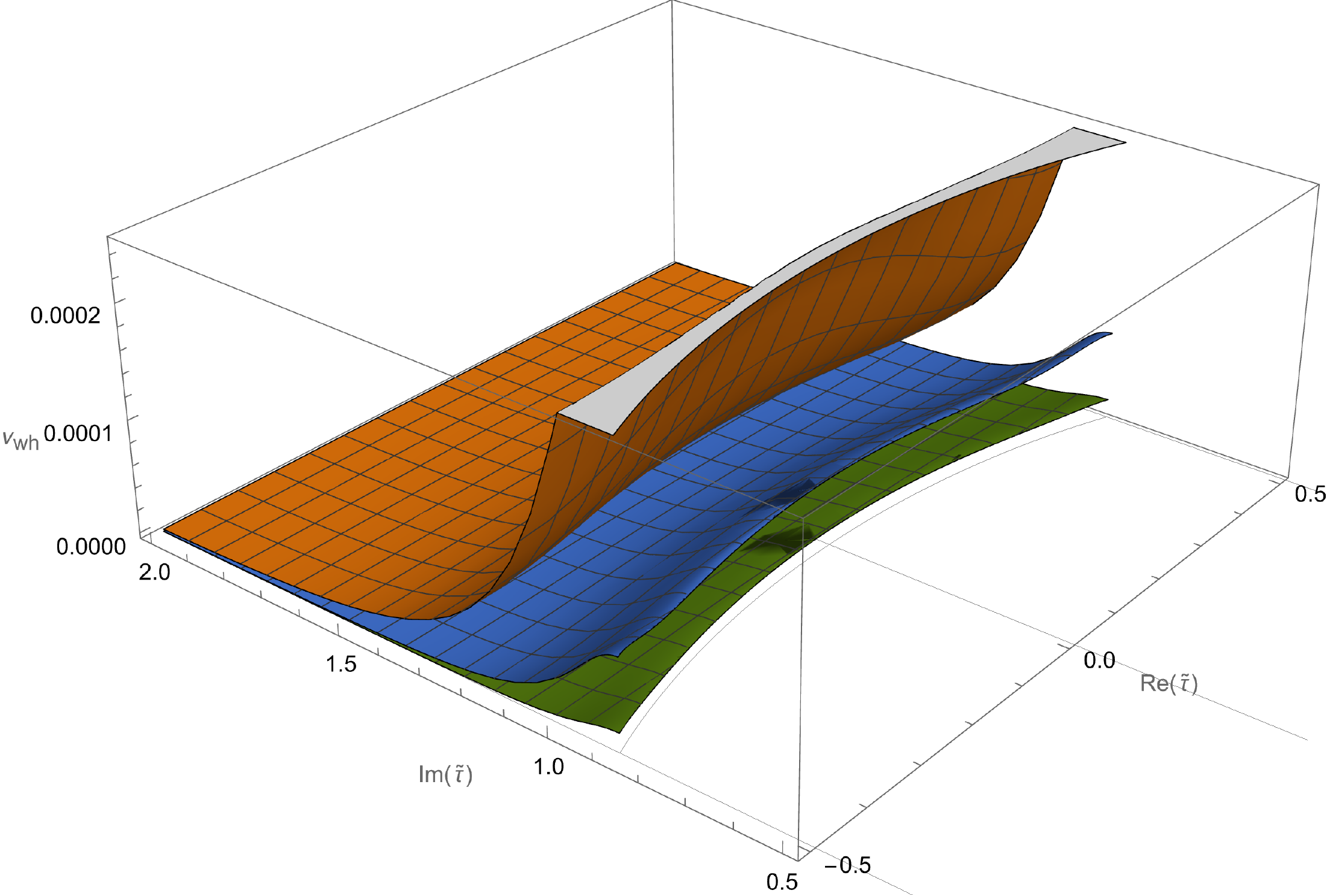}
    \caption{Wormholes contribution $\nu_{\mathrm{wh}}(i,\tilde\tau)$ for increasing values of $c=\bar c=n/2$. Orange / top: $n=40$, blue / middle: $n=48$, green / bottom $n=56$.}
    \label{fig:wormholes}
\end{figure}

In figure~\ref{fig:wormholes} we plot the wormhole contribution for $\tau=i$ and varying $\tilde \tau$ in the fundamental domain. Again, we consider the case $c=\bar c=n/2$.

\section{Discussion}
\label{sec:Other}

The goal of this paper has been to explore a particular ensemble of 2d CFTs. These may be defined as the set of lattice theories where all compact directions are at the self-dual radius so all $U(1)$ symmetries are enhanced to $SU(2)$, or equivalently (we conjecture) as the set of theories defined through a particular construction by self-dual binary error-correcting codes. These theories have a number of nice properties. Among these are the facts that they include a finite number of primaries, and that the characters for these primaries are known for arbitrary genus. This allows, in principle, a fairly simple higher-genus bootstrap approach for determining all of these theories. In section~\ref{sec:Symmetry}, we have carried out this method, and have confirmed that it does yield the correct number of $SU(2)$ theories for theories up to $c = \bar c = 7$ under the hypothesis that all such theories derive from error-correcting codes. Higher genus constraints will be required for larger values of the central charge.

One of the primary motivations for considering the ensemble of $SU(2)$ theories is that they provide a nice example where averaged observables may be computed. The enumerator polynomial averaged over all codes was known already in 1975 \cite{Pless1975}. After reviewing this formula, we showed that it arises from a Poincar\'e sum. The seed is taken to be the term in the enumerator polynomial, $x^n$ that arises from the $\vec 0$ codeword, which coincides with the vacuum character of the $SU(2)^n$ symmetry. We subsequently used the Poincar\'e sum to conjecture the average for several other relevant observables, including the average for the Lorentzian Construction A partition function, and the higher-genus partition functions. We checked these averages explicitly for several small examples of $n$. 

Finally, we found that the gap in $SU(2)^n$ primaries approaches $\Delta_{\text{gap}} \sim n / 36.354$ at large $n$. We also computed $\langle Z(\tau) Z(\tilde\tau) \rangle - \langle Z(\tau) \rangle \langle Z(\tilde \tau) \rangle$ for these theories -- the small but non-zero result we found is indicative of wormhole contributions to the path integral. Both results point to a possible bulk interpretation of the average over this ensemble. It is natural to speculate that the bulk is some sort of $SU(2)^n$ level-1 Chern--Simons theory. This is in direct analogy to the $U(1)^c \times U(1)^{ c}$ Chern--Simons theories postulated to be the bulk dual of the $U(1)^c \times U(1)^{c}$ ensemble of Narain CFTs. It would be very interesting to try to understand more about this putative bulk dual, but we will leave this until future work. Let us conclude by mentioning a few more directions which might be interesting to consider in the future.

\subsection{Twisted Construction~A}

There exists a simple procedure called ``twisting,'' considered in \cite{Dolan:1989vr}, which maps an even self-dual lattices $\Lambda$ into a new even self-dual lattice $\Lambda'$ (see equation (2.39) of \cite{Dymarsky:2020qom} for a full description). At the level of the lattice theta functions for even self-dual lattices, twisting (using the $(1,1,1\ldots,1)$ codeword), maps the lattice theta function into a new lattice theta function
\begin{align}
    \Theta_{\Lambda'(\mathcal{C})} = \frac{1}{2} \left(\Theta_{\Lambda(\mathcal{C})} + (\theta_2(q)\theta_3(q))^{c/2} + (\theta_3(q)\theta_4(q))^{c/2}  - (\theta_2(q)\theta_4(q))^{c/2} \right).
    \label{eq:thetatwist}
\end{align}
It is natural to ask if twisted Construction~A might lead to an interesting bulk theory. 
By applying the twisted construction to the type II average enumerator polynomial, we find that the number of currents approaches $c$ at large $c$, indicating that the twist breaks the $SU(2)^c$ symmetry. However, by expanding in $U(1)^c$ characters, we find that the gap does not grow with $c$ -- in fact, it is equal to 2. 
It would be interesting to understand if there is some symmetry enhancement in this case which leads to a large gap, and maybe even an interesting bulk dual.

\subsection{$SU(2)$ theories at higher level}

The symmetry group of the code theories considered here is $SU(2)^n$ at level 1. However, the formula for $SU(2)$ characters given in~\eqref{eq:su2chars1} applies for arbitrary level $k$, so it is natural to wonder if our modular bootstrap method can be extended to include theories with higher levels. 

To follow the approach of section~\ref{sec:Symmetry}, we need to determine the modular transformations of the characters in the symmetry group. In general, for level $k$ there are $k + 1$ unique characters when $\lambda = 0, \, 1, \, \ldots k$. Higher values of $\lambda$ give 0 or elements of this set of unique characters. Let us illustrate this with the example of level $k = 2$. Define now three characters, 
\begin{align}
    x_2  \ &= \ \chi_0^{k = 2} \ = \ \frac{1}{2} \left( \left( \frac{\theta_3(q)}{\eta(q)} \right)^{3/2}+ \left( \frac{\theta_4(q)}{\eta(q)} \right)^{3/2} \right) \, , \\ 
    y_2  \ &= \ \chi_1^{k = 2} \ = \ \frac{1}{\sqrt{2}} \left( \frac{\theta_2(q)}{\eta(q)} \right)^{3/2} \, ,\\  
    z_2  \ &= \ \chi_2^{k = 2} \ = \  \frac{1}{2} \left( \left( \frac{\theta_3(q)}{\eta(q)} \right)^{3/2}-  \left( \frac{\theta_4(q)}{\eta(q)} \right)^{3/2} \right) \, . 
\end{align}
The modular transformations are
\begin{align}
    S: & &x_2 \to \frac{x_2 + \sqrt{2} y_2 + z_2}{2} & & y_2 \to \frac{x_2 - z_2}{\sqrt{2}} & & z_2 \to \frac{x_2 - \sqrt{2} y_2 + z_2}{2}\, ,  \\
    T: & & x_2 \to e^{- \frac{2 \pi i}{16}} x_2 & & y_2 \to e^{ \frac{2 \pi i}{8}} y_2 & & z_2 \to e^{ 7 \frac{2 \pi i}{16}} z_2 \, .
\end{align}
With these transformations in hand, we can repeat many of the calculations in this paper for this case. For instance, we can try to make modular invariant partition functions from homogeneous degree-$n$ polynomials of these variables. Chiral partition functions can only be found when $n$ is divisible by $16$. At $n = 16$, we find that there are 121 invariant polynomials corresponding to potential partition functions of chiral theories. For non-chiral theories, we find a single partition function for $n = 2$:
\begin{align}
    Z^{k = 2}_{c = 1} = x_2 \bar{x}_2 + y_2 \bar{y}_2 + z_2 \bar{z}_2 \,.
\end{align}
This is the partition function of the level 2 $SU(2)$ WZW model, which has $c = 3/2$. This also has a realization as a model of $3$ fermions (see \cite{Ginsparg:1988ui}). Going on, we then find 2 polynomials at $n = 4$, 10 at $n = 6$, and so on. One could also consider the Poincar\'e sum of such theories. The vacuum character $x$ has $384$ unique modular images, leading to a large but manageable expression for the ``average'' partition function for any $n$. It is possible to realize some of these theories with higher level $SU(2)_2^n$ symmetry by taking $n$ copies of higher-level WZW models. It is interesting to consider that doing this (taking $n$ copies) for $U(1)^n$ gives a trivial square lattice and for $SU(2)^n$ at level one gives $n$ copies of the repetition code. However, more interesting ways of combining them lead to non-trivial lattices and codes. It would be very interesting to understand if there are non-trivial versions of $SU(2)^n$ at higher level, and if they are related to an interesting mathematical structure analogous to the lattices and binary codes considered so far. 

\subsection{Conclusion}

It has become clear that, at least in low dimensions, quantum gravity has some relation to ensemble averages over multiple boundary theories, but the nature of this correspondence is still unclear. The set of codes gives a simple and very concrete ensemble where this idea may be explored. A number of such questions have a direct interpretation as questions about the set of codes. For example it is clear that an average theory has a gap of at least $\delta^* n / 4$, which is directly related to the Gilbert--Varshamov lower bound on the maximal Hamming distance. Interestingly, it is not known if the Gilbert--Varshamov value is also an asymptotically tight \emph{upper} bound on the maximal Hamming distance of a code. If it is, then the distribution of Hamming distances at large $n$ should become infinitely sharply peaked at large $n$.  More generally, we have shown that the problem of finding a code with a maximal Hamming distance of a code is the same as the problem of finding a code CFT with maximal gap in $SU(2)^n$ primaries. It would interesting to push the modular bootstrap of $SU(2)^n$ theories further\footnote{See \cite{Dyer:2017rul} for a study of the modular bootstrap with a single $SU(2)$ symmetry.} and to learn if current knowledge from either side of this code-CFT correspondence can give new insights to the other.

A lot can be understood analytically about code theories due to the simple forms of their partition functions. This could allow them to be very useful examples for future studies of 2d CFTs. For example, partition functions of 2d CFTs can be decomposed into eigenfunctions of the Laplacian on the upper half-plane, including a continuous series -- the real analytic Eisenstein, and a discrete series -- the Maass cusp forms \cite{Benjamin:2021ygh} (see \cite{Collier:2022emf, Benjamin:2022pnx} for recent related work). It should be possible to check that particular code theories with small central charges fit into the patterns observed in that paper. Furthermore it is known that the average over all lattice theories is a real analytic Eisenstein series -- the more complicated Maass cusp forms average to zero and disappear from the average. The average over code theories differs from the average over lattices, and is much simpler (in the sense that it is computed from a finite Poincar\'e series). It would be interesting to see if the code average has any overlap with the Maass cusp forms, or if perhaps it is a combination of different Eisenstein series.

It would also be nice to enlarge the set of examples even further. Two different constructions of CFTs from codes over $\F_4$ were considered in \cite{Dymarsky:2020qom, Dymarsky:2020bps, Dymarsky:2021xfc}, and then generalized to codes over $\F_p \times \F_p$ in \cite{Yahagi:2022idq}. Finally a ``master construction'' for codes over $\F_p \times \F_q$ was given in \cite{Angelinos:2022umf} and shown to include all previous constructions.\footnote{In fact, it also captures our Lorentzian Construction~A for the case $c = \bar c$. We thank Anatoly Dymarsky for a discussion of this point.} Understanding the average over such theories could potentially provide an infinite family of bulk theories to study. 

It is natural to expect that the bulk theory dual to the ensemble average over code theories is an $SU(2)^c \times SU(2)^{\bar{c}}$ Chern--Simons theory analogous to the $U(1)^c \times U(1)^c$ theory conjectured in \cite{Maloney:2020nni, Afkhami-Jeddi:2020ezh}. In this paper, we have mostly avoided speculating about this theory, which we leave as an important open direction. It would be interesting, for example, to try to understand the bulk saddle points, analogous to the BTZ black holes and Brown--Henneaux \cite{Brown:1986nw} boundary excitations, that appear in the path integral and lead to the Poincar\'e sum formula in the bulk. 

To understand pure gravity, it appears that one should average over an ensemble of theories with many degrees of freedom and no currents. However, at the present time \emph{no such theory is known}. Error-correcting codes are rich mathematical structures and we hope that they might provide some insight into this question. It is clear that this requires going beyond Construction A, which yields only lattice theories that have large symmetry groups. Still, we believe that the beautiful connection between codes, lattices, and CFTs has more to teach us.

\section*{Acknowledgments}

We would like to thank Nathan Benjamin, Ilija Buri\'c, Alejandra Castro, Scott Collier, Tolya Dymarsky, Noam Elkies, Tom Hartman, Ashish Kakkar, Bob Knighton, Murat Kolo$\breve{\text{g}}$lu, and Gregoir\'e Mathys for a number of useful conversations and correspondences. This work has received funding from the European Research Council (ERC) under the European Union's Horizon 2020 research and innovation program (grant agreement no.~758903). 

\appendix
\addtocontents{toc}{\protect\setcounter{tocdepth}{1}}

\section{Example: Hamming code}
\label{app:Hamming}

The main text discussed a number of basic concepts from coding theory. Here will illustrate some of them in detail using the $[8,4,4]$ Hamming code, which we will hereafter call the Hamming code. This is a type II (doubly-even self-dual) code of length $n = 8$. Its generator matrix is given by
\begin{align}
\label{eq:GThamming}
    G^T =
\begin{pmatrix}
1 & 0 & 0 & 0 & 0 & 1 & 1 & 1 \\
 0 & 1 & 0 & 0 & 1 & 0 & 1 & 1 \\
 0 & 0 & 1 & 0 & 1 & 1 & 0 & 1 \\
 0 & 0 & 0 & 1 & 1 & 1 & 1 & 0
\end{pmatrix},
\end{align}
which yields the codewords
\begin{align}
\resizebox{\textwidth}{!}{$
    \begin{pmatrix}
    0 \\ 0 \\ 0 \\ 0 \\ 0 \\ 0 \\ 0 \\ 0
    \end{pmatrix}\, ,  
    \begin{pmatrix}
 0 \\
 0 \\
 0 \\
 1 \\
 1 \\
 1 \\
 1 \\
 0 
    \end{pmatrix}\, , 
    \begin{pmatrix}
 0 \\
 0 \\
 1 \\
 0 \\
 1 \\
 1 \\
 0 \\
 1 
    \end{pmatrix}\, , 
    \begin{pmatrix}
  0 \\
 0 \\
 1 \\
 1 \\
 0 \\
 0 \\
 1 \\
 1 
    \end{pmatrix}\, , 
    \begin{pmatrix}
  0 \\
 1 \\
 0 \\
 0 \\
 1 \\
 0 \\
 1 \\
 1
    \end{pmatrix}\, ,  
    \begin{pmatrix}
 0 \\
 1 \\
 0 \\
 1 \\
 0 \\
 1 \\
 0 \\
 1
    \end{pmatrix}\, , 
    \begin{pmatrix}
  0 \\
 1 \\
 1 \\
 0 \\
 0 \\
 1 \\
 1 \\
 0
    \end{pmatrix}\, , 
    \begin{pmatrix}
  0 \\
 1 \\
 1 \\
 1 \\
 1 \\
 0 \\
 0 \\
 0
    \end{pmatrix}\, , 
    \begin{pmatrix}
    1 \\
 0 \\
 0 \\
 0 \\
 0 \\
 1 \\
 1 \\
 1
    \end{pmatrix}\, ,  
    \begin{pmatrix}
    1 \\
 0 \\
 0 \\
 1 \\
 1 \\
 0 \\
 0 \\
 1
    \end{pmatrix}\, , 
    \begin{pmatrix}
   1 \\
 0 \\
 1 \\
 0 \\
 1 \\
 0 \\
 1 \\
 0 
    \end{pmatrix}\, , 
    \begin{pmatrix}
  1 \\
 0 \\
 1 \\
 1 \\
 0 \\
 1 \\
 0 \\
 0 
    \end{pmatrix}\, , 
    \begin{pmatrix}
   1 \\
 1 \\
 0 \\
 0 \\
 1 \\
 1 \\
 0 \\
 0
    \end{pmatrix}\, ,  
    \begin{pmatrix}
     1 \\
 1 \\
 0 \\
 1 \\
 0 \\
 0 \\
 1 \\
 0
    \end{pmatrix}\, , 
    \begin{pmatrix}
 1 \\
 1 \\
 1 \\
 0 \\
 0 \\
 0 \\
 0 \\
 1 
    \end{pmatrix}\, , 
    \begin{pmatrix}
    1 \\ 1 \\ 1 \\ 1 \\ 1 \\ 1 \\ 1 \\ 1
    \end{pmatrix}\, .$}
    \label{eq:e8codewords}
\end{align}
We see one codeword with weight 0, fourteen with weight 4, and one with weight 8. Therefore the enumerator polynomial is given by %
\begin{align}
    W_{e_8}(x, y) = x^8 + 14 x^4 y^4 + y^8 \, , 
    \label{eq:Hamming_EP}
\end{align}
We can see that the code is doubly-even and self-dual because the enumerator polynomial satisfies~\eqref{eq:dubevenSD}. From the classic relation between type II codes and chiral theories \cite{Dolan:1989kf, Dolan:1994st, Dolan1996}, this leads to a CFT whose genus 1 partition function is 
\begin{align}
    Z_{e_8}(q) = \frac{\theta_3(q^2)^8 + 14 \theta_2(q^2)^4 \theta_2(q^2)^4 + \theta_2(q^2)^8}{\eta(\tau)^8} \, .
\end{align}
The Hamming code is also a type I theory, so if we use the take the codewords in the order given in \eqref{eq:e8codewords} and embed them into $\Z^{(4,4)}$ with the metric $(++--++--)$, this defines a Lorentzian lattice theory whose partition function is 
\begin{align}
    Z_{\hat{e_8}} =\frac{\left( \theta_3(q^2)^4 + \theta_2(q^2)^4 \right) \left(\overline{\theta_3(q^2)}^4  + \overline{\theta_2(q^2)}^4 \right) +12 \theta_3(q^2)^2 \overline{\theta_3(q^2)}^2 \theta_2(q^2)^2 \overline{\theta_2(q^2)}^2 }{|\eta(\tau)|^8}\,.
\end{align}
This theory is in fact the $c =\bar c= 4$ theory with extended $SO(8)$ symmetry which is known to saturate the modular bootstrap bounds on the spectral gap (in Virasoro primaries) \cite{Collier:2016cls}. Interestingly, it also arises from codes using a completely different construction involving codes over $\F_4$ \cite{Dymarsky:2020qom}.

\paragraph{Averaging $n = 8$ theories}

The Hamming code is the only type II code, and indeed we see that the average for type II codes~\eqref{eq:typeIIaver} reduces to~\eqref{eq:Hamming_EP} when $n = 8$. There is another type I theory-- four copies of the repetition code, $(i_2)^4$. The Hamming code has 1344 automorphisms and $(i_2)^4$ has $4!\cdot 2^4 = 384$. Weighting both theories accordingly, we find 
\begin{align}
    \overline{W}^I_8 (x, y) = x^8 + \frac{28}{9} x^6 y^2 + \frac{70}{9} x^4 y^4 + \frac{28}{9} x^2 y^6 + y^8 \, ,   
\end{align}
in accordance with the type I average given in~\eqref{eq:typeIaver}. 

Finally, let us consider the average of non-chiral theories given by~\eqref{eq:typeIaverNew}.
Because these are defined with a fixed metric signature, not all of the $8!$ permutations relates an non-chiral theory to another non-chiral theory. The number of automorphisms will be the same, so the number of equivalent but distinct ordered codes is smaller, as reflected by the $r_{\mathcal{C}, i}$ used in table~\ref{tab:typeIaver}. For the repetition code $(i_2)^4$, we find that only $16\cdot (4!)^2$ of the permutations preserve the $w^+(c)-w^-(c) = 0 \ \text{(mod $4$)}$ condition on all codewords, leading to $r_{\mathcal{C}, i} = 16\cdot (4!)^2 / 8! = \tfrac{24}{105}$. For the Hamming code $e_8$ we checked using Mathematica that 8064 of the permutations lead preserve the type $\hat{\mathrm{I}}$ structure, leading to  $r_{\mathcal{C}, i} = \tfrac{1}{5}$. Using these results, which are in agreement with table~\ref{tab:typeIaver}, we confirm the average formulas \eqref{eq:typeIIaver} and \eqref{eq:typeIaverNew} for $n=8$.

\section{Averaging over chiral lattice theories}
\label{app:lattices}

A number of the concepts and techniques in this paper may be more familiar in the context of lattice CFTs, of which code CFTs are a subset. The case of chiral CFTs defined from Euclidean self-dual lattices is a very simple version of the idea of \cite{Maloney:2020nni, Afkhami-Jeddi:2020ezh}, though to our knowledge the chiral version of it has not been explored in detail, so we will review some of these topics in this appendix.

\subsection{Averaging and the (discrete) Siegel--Weil formula}

The set of chiral CFTs defined by even self-dual lattices is discrete since there are no continuous transformations relating even self-dual lattices. The formula which is analogous to \cite{Maloney:2020nni, Afkhami-Jeddi:2020ezh} gives the average theta function over all lattices with dimension $c$ \cite{Siegel1935}
\begin{align}
    \sum_\Lambda \frac{1}{|\text{Aut}(\Lambda)|} \Theta_\Lambda(\tau) = m_{\frac{c}{2}} E_{\frac{c}{2}}(\tau)\,,
    \label{eq:latticeaverage}
\end{align}
where $\Theta_\Lambda$ is the lattice theta function and $|\text{Aut}(\Lambda)|$ is the size of the automorphism group of the lattice $\Lambda$. $m_{\frac{c}{2}}$ is a rational number given by a product of Bernoulli numbers $B_k$: 
\begin{align}
    m_{k} = \frac{B_k}{2k} \frac{B_2}{4}\frac{B_4}{8}\cdots \frac{B_{2k-2}}{4k-4}   \,.
\end{align}
The formula~\eqref{eq:latticeaverage} holds for arbitrary genus, with $\tau$ being replaced with the period matrix $\Omega$, and the holomorphic Eisenstein series $E_k(\Omega)$ being defined for any genus by the Poincar\'e sum
\begin{equation}
    G_k(\Omega)
\label{eq:Eisensteingeneral}=\sum_{C,D}\frac1{\det(C\Omega+D)^k} \, ,
\end{equation}
where $G_k = 2 \zeta(k) E_k$. The sum is over all matrices $C$ and $D$ such that one can form an $\mathrm{Sp}(2g,\Z)$ matrix $\left(\begin{smallmatrix}A&B\\C&D\end{smallmatrix}\right)$.
%

In fact,~\eqref{eq:latticeaverage} is a refinement of the Smith--Siegel--Minkowski mass formula,
\begin{align}
    \sum_\Lambda \frac{1}{|\text{Aut}(\Lambda)|}  = m_{\frac{c}{2}}  \, .
    \label{eq:SmithSiegelMinkowski}
\end{align}
This formula is useful in the study of lattices. For example, it was used in \cite{Conway1982} to verify that Niemeier's list of 24 even self-dual lattices \cite{Niemeier1973} is complete. Furthermore, since the size of the automorphism group must be at least 1, the constant $m_{\frac{c}{2}}$ directly implies a (weak) lower bound on the number of lattices at dimension $c$. For example $m_{12} \sim 7.93 \times 10^{-15}$ implies a lower bound of 0 on the number of even self-dual lattices at $c = 24$. However the formula quickly becomes nontrivial. For $c = 32$ we have $m_{16} \sim 4 \times 10^7$. This implies that there are at least $4 \times 10^7$ even self-dual lattices in 32 dimensions. 

In \cite{Schellekens:2016hhf}, also an upper bound is given, based on an assumption on the maximal size of the automorphism group given by the Weyl group of $D_{8k}^+$ for $k\geqslant3$. In table~\ref{tab:latticesEstimates} we quote the results for lower and upper bounds for the first few values of $c$.

\begin{table}
\begin{tabular}{|c|l|c|}
\hline
    $c$ & \multicolumn{1}{c}{bound} & actual number \\\hline
    $8$ & $2.87055\cdot10^{-9}\leqslant N \leqslant1$ & $1$
    \\
    $16$ & $4.97718\cdot10^{-18}\leqslant N\leqslant 2.41608$ & $2$
    \\
$24$ &   $1.58735\cdot10^{-14}\leqslant N\leqslant 4.13085\cdot 10^{16}$ & $24$
\\
$32$ & $8.06185\cdot10^7\leqslant N\leqslant 2.27775\cdot10^{52}$ & ?
\\
$40$ & $8.78616\cdot 10^{51}\leqslant N\leqslant 1.970535004851803\cdot10^{111}$ & ?
\\\hline
\end{tabular}
\caption{Estimates for number of even self-dual lattices.}\label{tab:latticesEstimates}
\end{table}

\subsubsection{Example: $c = 24$ Niemeier lattices}

We can check the formula explicitly for the case of $c = 24$, where it is known that there are exactly 24 even self-dual lattices (the so-called Niemeier lattices \cite{Niemeier1973}). In this case, the constant $m_k$ is equal to 
\begin{align}
    m_{12} = \frac{1027637932586061520960267}{129477933340026851560636148613120000000} \simeq 7.93678 \times 10^{-15} \, .
\end{align}
Using the list of lattices and the size of their automorphism groups given in table~\ref{tab:lats}, it is easy to verify the averaging formula~\eqref{eq:latticeaverage}
\begin{align}
    \frac{1}{m_{12}} \sum_\Lambda \frac{1}{|\text{Aut}(\Lambda)|} \Theta_\Lambda(\tau) =  E_{12}(\tau) = 1 + 94.819 \,  q + 194284. \, q^2 + \ldots.
\end{align}

\begin{table}
\centering
\begin{tabular}{| c | c | c | }
\hline
 $\Lambda$ & $|\text{Aut}(\Lambda)|$ & $N_\text{currents}$ \\ 
 \hline \hline
 Leech &  $2^{22} 3^9 5^4 7^2 11 \cdot 13 \cdot 23$  &  24 \\ \hline
 $(a1)^{24}$ & $2^{34} 3^3 \cdot 5 \cdot 7 \cdot 11 \cdot 23 $ & 72 \\ \hline
 $(a2)^{12}$ & $2^{19} 3^{15} 5 \cdot 11 $ & 96\\ \hline
 $(a3)^{8}$ & $2^{31} 3^9 7$ & 120 \\ \hline
 $(a4)^{6}$ & $2^{22} 3^7 5^7$ & 144 \\ \hline
 $(d4)^{6}$ & $2^{40} 3^9 5$ & 168 \\ \hline
 $(a5)^{4} (d4)$ & $2^{26} 3^{10} 5^4$ & 168 \\ \hline
 $(a6)^{4}$ & $2^{19} 3^{9} 5^4 7^4$ & 192 \\ \hline
 $(a7)^2(d5)^2 $ & $2^{31} 3^{6} 5^4 7^2$ & 216 \\ \hline
 $(a8)^3$ & $2^{23} 3^{13} 5^3 7^3$ & 240 \\ \hline 
 $(d6)^4$ & $2^{39} 3^{9} 5^4$ & 264  \\ \hline
 $(a9)^2 (d6)$ & $2^{27} 3^{10} 5^5 7^2$ & 264  \\ \hline
 $(e6)^4$ & $2^{32} 3^{17} 5^4 $ & 312 \\ \hline 
 $(a11)(d7)(e6)$ & $2^{28} 3^{11} 5^4 7^2 11$ & 312 \\ \hline
 $(a12)^2$ & $2^{22} 3^{10} 5^4 7^2 11^2 13^2$ & 336 \\ \hline
 $(d8)^3$ & $2^{43} 3^{7} 5^3 7^3$ & 360 \\ \hline
 $(a15)(d9)$ & $2^{31} 3^{10} 5^4 7^3 11 \cdot 13 $ & 408\\ \hline
 $(a17)(e7)$ & $2^{27} 3^{12} 5^4 7^3 11 \cdot 13 \cdot 17$ &  456 \\\hline
 $(d10)(e7)^{2}$ & $2^{38} 3^{12} 5^4 7^3$ & 456 \\ \hline 
 $(d12)^{2}$ &  $2^{43} 3^{10} 5^4 7^2 11^2 $ & 552\\ \hline
 $(a24)$ & $2^{23} 3^{10} 5^6 7^3 11^2 13 \cdot 17 \cdot 19 \cdot 23$ & 624 \\ \hline
 $(d16)(e8)$ & $2^{44} 3^{11} 5^5 7^3 11 \cdot 13 $ & 744 \\ \hline 
 $(e8)^{3}$ & $2^{43} 3^{16} 5^6 7^3$ &  744 \\ \hline
 $(d24)$  & $2^{45} 3^{10} 5^4 7^3 11^2 13 \cdot 17 \cdot 19 \cdot 23$ & 1128 \\
 \hline \hline
\end{tabular}
\caption{24-dimensional even self-dual lattices, and the size of their automorphism groups (see \cite{Conway1982}), and number of currents in the corresponding CFT.}
    \label{tab:lats}
\end{table}

With the data of table~\ref{tab:lats}, it is possible to confirm also the genus-2 average formula. This check can be done using the form \eqref{eq:paramintermsofcurrents} for the genus-2 partition function in terms of $N_c$. We find
\begin{equation}
   \frac1{m_{12}} \sum_\Lambda\frac{\Theta_{\Lambda}^{g=2}}{|\mathrm{Aut(\Lambda)}|} = \frac{441}{691}E_4^3+\frac{250}{691}E_6^2-\frac{36980665344000}{53678953}\chi_{12}\,.
\end{equation}
This agrees exactly with the known expression for $E_{12}$, see for instance (B.26) of \cite{Henriksson:2021qkt}.

\subsection{Discrete lattice average at large $c$}
\label{sec:lattice_average}

To study the lattice average at general dimension $c$, we would like to evaluate the genus 1 Eisenstein series $E_k$ for large values of $k$. A rather effective way to do this in practice is through the recurrence relation, which allows all $E_k$ to be written in terms of $E_4$ and $E_6$ (the fact that this is possible is a consequence of the fact that the ring of genus 1 modular forms has two generators):
\begin{align}
    d_{k} = 2 \zeta(2k + 4) (2k+3) k! E_{2k+4} \, , \qquad \sum_{k = 0}^n \binom{n}{k}d_k d_{n-k} = \frac{2n + 9}{3n+6} d_{n+2} \, .
\end{align}
\paragraph{Virasoro state expansion}
We can use this formula to compute some of the low-lying Eisenstein series. We find\footnote{We have pulled out powers of $q$ and $\eta(q)$ so that the degeneracies of Virasoro primaries may be easily read off from the expansion in parentheses.}
\begin{align}
    E_4 \quad &= \quad  \frac{\eta(q)^7}{q^{7/24}} \left(1-q + 248 q + 3875 q^2 + 30380 q^3\ldots \right)
  ,  \nonumber 
    \\
    E_8 \quad &= \quad  \frac{\eta(q)^{15}}{q^{15/24}} \left(1-q + 496 q + 69255 q^2 +2044760 q^3 \ldots \right) 
,    \nonumber \\
    E_{12} \quad &= \quad  \frac{\eta(q)^{23}}{q^{23/24}} \left(1-q + 118.819 q + 196764 q^2 + \ldots \right) 
 ,   \nonumber \\
    E_{16} \quad &= \quad  \frac{\eta(q)^{31}}{q^{31/24}} \left(1-q + 36.512 q + 148521.4 q^2 + \ldots \right)
,    \nonumber  \\
    E_{24} \quad &= \quad  \frac{\eta(q)^{47}}{q^{47/24}} \left(1-q + 48.0006 q + 5825.66 q^2 + \ldots \right) 
 ,   \nonumber \\
    E_{120} \quad &= \quad  \frac{\eta(q)^{239}}{q^{239/24}} \left(1-q + 240.00000000000003 q + 28919.000000000004 q^2 + \ldots \right) 
 ,   \nonumber \\
    E_{1200} \quad &= \quad  \frac{\eta(q)^{2399}}{q^{2399/24}} \left(1-q + 2400. \,  q + 2881199. \,  q^2 + \ldots \right) .
\label{eq:eisensteinexamples}
\end{align}
We see that there is a single $c = 8$ lattice -- the root lattice of $e_8$. Its lattice theta-function is proportional to $E_4$, as is actually required by~\eqref{eq:latticeaverage}. For $c = 16$ there are two lattices: $e_8^2$ and $d_{16}$. However they are isospectral, so again we find a single lattice theta function appears in the average, and as a result we have $E_8 \sim E_4^2$. Only at $c = 24$ do we actually need to average over different lattice theta functions, so we begin to find fractional coefficients. However, we see that as $c$ increases, the degeneracies of the average theory approach integral values. 

\subsection{The gap in $U(1)^n$ primaries}

The expansion of the Eisenstein series in section~\ref{sec:lattice_average} was in terms of Virasoro primaries, so we did not find a large gap in the spectrum. Let us now perform the corresponding calculation for the $U(1)^c$ primaries, which are given by 
\begin{align}
    \chi^{U(1)}_h(q) = \frac{q^h}{\eta(q)^c} \, .
\end{align}
Since $Z_{c} = E_{c/2} / \eta(q)^c$, we find that the $U(1)^c$ primary expansion is just the simple $q$ expansion of $E_{c/2}$. Let us first compute a few examples:
\begin{align}
    \begin{split}
        E_{12} \ &= \ 1 + 94.819\,  q + 194284 \, q^2 + \ldots \,, \\
        E_{120} \ &= \ 1 + \ldots + 0.400 q^7 + 3.185 q^8 + 3.893 \times 10^{12} q^9 + \ldots \,,\\
        E_{1200} \ &= \ 1 + \ldots + 7.451 \times 10^{-11} q^{69} + 0.00232 \,  q^{70}  + 56358.01 q^{71}+\ldots\,.
    \end{split}
\end{align}
We find that the degeneracy is tiny for small $h$, but it grows exponentially with $h$ so it goes from very small to very large. For $c = 2400$, we see that this happens at about $70$, which leads to an estimate of 
\begin{align}
    \frac{\Delta_{\text{gap}}}{c} \simeq \frac{70}{2400} \simeq \frac{1}{34.28}
\end{align}
for the gap. This is very close to the known asymptotic value
\begin{align}
    \frac{\Delta_{\text{gap}}}{c} = \frac{1}{4 \pi e},
\end{align}
the chiral version of the gap in $U(1)$ gravity discussed in \cite{Maloney:2020nni, Afkhami-Jeddi:2020ezh}.

\subsection{Contributions from disconnected boundaries}

Consider now the formula \eqref{eq:wormholecontr} for the wormhole contribution. For the case at hand, we replace the average enumerator polynomial by the average lattice theta function, which by the considerations above equals the holomorphic Eisenstein series,
\begin{equation}
    \nu_{\mathrm{wh}}(\tau,\tilde \tau)=\frac{ E_k^{(2)}(\tau\oplus\tilde\tau)-E_k^{(1)}(\tau) E_k^{(1)}(\tilde\tau)}{ E_k^{(1)}(\tau)E_k^{(1)}(\tilde\tau)}\,.
\end{equation}
We will consider the special value $\tau=\tilde\tau=i$, and consider the lattice average $\overline W_c^{(g)}=n_{\frac c2}E_{\frac c2}^{(g)}$ of \eqref{eq:latticeaverage}. In table~\ref{tab:Eisensteinconnected} we give the results of a numerical calculation for $c$ up to $72$. 
\begin{table}
    \centering
    \begin{tabular}{|ccccc|}
    \hline
           $c$ & $E_{\frac c2}^{(2)}(\diag(i,i))$ & $E_{\frac c2}^{(1)}(i)^2$ & $\nu_{\mathrm{wh}}$  & $\nu_{\mathrm{wh}}/2^{-\frac c2}$ 
           \\\hline
           $8$ & $2.11924560$& $2.11924560$ & $0$ & $0$
           \\
           $16$ & $4.49120191$& $4.49120191$ & $0$ & $0$
           \\
           $24$ & $3.87844092$ & $3.87672463$ & $4.427\cdot 10^{-4}$ & $1.813$
           \\
           $32$ & $4.03140806$  & $4.03132843$  & $1.975\cdot 10^{-5}$  &$1.294$
           \\
           $40$ & $3.99219188$ & $3.99218969$ & $5.472\cdot 10^{-7}$ & $0.574$
           \\
           $48$ & $4.00195361$ & $4.00195337$ & $6.180\cdot 10^{-8}$ & $1.037$
           \\
           $56$ & $3.99951175$ & $3.99951174$ & $4.325\cdot 10^{-9}$ & $1.161$
           \\
           $64$ & $4.00012207$ & $4.00012207$ & $2.159\cdot 10^{-10}$ & $0.928$
           \\
           $72$ & $3.99996948$ & $3.99996948$  & $1.397\cdot 10^{-11}$ &  $0.960$
           \\\hline
    \end{tabular}
    \caption{Wormhole contribution $\nu_{\mathrm{wh}}$ at $\tau=\tilde \tau=i$.}
    \label{tab:Eisensteinconnected}
\end{table}

\begin{figure}
    \centering
    \includegraphics[width=0.65\textwidth]{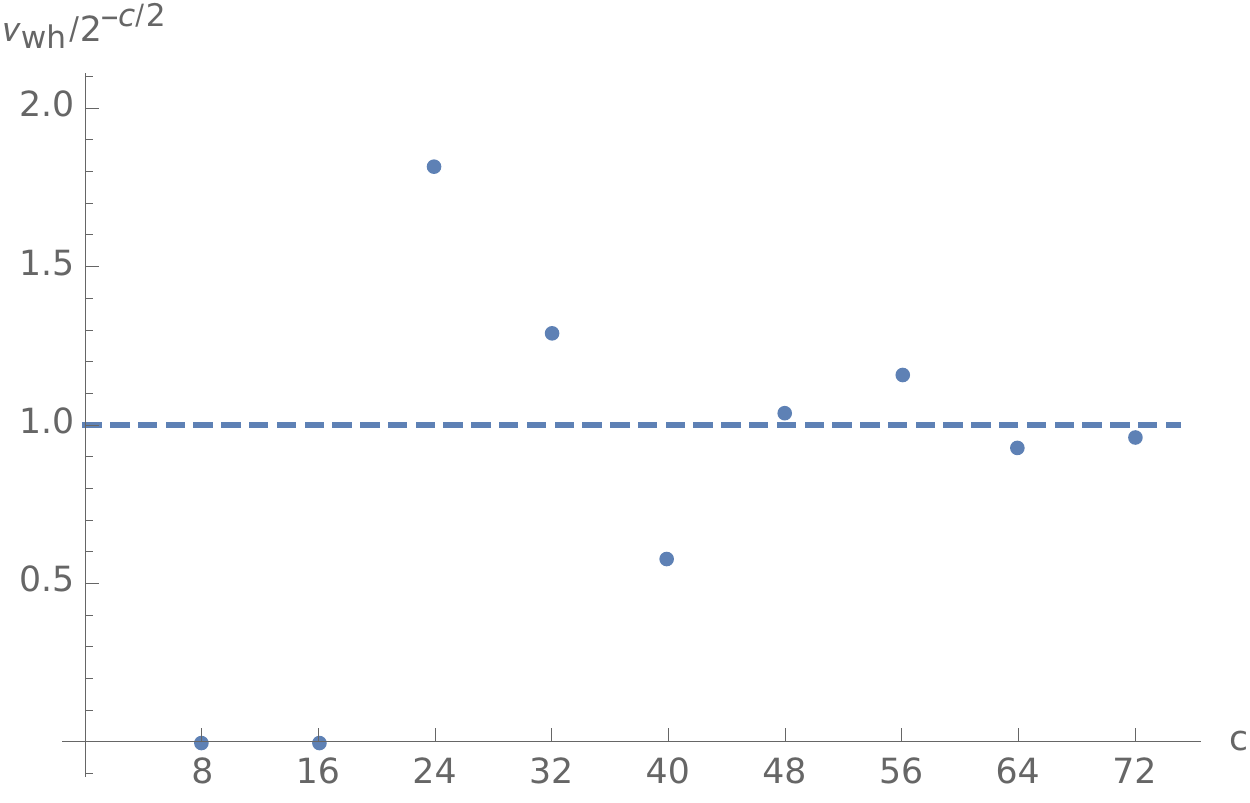}
    \caption{The wormhole contribution at $\tau=\tilde\tau=1$ for the lattice average, compared to the asymptotic value $2^{-c/2}$.}
    \label{fig:EisensteinWormhole}
\end{figure}

Our numerical investigations indicate that at $\tau=\tilde\tau=i$, the wormhole contribution approaches the value $2^{-c/2}$,\begin{equation}
\lim_{c\to\infty}  \frac{  \nu_{\mathrm{wh}}(i,i)}{2^{-\frac c2}}\to 1 ,
\end{equation} 
see figure~\ref{fig:EisensteinWormhole}. 

\section{Classification and data for of self-dual binary codes}
\label{app:data}

In this appendix we will provide some data and other details relevant to calculations in the main text. Table~\ref{tab:sdcodeslist1} contains a summary of the classification of type I and type II codes. The counting and references are taken from the database of Harada and Munemasa, which contains data files and tabulated results for many types of error-correcting codes \cite{Harada2015}.

\begin{table}
    \centering
    \begin{tabular}{|c|cr|cr|}\hline
         $n$ & \multicolumn{2}{c|}{Type I}  & \multicolumn{2}{c|}{Type II}
         \\
    \hline\hline
    $2$ & $1$ && &
    \\
        $4$ & $1$ && &
    \\
        $6$ & $1$ && &
    \\
    $8$ & $2$ && $1$ &
    \\
    $10$ & $2$ && &
    \\
        $12$ & $3$ && &
    \\
        $14$ & $4$ && &
    \\
    $16$ & $7$ && $2$ &
\\
    $18$ & $9$ && &
    \\
        $20$ & $16$ & \cite{Pless1972} & &
    \\
        $22$ & $25$ &\cite{Pless1975}  & &
    \\
$24$ & $55$ & \cite{Pless1975} & $9$ &\cite{Pless1975}
\\
    $26$ & $103$ &\cite{Conway1992} & &
    \\
        $28$ & $261$ &\cite{Conway1992} & &
    \\
        $30$ & $713$ & \cite{Conway1992} & &
    \\
$32 $& $3295$ & \cite{Bilous2002}  &  $85$ & \cite{Conway1992}
\\
    $34$ & $24147$ & \cite{Bilous2006} & &
    \\
        $36$ & $519492$ & \cite{Harada2012} & &
    \\
        $38$ & $38682183$ & \cite{Bouyuklieva2012} & &
    \\
$40$  & ? & &  $94343$ & \cite{Betsumiya2012} \\\hline
    \end{tabular}
    \caption{Number of self-dual binary codes, summary of results in \cite{Harada2015}. Note that type II is a subset of type I.}
    \label{tab:sdcodeslist1}
\end{table}

This database is very useful for explicit computations with codes of low and moderate length, so let us briefly outline the relevant content. The main category of codes relevant to this paper is found on the page \texttt{Binary self-dual codes}. This page in turn contains a table of the relevant enumeration, and links to \texttt{Magma} files containing the (transposed) generator matrices of the codes. For instance, the generator matrices for the Hamming \eqref{eq:GThamming} code is the only code listed in \texttt{8-II.magma}, and the generator matrix \eqref{eq:GTnew} for the $B_{12}$ code is found as the last of the three codes in \texttt{12.magma}.

To check the average formula \eqref{eq:typeIaver}, one needs access to the weights of the form $\frac1{|\mathrm{Aut}|}$, which for $n\leqslant 20$ can be extracted from \cite{Pless1972}. Unfortunately, the codes are not listed in the same order in the two references. The two orderings can be related for instance by computing the enumerator polynomial of the codes in the database, and compare with those given in \cite{Pless1972}. See table~\ref{tab:sdcodeslist2} for the first few cases.
\begin{table}
    \centering
    \begin{tabular}{|c|l|l|}\hline
         $n$ & Codes & $|\mathrm{Aut}|$
         \\
    \hline\hline
    $8$ & $\{i_2^4\}_{\text I}$, $\{e_8\}_{\text{II}}$ & $\{384\}_{\text I}$, $\{1344\}_{\text{II}}$
    \\\hline
    $10$ & $\{i_2^4e_8,\, i_2^5\}$ &  $\{2688,\, 3840\}$
    \\\hline
        $12$ & $\{i_2^6,\,i_2^2e_8,\,b_{12}\}$ & $\{10752,\, 46080,\, 23040\}$
    \\\hline
        $14$ & $\{i_2b_{12},\,i_2^7,\,i_2^3e_8,\,d_{14}\}$ & $\{46080,\, 645120,\, 64512,\, 56448\}$
    \\\hline
    \multirow{2}{*}{$16$} & $\{i_2^4e_8,\,i_2^2b_{12},\,i_2^8,\,i_2d_{14},\,f_{16}\}_{\text I}$ & $\{516096,\, 184320,\, 10321920,\, 112896,\, 73728\}_{\text{I}}$
    \\  &   $\{e_8^2,\,d^+_{16}\}_{\text{II}}$& $\{3612672,\, 5160960\}_{\text{II}}$
\\\hline
  \multirow{2}{*}{ $18$ }& $\{i_2^8,\,i_2e_8^2,\,i_2^2d_{14},\,i_2^5e_8,$ & $\{185794560,\, 7225344,\, 451584,\, 5160960$,
    \\
    &$i_2d_{16}^+,\,i_2f_{16},\,i_2^3b_{12},\,h_{18},\,I_{18} \}$& $10321920, 147456, 1105920, 82944, 322560\}$
     \\\hline
    \end{tabular}
    \caption{Codes and sizes of the automorphism groups. Here $i_2$ represents the repetition code, and product symbols are suppressed. }
    \label{tab:sdcodeslist2}
\end{table}

\subsection{9 type II codes at $c = 24$}
\label{app:typeII24data}

In section \ref{sec:codeaverage} of the main text, we considered the example of the average over the type II theories of length 24. The data needed to compute this average are presented in table \ref{tab:codes}.

\begin{table}
\centering
\begin{tabular}{| c | c | c |}
\hline
 $\mathcal{C}$ & $|\text{Aut}(\mathcal{C})|$ & $N_c$   \\
 \hline \hline
 $(a1)^{24}$ & $2^{10} 3^3 \cdot 5 \cdot 7 \cdot 11 \cdot 23 $ & 72 \\ \hline
 $(d4)^{6}$ & $2^{16} 3^3 5$ & 168 \\ \hline
 $(d6)^4$ & $2^{15} 3^{5}$ & 264 \\ \hline
 $(d8)^3$ & $2^{19} 3^{4} $  & 360\\ \hline
 $(d10)(e7)^{2}$ & $2^{14} 3^{3} 5 \cdot 7^2$ & 456 \\ \hline $(d12)^{2}$ &  $2^{19} 3^{4} 5^2 $ & 552 \\ \hline
 $(d16)(e8)$ & $2^{20} 3^{3} 5  \cdot 7^2$  & 744 \\ \hline 
 $(e8)^{3}$ & $2^{19} 3^{4} 7^3$ & 744 \\ \hline
 $(d24)$  & $2^{21} 3^{5} 5^2 7 \cdot 11$ & 1128 \\
 \hline \hline
\end{tabular}
\caption{24 dimensional doubly-even self-dual codes, the size of their automorphism groups, and the number of currents $N_c$ in the corresponding CFT.}
    \label{tab:codes}
\end{table}

\subsection{Spectrum of the code average}

The enumerator polynomial of an error-correcting code defines a lattice theta function via the ``Construction~A'' of Leech and Sloane,
\begin{align}
    \Theta_{\Lambda(\mathcal C)}(\tau) = W_{\mathcal C}\big( \theta_3(q^2), \theta_2(q^2) \big) \, .
\end{align}
We can use this to compute the spectra corresponding to the averaged CFT partition functions. Let us do this for a few explicit values of $c$ just as we did for the lattices in~\eqref{eq:eisensteinexamples}. Again, we pull out factors so that the coefficients are degeneracies of Virasoro characters:
\begin{align}
\nonumber
    \overline{W}_8 \quad &= \quad  \frac{\eta(q)^7}{q^{7/24}} \left(1-q + 248 q + 3875 q^2 + 30380 q^3\ldots \right) 
    ,\\
\nonumber
    \overline{W}_{16} \quad &= \quad  \frac{\eta(q)^{15}}{q^{15/24}} \left(1-q + 496 q + 69255 q^2 +2044760 q^3 \ldots \right) 
    ,\\
\nonumber
    \overline{W}_{24}  \quad &= \quad  \frac{\eta(q)^{23}}{q^{23/24}} \left(1-q + 237.869\, q + 196645.13 \,  q^2 + \ldots \right)
    , \\
\nonumber
    \overline{W}_{32}  \quad &= \quad  \frac{\eta(q)^{31}}{q^{31/24}} \left(1-q + 131.115 \, q + 171888.42 \, q^2 + \ldots \right) 
    ,\\
\nonumber
    \overline{W}_{48}  \quad &= \quad  \frac{\eta(q)^{47}}{q^{47/24}} \left(1-q + 144.742\,  q + 33330.8 \,  q^2 + \ldots \right) 
\nonumber
    ,\\
    \overline{W}_{240}  \quad &= \quad  \frac{\eta(q)^{239}}{q^{239/24}} \left(1-q + 720.000000000000000000000000006 \, q + 258359. q^2 + \ldots \right) 
\nonumber
    ,\\
    \overline{W}_{2400}  \quad &= \quad  \frac{\eta(q)^{2399}}{q^{2399/24}} \left(1-q + 7200. \,  q + 25911599. \,  q^2 + \ldots \right) .
\label{eq:averagecodeexamples}
\end{align}
For $c = 8$ and $c = 16$, the results are the same as in \eqref{eq:eisensteinexamples}. That is due to the fact that the lattices for these dimensions come from codes. Again, we see the enhanced $e_8$ symmetry, and the corresponding 248 currents for $c = 8$. However as $c$ increases, a strikingly different pattern appears: the number of currents is three times the central charge.\footnote{More precisely, the number of currents in the average theory is $3c+\frac{8c(c-1)(c-2)(c-3)}{3(2^{c/2}+4)}$.} This is a signature of the $SU(2)^c$ enhanced symmetry. Also, based purely the number of zeros (26) following the decimal in the expression for the $q$ term in $\overline{W}_{240}$, we speculate that this ensemble may be approaching its asymptotic value ``faster'' than the Narain lattice ensemble, where there are only 13 zeros after the decimal. In fact, perhaps it is approaching it ``twice as fast'' in some sense. If this is true, we do not know the meaning of it.

\section{Bounds on the Hamming Distance}
\label{app:distance bounds}

The gap in $SU(2)$ primaries is directly related to the Hamming distance of the code by $\Delta_{\text{gap}} = d / 4$. Proving bounds on the maximal Hamming distance obtainable by a code with given properties is an important general problem in coding theory. In general, there is a tradeoff between the efficiency of encoding versus the amount of error-correction -- that is, information may be encoded more densely in codes with a large number of codewords, but increasing the number of codewords may force you to accept a lower Hamming distance. In this paper we are always interested in self-dual codes, whose size is given by $2^{n/2}$. Therefore the relevant bounds will be upper bounds on the relative Hamming distance $\delta = d / n$. Let us briefly discuss a few simple bounds.\footnote{We found \cite{Guruswami2010} to be a nice pedagogical introduction.}

\paragraph{Gilbert-Varshamov bound}

A bound can be proven for general (\emph{i.e.} non-self-dual) codes, which have $2^k$ elements. Consider the code which has the maximal number of codewords for any code of length $n$ and Hamming distance $d$. We first define a \emph{ball} with radius $\ell$ centered at a codeword $c$ as the set of codewords $c'$ with Hamming distance satisfying $d(c, c')<\ell$. With this definition, one can show that the volume of a ball with radius $\ell$ is given by
\begin{align}
    \text{Vol}(n, \ell) \ = \ \sum_{j = 0}^\ell \begin{pmatrix} n \\ j \end{pmatrix}  \ = \ 2^n -   \begin{pmatrix}n \\ \ell + 1 \end{pmatrix} {}_2 F_1(1, \ell + 1 -n; \ell + 2; -1)  \, .
\end{align}
This volume is the number of other possible codewords -- elements of the unit $n$-cube -- contained in the ball.

Now center a ball with radius $d - 1$ at each codeword. The union of such balls must cover all the corners of unit $n$-cube -- if not, then we could add a codeword to the uncovered corner to make the code bigger without decreasing the Hamming distance $d$. Then the volume of $2^k$ such balls must exceed the volume of the unit cube. This leads to 
\begin{align}
    2^k \text{Vol}(n, d-1)  \geqslant 2^n,
\end{align}
If $k = n / 2$, as it does for self-dual codes, we can compute this to get bound on $n$ and $d$
\begin{align}
    \text{Vol}(n, d-1)  > 2^{n/2}\, .
\end{align}
Using $d = \delta n$ we find the asymptotic bound $ h(\delta) < \frac{1}{2}$. This is the so-called Gilbert--Varshamov bound, and it is the same bound we find from the averaging argument above. However, this argument does not apply to self-dual codes, because the process of adding a codeword may spoil the self-duality condition of the code. It is interesting that it yields the same result as the averaging argument given above.

\paragraph{Hamming bound}

The Hamming bound is a complementary bound to the Gilbert-Varshamov bound, and tells you that you can't have any code with relative distance greater than $2 \delta^*$. Its proof is conceptually similar. Consider a code with length $n$, $2^k$ codewords, and distance $d$. Center a ball of radius $\lfloor \frac{d-1}{2} \rfloor$ at each codeword. Because the distance is $d$, it is impossible that any such ball overlaps with any other ball. Therefore the total volume of all balls must be less than or equal to the total volume of the unit cube. Applying it to self-dual codes, we choose $n = 2k$ and find the bound
\begin{align}
    \text{Vol}\left(n, \lfloor \tfrac{d-1}{2} \rfloor \right)  \leqslant 2^{n/2}\, .
\end{align}

Since this bound simply arises from considering the maximum number of codewords that could fit into the unit cube, it is also called the volume bound. Codes that saturate this bound are called perfect codes. It turns out that the only perfect codes are the Hamming code, the Golay code, and trivial codes (empty code, code with all codewords, and $n$-repetition code). Asymptotically, it gives a bound on the relative distance of any possible code $ \delta_{\text{max}} \leqslant 2 \delta^* $.

\paragraph{Further bounds}

A number of stronger bounds, using more complicated arguments, are known in the literature. Let us list a few, specialized to the case where $n = 2k$ (see \cite{Guruswami2010} for the proofs):
\begin{itemize}
    \item Singleton bound: $\delta \leqslant \frac{1}{2}$,
    \item Elias-Bassalygo bound: $\delta \leqslant 0.196$,
    \item first MRRW bound: $\delta \leqslant 0.187$,
    \item second MRRW bound: $\delta \leqslant 0.1824$.
\end{itemize}
The second MRRW bound, given in \cite{McEliece1977}, is currently the best known asymptotic bound for general codes.

\paragraph{A bound from a positive functional}

Now we will formulate a simple bound in a language more familiar from the modular bootstrap. Consider a type I enumerator polynomial at generic $n$. Then we have
\begin{align}
    W(x, y) = x^n + a_2 x^{n-2} y + \ldots + a_n y^n \, .
\end{align}
Then modular invariance implies that this must be invariant under $x \to (x + y) / \sqrt{2}$, $y \to (x-y)/
\sqrt{2}$. Therefore we can subtract two equivalent representations of $W$ to write a bootstrap equation:
\begin{align}
    0 = \left(x^n - \left(\frac{x+y}{\sqrt{2}} \right)^n \right) + \sum_m  a_{2m} \left(x^{n-2m}y^{2m} - \left(\frac{x+y}{\sqrt{2}} \right)^{n-2m}\left(\frac{x-y}{\sqrt{2}} \right)^{2m} \right) \, ,
\end{align}
where we have broken the sum into its ``identity'' component, and the rest. It is clear that the vacuum part will be annihilated by the operator 
\begin{align}
    \alpha[f] = \left(\partial_y^2 - \frac{n - 1}{\sqrt{2}} \partial_y \right) f|_{x = 1, y = \sqrt{2} - 1 } \, ,
\end{align}
where $x = 1$, $y = \sqrt{2} - 1$ is the ``self-dual point.''  Then acting with $\alpha$ on the sum part gives us a sum rule:
\begin{align}
    0 = \sum_{m=0}^n a_{2m} (\sqrt{2} - 1)^{2 m} (2^m - 1) m \left(3 - 2 \sqrt{2} - (  6 \sqrt{2}-8) m + ( 5 \sqrt{2}-7)n \right)  \, .
\end{align}
For large $n$, this implies that there must be non-zero $a_{2m}$ for $m$ satisfying
\begin{align}
m < \frac{ 5 \sqrt{2}-7 }{  6 \sqrt{2}-8}n = \frac{ 2- \sqrt{2} }4n\,.
\end{align}
This implies that the relative gap must satisfy
\begin{align}
    \delta < \frac{2-\sqrt{2}}{2} \simeq .292893\,.
\end{align}
This is weaker than the Hamming bound.

\section{A theory with $c=10$ and $\bar c=2$}
\label{app:cc12theory}

In this paper we have focused on the case where $c = \bar c$ for simplicity. Here we will use the example of $ c = 10$, $\bar c = 2$ to demonstrate that there may be interesting theories for unequal values of $c$ and $\bar c$.\footnote{To be truly modular invariant (at genus 1), these theories should have $c - \bar c \equiv 0 \text{ (mod $24$)}$. The discussion here will yield partition functions which are modular invariant in the same sense that the Hamming code partition function is modular invariant. That is, they transform with phases under $T$.} First, we can write the most general genus 1 polynomial 
\begin{align}
    P_\text{gen} = x^{10} \bar x^2 + \ldots
\end{align}
and then impose modular invariance and positive integer coefficients. The result is that there are three solutions. One of the solutions is
\begin{align}
    P_1 = (x^8 + 14 x^4 y^4 + y^8) (x \bar x + y \bar y) \, ,
\end{align}
which we can immediately recognize as the theory arising from the Hamming code times the repetition code. Of the other two partition functions, one of them does not arise from factorizing a genus 2 partition function, so we can eliminate it. The one we are left with is
\begin{align}
    P_2 = x^2 \bar x^2 (x^8 + 10 x^4 y^4 + 5 y^8) + 32 x^5 \bar x y^5 \bar y + y^2 \bar y^2(5 x^8 + 10 x^4 y^4 + y^8).
    \label{eq:c10cb2PF}
\end{align}
In fact, we can find a lattice theory that gives this very partition function using the construction in section~\ref{sec:Symmetry}. One starts with the $b_{12}$ code from table~\ref{tab:typeIaver}, and then chooses 10 plus signs and 2 minus signs for the metric signature such that all codewords satisfy $w_+(c)-w_-(c)\equiv 0\ (\mathrm{mod}\ 4)$, \emph{i.e.} such that the code is type $\hat{\mathrm{I}}$.
This leads to the partition function~\eqref{eq:c10cb2PF}, verifying that this invariant polynomial indeed corresponds to an actual theory.

From the partition function we can extract the number of Virasoro primaries for low spin $\ell$ and scaling dimension $\Delta$. We find that there are $6$ states with $(\Delta, \ell) = (1, -1)$, which is what we would expect given that there are 2 (right-moving) copies of $SU(2)$ in the symmetry group. At $(\Delta, \ell) = (1, 1)$, we find 190 states, suggesting a large enhanced symmetry in the left-moving sector.

\section{Chiral partition functions}
\label{app:denominator}

In the main text of this paper, we encountered genus-$g$ partition functions for CFTs with central charges $c,\bar c$ taking the general form
\begin{equation}
\label{eq:pfApp}
    Z(\Omega,\bar \Omega)=\frac{W^{(g)}(\vartheta_i(\Omega),\overline{ \vartheta_i(\Omega)}\,)}{{\Phi_g(\Omega)}^{c}\overline{\Phi_g(\Omega)}^{ \bar c}}\,.
\end{equation}
Here $W^{(g)}(x_i,\bar x_i)$ is the weight-$g$ enumerator polynomial, which is of homogeneous degree $c$ in the $x_i$ and $\bar c$ in the $\bar x_i$.
The numerator represents the (theory dependent) sum over winding modes, or equivalently a sum over the momentum lattice. The denominator represents a one-loop determinant and is universal for all our theories, which are free theories of $c$ chiral and $\bar c$ antichiral bosons. 
In the case of $c=\bar c=\frac n2$, we can write this denominator as
\begin{equation}
\label{eq:denominatorChiral}
|\Phi_g|^n=\frac{({\det}'\Delta_0)^{n/2}}{(\det\IM\Omega)^{n/2}}\,,
\end{equation}
where ${\det}'\Delta_0$ represents the determinant of the Laplacian on the Riemann surface with zero modes removed.\footnote{The factor $(\det\IM\Omega)^{-n/2}$ is the would-be contribution from momenta if the bosons were not compactified.} Such determinants appear naturally in the quantization of the Polyakov string \cite{Polyakov:1981rd}, and $|\Phi_g|$ was evaluated at arbitrary genus in \cite{Belavin:1986cy} in terms of the Selberg zeta function, see also \cite{DHoker:1986eaw,Voros:1986vw,Sarnak1987}. The expression \eqref{eq:denominatorChiral} has the correct form to make the partition function \eqref{eq:pfApp} modular invariant, or in other words, for $c=\bar c$, \eqref{eq:pfApp} is a (single-valued) function on the moduli space $\mathcal M_g$ of genus-$g$ Riemann surfaces.\footnote{More precisely, $\mathcal M_g=\mathfrak M_{g,1}$ is the moduli space of genus-$g$ Riemann surfaces with one marked point.}

For $c\neq\bar c$, the situation is more complicated, and requires some further comments on the holomorphic dependence on $\Omega$. It is sufficient to consider the case $\bar c=0$ and $c$ divisible by $8$. In this case, we would like to find an expression for the partition function of the form
\begin{equation}
\label{eq:pfHolo}
    Z(\Omega)=\frac{W^{(g)}(\vartheta_i(\Omega))}{{\Phi_g(\Omega)}^{c}}\,,
\end{equation}
where $Z(\Omega)$ is a (single-valued) holomorphic function on $\mathcal M_g$. 
There are some obstructions to the existence of such functions, which we would like to briefly review. For more details and precise statements, we refer to the references \cite{Krasnov:2000zq}, \cite{Gaberdiel:2010jf} and \cite{Tan:2014mba}.

By construction, the numerator $W^{(g)}(\vartheta_i(\Omega))$ is a Siegel modular form of degree $k=\frac c2$. Siegel modular forms are functions $f_k$ on the Siegel upper half plane
\begin{equation}
    \mathcal H_g=\left\{\Omega\in \mathrm{Mat}_{g\times c}(\C)\middle|\IM\Omega\succcurlyeq0,\,\Omega=\Omega^T\right\},
\end{equation}
which transform as 
\begin{equation}
\label{eq:modularformTF}
	f_k(\Lambda\Omega)=\det(C\Omega+D)^kf_k(\Omega), 
\end{equation}
where 
\begin{equation}
 \Lambda\Omega=(A\Omega+B)(C\Omega+D)^{-1},   \qquad \Lambda=\begin{pmatrix}
A&B\\C&D
\end{pmatrix}\in \mathrm{Sp}(2g,\Z).
\end{equation}
The relation of $ \mathcal H_g$ to the moduli space is as follows. The space of inequivalent period matrices $\Omega$ is corresponds to the quotient $\mathcal A_g=\mathrm{Sp}(2g,\Z) \backslash\mathcal H_g$, known as the moduli space if principally polarized Abelian varieties. The Jacobian map $J:\mathcal M_g\longrightarrow \mathcal A_g$ maps a Riemann surface to its period matrix, or more precisely to the quotient space $\C^g/\{\vek m+\Omega\vek n,\ \vek m,\,\vek n\in \Z^g\}$. For $g\leqslant3$, $ J(\mathcal M_g)$ is dense inside $\mathcal A_g$.\footnote{Recall the dimension formulas $\dim_{\C}\mathcal M_g=3g-3$ and $\dim_{\C}\mathcal A_g=\dim_{\C}\mathcal H_g=\frac{g(g+1)}2$ for $g\geqslant2$ ($\dim_{\C}\mathcal A_1=\dim_{\C}\mathcal M_1=1$). For $g\geqslant4$, we have $\dim_{\C}\mathcal A_g>\dim_{\C}\mathcal M_g$.}
Functions on $\mathcal H_g$ satisfying \eqref{eq:modularformTF} descend to sections of $\lambda^k(\mathcal A_g)$, the $k$th power of the determinant line bundle on $\mathcal A_g$, and via the Jacobian map to sections of $\lambda^k(\mathcal M_g)$.

To make $Z^{(g)}(\Omega)$ modular invariant and well-defined, we would like $\Phi_g(\Omega)^c$ to be a nowhere-vanishing modular form of degree $\frac c2$. At genus $1$, such modular forms exist for degree $12k$ and are proportional to powers of the Dedekind $\eta$ function. This precisely agrees with the known formula for the genus-1 denominator,
\begin{equation}
\label{eq:Phi1holo}
    \Phi_1(\tau)=\eta(\tau)=q^{\frac1{24}}\prod_{m=1}^\infty(1-q^m)\,.
\end{equation}
This conclusion is the well-known result that chiral genus-1 modular invariant partition functions only exist for $c=24k$. For other values of $c$, the partition function is only invariant up to phases induced by the modular $T$ transformations.

The situation at higher genus is not that simple, since for $g\geqslant 3$, there are no nowhere-vanishing Siegel modular forms.\footnote{For genus 2, there is the degree-10 modular form $\chi_{10}$, proportional to the product over ten theta functions. It appears however to be unrelated to $\Phi_2$, as can be seen from the expansion \eqref{eq:F2explicit} below.}
To discuss this further, we need to introduce some additional spaces, namely Teichm\"uller space $\mathcal T_g$, which is the universal cover of $\mathcal M_g$, and Schottky space $\mathfrak S_g$, which is a partial cover. In summary, the spaces we consider are related as follows
\begin{equation}
\label{eq:spaces}
\begin{tikzcd}
\mathcal H_g \arrow[rr,twoheadrightarrow]  &  & \mathcal A_g \arrow[d,hookleftarrow,"\,J"] 
\\
\mathcal T_g \arrow[r,twoheadrightarrow] & \mathfrak S_g \arrow[r,twoheadrightarrow] & \mathcal M_g ,
\end{tikzcd}
\end{equation}
At genus 1, $\mathcal M_1$ and $\mathcal A_1$ correspond to the region $|\tau|>1$, $-\frac12<\RE\tau<\frac12$, both $\mathcal H_1$ and $\mathcal T_1$ can be seen as the upper half-plane $\IM\tau>0$ and $\mathfrak S_1$ is the region $-\frac12<\RE\tau<\frac12$. At higher genus, Schottky space is rather complicated, see e.g. \cite{Gaberdiel:2010jf}.

A key result is the holomorphic factorization theorem by Zograf and Takhtajan \cite{Zograf1987,Zograf1987b}, which says that $|\Phi_g|$ can be written as\footnote{The global existence of $T_g$ on Teichm\"uller space was proven by Zograf \cite{Zograf1989}, see discussion and references in \cite{McIntyre2002,McIntyre:2004xs}.}
\begin{equation}
\label{eq:PhigFactorization}
|\Phi_g|^2=\exp\left(\frac{S}{12\pi}\right)|F_g|^2
\end{equation}
where $F_g$ is a holomorphic function on Schottky space $\mathfrak S_g$. 
It is the direct generalization of the infinite product in \eqref{eq:Phi1holo}, taking the form
\begin{equation}
\label{eq:Fprod}
F_g=\prod_\gamma\prod_{m=1}^\infty(1-p_\gamma^m).
\end{equation}
Here the outer product is over primitive elements $\gamma$ in the Schottky uniformization group. For details on this formula and coordinates on Schottky space, see for instance the appendices of \cite{Gaberdiel:2010jf}. Product formulas like \eqref{eq:Fprod}, but without the ``anomaly'' term $\exp(\frac S{12\pi})$, were known from the early days of string theory \cite{Alessandrini:1971dd,Montonen:1974jj}, see \cite{DiVecchia:1987uf}.
The exponential factor of \eqref{eq:PhigFactorization} contains the classical Liouville action $S$, which is a real-valued function that was determined in \cite{Zograf1987,Zograf1987b}. In fact, for $g\geqslant2$, it is the K\"ahler potential of the Weil--Petersson metric on $\mathcal T_g$, $\partial\bar\partial S=-2i\omega_{\mathrm{WP}}$. 

At genus 1, $S=2\pi\log |q|$, and \eqref{eq:PhigFactorization} reduces to \eqref{eq:Phi1holo}. At higher genus, on the other hand, the factor $\exp(\frac S{24\pi})$ provides an obstacle to holomorphicity of $\Phi_g$. One option is to follow \cite{Tan:2014mba} and include this real factor in $\Phi_g$. This does not solve the problem of creating a (single-valued holomorphic) function on $\mathcal M_g$.
Note that this problem does not occur for the case $c=\bar c$, where all phases cancel.

Another approach would be to only work with the numerator $W(\vartheta_i(\Omega))$, and treat it as the partition function. It would then be a section of the $\frac c2$th power of the determinant line bundle of $\mathcal M_g$.

\paragraph{Example} 
We conclude by giving some explicit expressions in the case of genus $g=2$, extracted from \cite{Gaberdiel:2010jf}. They gave an expansion for $F_2$ in terms of coordinates on Schottky space, and the relation to the  usual modular parameters $q_1=e^{2\pi i\Omega_{11}}$, $q_2=e^{2\pi i\Omega_{22}}$ and $r=e^{2\pi i\Omega_{12}}$:
\begin{align}
F_2(q_1,q_2,r)&=1-\left(q_1+q_2\right)  + \left(-q_2 q_1 \left(r^2+\tfrac{1}{r^2}\right)+6 q_2 q_1 \left(r+\tfrac{1}{r}\right)-q_1^2-9 q_2
   q_1-q_2^2\right)
\nonumber\\&\quad+ q_1q_2\left(-9 \left(q_1+q_2\right) \left(r^2+\tfrac{1}{r^2}\right)+40 \left(q_1+q_2\right)
   \left(r+\tfrac{1}{r}\right)-61 \left(q_1+q_2\right)\right)
\nonumber\\&\quad +q_1q_2 \bigg[-q_2 q_1 \left(r^4+\tfrac{1}{r^4}\right)+2 \left(3 (q_1^2+q_2^2)+20 q_2 q_1 \right) \left(r^3+\tfrac{1}{r^3}\right)
\nonumber\\&\quad\qquad\quad-\left(61 (q_1^2+q_2^2)+296 q_2 q_1\right) \left(r^2+\tfrac{1}{r^2}\right)-270 (q_1^2+q_2^2)-1181 q_2 q_1
\nonumber
\\&\quad\qquad\quad+2 \left(95( q_1^2+q_2^2)+424 q_2  q_1\right) \left(r+\tfrac{1}{r}\right)
\bigg]+O(q_i^5)
\label{eq:F2explicit}
\end{align}
One can check that in the factorization limit $r\to 1$, 
\begin{equation}
F_2(q_1,q_2,r)\to F_1(q_1)F_1(q_2)= q_1^{-\frac1{24}}q_2^{-\frac1{24}}\Phi_1(q_1)\Phi_1(q_2) \,.
\end{equation}

\bibliography{cite.bib}

\providecommand{\href}[2]{#2}\begingroup\raggedright\begin{thebibliography}{10}

\bibitem{LeechSloane1971}
J.~Leech and N.J.A.~Sloane, \emph{{Sphere packings and error-correcting
  codes}}, \href{https://doi.org/10.4153/CJM-1971-081-3}{\emph{Canad. J. Math.}
  {\bfseries 23} (1971) 718}.

\bibitem{Narain:1985jj}
K.S.~Narain, \emph{{New heterotic string theories in uncompactified dimensions
  $< 10$}}, \href{https://doi.org/10.1016/0370-2693(86)90682-9}{\emph{Phys.
  Lett. B} {\bfseries 169} (1986) 41}.

\bibitem{Dolan:1989kf}
L.~Dolan, P.~Goddard and P.~Montague, \emph{{Conformal field theory, triality
  and the Monster group}},
  \href{https://doi.org/10.1016/0370-2693(90)90821-M}{\emph{Phys. Lett. B}
  {\bfseries 236} (1990) 165}.

\bibitem{Dolan:1994st}
L.~Dolan, P.~Goddard and P.~Montague, \emph{{Conformal field theories,
  representations and lattice constructions}},
  \href{https://doi.org/10.1007/BF02103716}{\emph{Commun. Math. Phys.}
  {\bfseries 179} (1996) 61}
  [\href{https://arxiv.org/abs/hep-th/9410029}{{\ttfamily hep-th/9410029}}].

\bibitem{Gaiotto:2018ypj}
D.~Gaiotto and T.~Johnson-Freyd, \emph{{Holomorphic SCFTs with small index}},
  \href{https://doi.org/10.4153/S0008414X2100002X}{\emph{Canad. J. Math.}
  {\bfseries 74} (2022) 573}
  [\href{https://arxiv.org/abs/1811.00589}{{\ttfamily 1811.00589}}].

\bibitem{Dymarsky:2020qom}
A.~Dymarsky and A.~Shapere, \emph{{Quantum stabilizer codes, lattices, and
  CFTs}}, \href{https://doi.org/10.1007/JHEP03(2021)160}{\emph{JHEP} {\bfseries
  21} (2020) 160} [\href{https://arxiv.org/abs/2009.01244}{{\ttfamily
  2009.01244}}].

\bibitem{Dymarsky:2020bps}
A.~Dymarsky and A.~Shapere, \emph{{Solutions of modular bootstrap constraints
  from quantum codes}},
  \href{https://doi.org/10.1103/PhysRevLett.126.161602}{\emph{Phys. Rev. Lett.}
  {\bfseries 126} (2021) 161602}
  [\href{https://arxiv.org/abs/2009.01236}{{\ttfamily 2009.01236}}].

\bibitem{Dymarsky:2021xfc}
A.~Dymarsky and A.~Sharon, \emph{{Non-rational Narain CFTs from codes over
  $F_{4}$}}, \href{https://doi.org/10.1007/JHEP11(2021)016}{\emph{JHEP}
  {\bfseries 11} (2021) 016}
  [\href{https://arxiv.org/abs/2107.02816}{{\ttfamily 2107.02816}}].

\bibitem{Yahagi:2022idq}
S.~Yahagi, \emph{{Narain CFTs and error-correcting codes on finite fields}},
  \href{https://doi.org/10.1007/JHEP08(2022)058}{\emph{JHEP} {\bfseries 08}
  (2022) 058} [\href{https://arxiv.org/abs/2203.10848}{{\ttfamily
  2203.10848}}].

\bibitem{Angelinos:2022umf}
N.~Angelinos, D.~Chakraborty and A.~Dymarsky, \emph{{Optimal Narain CFTs from
  codes}}, \href{https://doi.org/10.1007/JHEP11(2022)118}{\emph{JHEP}
  {\bfseries 11} (2022) 118}
  [\href{https://arxiv.org/abs/2206.14825}{{\ttfamily 2206.14825}}].

\bibitem{Cotler:2016fpe}
J.S.~Cotler, G.~Gur-Ari, M.~Hanada, J.~Polchinski, P.~Saad, S.H.~Shenker
  et~al., \emph{{Black holes and random matrices}},
  \href{https://doi.org/10.1007/JHEP05(2017)118}{\emph{JHEP} {\bfseries 05}
  (2017) 118} [\href{https://arxiv.org/abs/1611.04650}{{\ttfamily
  1611.04650}}]. [Erratum: {\em JHEP} {\bf 09} (2018) 002].

\bibitem{Saad:2018bqo}
P.~Saad, S.H.~Shenker and D.~Stanford, \emph{{A semiclassical ramp in SYK and
  in gravity}},  \href{https://arxiv.org/abs/1806.06840}{{\ttfamily
  1806.06840}}.

\bibitem{Saad:2019lba}
P.~Saad, S.H.~Shenker and D.~Stanford, \emph{{JT gravity as a matrix
  integral}},  \href{https://arxiv.org/abs/1903.11115}{{\ttfamily 1903.11115}}.

\bibitem{Maloney:2020nni}
A.~Maloney and E.~Witten, \emph{{Averaging over Narain moduli space}},
  \href{https://doi.org/10.1007/JHEP10(2020)187}{\emph{JHEP} {\bfseries 10}
  (2020) 187} [\href{https://arxiv.org/abs/2006.04855}{{\ttfamily
  2006.04855}}].

\bibitem{Afkhami-Jeddi:2020ezh}
N.~Afkhami-Jeddi, H.~Cohn, T.~Hartman and A.~Tajdini, \emph{{Free partition
  functions and an averaged holographic duality}},
  \href{https://doi.org/10.1007/JHEP01(2021)130}{\emph{JHEP} {\bfseries 01}
  (2021) 130} [\href{https://arxiv.org/abs/2006.04839}{{\ttfamily
  2006.04839}}].

\bibitem{Dijkgraaf:2000fq}
R.~Dijkgraaf, J.M.~Maldacena, G.W.~Moore and E.P.~Verlinde, \emph{{A black hole
  Farey tail}},  \href{https://arxiv.org/abs/hep-th/0005003}{{\ttfamily
  hep-th/0005003}}.

\bibitem{Maloney:2007ud}
A.~Maloney and E.~Witten, \emph{{Quantum gravity partition functions in three
  dimensions}}, \href{https://doi.org/10.1007/JHEP02(2010)029}{\emph{JHEP}
  {\bfseries 02} (2010) 029} [\href{https://arxiv.org/abs/0712.0155}{{\ttfamily
  0712.0155}}].

\bibitem{Keller:2014xba}
C.A.~Keller and A.~Maloney, \emph{{Poincare series, 3D gravity and CFT
  spectroscopy}}, \href{https://doi.org/10.1007/JHEP02(2015)080}{\emph{JHEP}
  {\bfseries 02} (2015) 080} [\href{https://arxiv.org/abs/1407.6008}{{\ttfamily
  1407.6008}}].

\bibitem{Dymarsky:2020pzc}
A.~Dymarsky and A.~Shapere, \emph{{Comments on the holographic description of
  Narain theories}}, \href{https://doi.org/10.1007/JHEP10(2021)197}{\emph{JHEP}
  {\bfseries 10} (2021) 197}
  [\href{https://arxiv.org/abs/2012.15830}{{\ttfamily 2012.15830}}].

\bibitem{Henriksson:2021qkt}
J.~Henriksson, A.~Kakkar and B.~McPeak, \emph{{Classical codes and chiral CFTs
  at higher genus}}, \href{https://doi.org/10.1007/JHEP05(2022)159}{\emph{JHEP}
  {\bfseries 05} (2022) 159}
  [\href{https://arxiv.org/abs/2112.05168}{{\ttfamily 2112.05168}}].

\bibitem{Henriksson:2022dnu}
J.~Henriksson, A.~Kakkar and B.~McPeak, \emph{{Narain CFTs and quantum codes at
  higher genus}}, \href{https://doi.org/10.1007/JHEP04(2023)011}{\emph{JHEP}
  {\bfseries 04} (2023) 011}
  [\href{https://arxiv.org/abs/2205.00025}{{\ttfamily 2205.00025}}].

\bibitem{Harada2015}
M.~Harada and A.~Munemasa, \emph{{Database of self-dual codes}}, .
  \href{https://www.math.is.tohoku.ac.jp/~munemasa/selfdualcodes.htm}{URL}.
  Accessed 25 January 2022.

\bibitem{Hartman:2019pcd}
T.~Hartman, D.~Maz\'a\v{c} and L.~Rastelli, \emph{{Sphere packing and quantum
  gravity}}, \href{https://doi.org/10.1007/JHEP12(2019)048}{\emph{JHEP}
  {\bfseries 12} (2019) 048}
  [\href{https://arxiv.org/abs/1905.01319}{{\ttfamily 1905.01319}}].

\bibitem{Viazovska2017}
M.~Viazovska, \emph{{The sphere packing problem in dimension 8}},
  \href{https://doi.org/10.4007/annals.2017.185.3.7}{\emph{Annals of Math.}
  {\bfseries 185} (2017) 991}
  [\href{https://arxiv.org/abs/1603.04246}{{\ttfamily 1603.04246}}].

\bibitem{Cohn2017}
H.~Cohn, A.~Kumar, S.~Miller, D.~Radchenko and M.~Viazovska, \emph{{The sphere
  packing problem in dimension 24}},
  \href{https://doi.org/10.4007/annals.2017.185.3.8}{\emph{Annals of Math.}
  {\bfseries 185} (2017) 1017}
  [\href{https://arxiv.org/abs/1603.06518}{{\ttfamily 1603.06518}}].

\bibitem{Pless1975}
V.~Pless and N.J.A.~Sloane, \emph{{On the classification and enumeration of
  self-dual codes}},
  \href{https://doi.org/10.1016/0097-3165(75)90042-4}{\emph{J. Combin. Theory
  Ser. A} {\bfseries 18} (1975) 313}.

\bibitem{Krasikov2000}
I.~Krasikov and S.~Litsyn, \emph{{An improved upper bound on the minimum
  distance of doubly-even self-dual codes}},
  \href{https://doi.org/10.1109/18.817527}{\emph{IEEE Transactions on
  Information Theory} {\bfseries 46} (2000) 274}.

\bibitem{Rains2003}
E.~Rains, \emph{{New asymptotic bounds for self-dual codes and lattices}},
  \href{https://doi.org/10.1109/TIT.2003.810623}{\emph{IEEE Transactions on
  Information Theory} {\bfseries 49} (2003) 1261}.

\bibitem{Maldacena:2004rf}
J.M.~Maldacena and L.~Maoz, \emph{{Wormholes in AdS}},
  \href{https://doi.org/10.1088/1126-6708/2004/02/053}{\emph{JHEP} {\bfseries
  02} (2004) 053} [\href{https://arxiv.org/abs/hep-th/0401024}{{\ttfamily
  hep-th/0401024}}].

\bibitem{Witten:1999xp}
E.~Witten and S.-T.~Yau, \emph{{Connectedness of the boundary in the AdS / CFT
  correspondence}},
  \href{https://doi.org/10.4310/ATMP.1999.v3.n6.a1}{\emph{Adv. Theor. Math.
  Phys.} {\bfseries 3} (1999) 1635}
  [\href{https://arxiv.org/abs/hep-th/9910245}{{\ttfamily hep-th/9910245}}].

\bibitem{Dolan:1989vr}
L.~Dolan, P.~Goddard and P.~Montague, \emph{{Conformal field theory of twisted
  vertex operators}},
  \href{https://doi.org/10.1016/0550-3213(90)90644-S}{\emph{Nucl. Phys. B}
  {\bfseries 338} (1990) 529}.

\bibitem{DiFrancesco:1997nk}
P.~Di~Francesco, P.~Mathieu and D.~Senechal, \emph{{Conformal Field Theory}},
  Graduate Texts in Contemporary Physics, Springer-Verlag, New York (1997),
  \href{https://doi.org/10.1007/978-1-4612-2256-9}{10.1007/978-1-4612-2256-9}.

\bibitem{Dijkgraaf:1987vp}
R.~Dijkgraaf, E.P.~Verlinde and H.L.~Verlinde, \emph{{$C = 1$ conformal field
  theories on Riemann surfaces}},
  \href{https://doi.org/10.1007/BF01224132}{\emph{Commun. Math. Phys.}
  {\bfseries 115} (1988) 649}.

\bibitem{Narain:1986am}
K.S.~Narain, M.H.~Sarmadi and E.~Witten, \emph{{A note on toroidal
  compactification of heterotic string theory}},
  \href{https://doi.org/10.1016/0550-3213(87)90001-0}{\emph{Nucl. Phys. B}
  {\bfseries 279} (1987) 369}.

\bibitem{Runge1996}
B.~Runge, \emph{{Codes and Siegel modular forms}},
  \href{https://doi.org/10.1016/0012-365X(94)00271-J}{\emph{Discr. Math.}
  {\bfseries 148} (1996) 175}.

\bibitem{Betsumiya2012}
K.~Betsumiya, M.~Harada and A.~Munemasa, \emph{{A complete classification of
  doubly even self-dual codes of length 40}},
  \href{https://doi.org/10.37236/2593}{\emph{Electronic J. Combin.} {\bfseries
  19} (2012) P18} [\href{https://arxiv.org/abs/1104.3727}{{\ttfamily
  1104.3727}}].

\bibitem{Bouyuklieva2012}
S.~Bouyuklieva and I.~Bouyukliev, \emph{{An algorithm for classification of
  binary self-dual codes}},
  \href{https://doi.org/10.1109/TIT.2012.2190134}{\emph{IEEE Transactions on
  Information Theory} {\bfseries 58} (2012) 3933}
  [\href{https://arxiv.org/abs/1106.5930}{{\ttfamily 1106.5930}}].

\bibitem{Pless1972}
V.~Pless, \emph{{A classification of self-orthogonal codes over $GF(2)$}},
  \href{https://doi.org/https://doi.org/10.1016/0012-365X(72)90034-9}{\emph{Discr.
  Math.} {\bfseries 3} (1972) 209}.

\bibitem{Gaberdiel:2013psa}
M.R.~Gaberdiel, A.~Taormina, R.~Volpato and K.~Wendland, \emph{{A K3 sigma
  model with $\mathbb{Z}^8_2$ : $\mathbb{M}_{20}$ symmetry}},
  \href{https://doi.org/10.1007/JHEP02(2014)022}{\emph{JHEP} {\bfseries 02}
  (2014) 022} [\href{https://arxiv.org/abs/1309.4127}{{\ttfamily 1309.4127}}].

\bibitem{Harvey:2020jvu}
J.A.~Harvey and G.W.~Moore, \emph{{Moonshine, superconformal symmetry, and
  quantum error correction}},
  \href{https://doi.org/10.1007/JHEP05(2020)146}{\emph{JHEP} {\bfseries 05}
  (2020) 146} [\href{https://arxiv.org/abs/2003.13700}{{\ttfamily
  2003.13700}}].

\bibitem{Marolf:2020xie}
D.~Marolf and H.~Maxfield, \emph{{Transcending the ensemble: baby universes,
  spacetime wormholes, and the order and disorder of black hole information}},
  \href{https://doi.org/10.1007/JHEP08(2020)044}{\emph{JHEP} {\bfseries 08}
  (2020) 044} [\href{https://arxiv.org/abs/2002.08950}{{\ttfamily
  2002.08950}}].

\bibitem{Cotler:2020ugk}
J.~Cotler and K.~Jensen, \emph{{AdS$_{3}$ gravity and random CFT}},
  \href{https://doi.org/10.1007/JHEP04(2021)033}{\emph{JHEP} {\bfseries 04}
  (2021) 033} [\href{https://arxiv.org/abs/2006.08648}{{\ttfamily
  2006.08648}}].

\bibitem{Chandra:2022bqq}
J.~Chandra, S.~Collier, T.~Hartman and A.~Maloney, \emph{{Semiclassical 3D
  gravity as an average of large-c CFTs}},
  \href{https://doi.org/10.1007/JHEP12(2022)069}{\emph{JHEP} {\bfseries 12}
  (2022) 069} [\href{https://arxiv.org/abs/2203.06511}{{\ttfamily
  2203.06511}}].

\bibitem{Schlenker:2022dyo}
J.-M.~Schlenker and E.~Witten, \emph{{No ensemble averaging below the black
  hole threshold}}, \href{https://doi.org/10.1007/JHEP07(2022)143}{\emph{JHEP}
  {\bfseries 07} (2022) 143}
  [\href{https://arxiv.org/abs/2202.01372}{{\ttfamily 2202.01372}}].

\bibitem{Castro:2011zq}
A.~Castro, M.R.~Gaberdiel, T.~Hartman, A.~Maloney and R.~Volpato, \emph{{The
  gravity dual of the Ising Model}},
  \href{https://doi.org/10.1103/PhysRevD.85.024032}{\emph{Phys. Rev. D}
  {\bfseries 85} (2012) 024032}
  [\href{https://arxiv.org/abs/1111.1987}{{\ttfamily 1111.1987}}].

\bibitem{Jian:2019ubz}
C.-M.~Jian, A.W.W.~Ludwig, Z.-X.~Luo, H.-Y.~Sun and Z.~Wang,
  \emph{{Establishing strongly-coupled $3$D AdS quantum gravity with Ising dual
  using all-genus partition functions}},
  \href{https://doi.org/10.1007/JHEP10(2020)129}{\emph{JHEP} {\bfseries 10}
  (2020) 129} [\href{https://arxiv.org/abs/1907.06656}{{\ttfamily
  1907.06656}}].

\bibitem{Keller:2017iql}
C.A.~Keller, G.~Mathys and I.G.~Zadeh, \emph{{Bootstrapping chiral CFTs at
  genus two}}, \href{https://doi.org/10.4310/ATMP.2018.v22.n6.a3}{\emph{Adv.
  Theor. Math. Phys.} {\bfseries 22} (2018) 1447}
  [\href{https://arxiv.org/abs/1705.05862}{{\ttfamily 1705.05862}}].

\bibitem{Conway1992}
J.H.~Conway, V.~Pless and N.J.A.~Sloane, \emph{The binary self-dual codes of
  length up to 32: A revised enumeration},
  \href{https://doi.org/10.1016/0097-3165(92)90003-D}{\emph{J. Combin. Theory
  Ser. A} {\bfseries 60} (1992) 183}.

\bibitem{Mallows1973}
C.~Mallows and N.~Sloane, \emph{{An upper bound for self-dual codes}},
  \href{https://doi.org/10.1016/S0019-9958(73)90273-8}{\emph{Information and
  Control} {\bfseries 22} (1973) 188}.

\bibitem{Delsarte1973}
P.~Delsarte, \emph{{An algebraic approach to the association schemes of coding
  theory}}, Ph.D. thesis, Universit\'e Catholique de Louvain, 1973.
\newblock [{\em Philips Res. Rep. Suppl.} (1973) {\bf 10}].

\bibitem{Manschot:2007ha}
J.~Manschot and G.W.~Moore, \emph{{A modern Farey tail}},
  \href{https://doi.org/10.4310/CNTP.2010.v4.n1.a3}{\emph{Commun. Num. Theor.
  Phys.} {\bfseries 4} (2010) 103}
  [\href{https://arxiv.org/abs/0712.0573}{{\ttfamily 0712.0573}}].

\bibitem{Ginsparg:1988ui}
P.H.~Ginsparg, \emph{{Applied conformal field theory}},  in \emph{{Les Houches
  Summer School in Theoretical Physics: Fields, Strings and Critical Phenomena.
  Les Houches, France, June 28-August 5, 1988}}, E.~{Br\'ezin} and
  J.~Zinn-Justin, eds., vol.~49, pp.~1--168, 1988
  [\href{https://arxiv.org/abs/hep-th/9108028}{{\ttfamily hep-th/9108028}}].

\bibitem{Dyer:2017rul}
E.~Dyer, A.L.~Fitzpatrick and Y.~Xin, \emph{{Constraints on flavored 2d CFT
  partition functions}},
  \href{https://doi.org/10.1007/JHEP02(2018)148}{\emph{JHEP} {\bfseries 02}
  (2018) 148} [\href{https://arxiv.org/abs/1709.01533}{{\ttfamily
  1709.01533}}].

\bibitem{Benjamin:2021ygh}
N.~Benjamin, S.~Collier, A.L.~Fitzpatrick, A.~Maloney and E.~Perlmutter,
  \emph{{Harmonic analysis of 2d CFT partition functions}},
  \href{https://doi.org/10.1007/JHEP09(2021)174}{\emph{JHEP} {\bfseries 09}
  (2021) 174} [\href{https://arxiv.org/abs/2107.10744}{{\ttfamily
  2107.10744}}].

\bibitem{Collier:2022emf}
S.~Collier and E.~Perlmutter, \emph{{Harnessing S-duality in $ \mathcal{N} $ =
  4 SYM \& supergravity as $SL(2,\mathbb{Z})$-averaged strings}},
  \href{https://doi.org/10.1007/JHEP08(2022)195}{\emph{JHEP} {\bfseries 08}
  (2022) 195} [\href{https://arxiv.org/abs/2201.05093}{{\ttfamily
  2201.05093}}].

\bibitem{Benjamin:2022pnx}
N.~Benjamin and C.-H.~Chang, \emph{{Scalar modular bootstrap and zeros of the
  Riemann zeta function}},
  \href{https://doi.org/10.1007/JHEP11(2022)143}{\emph{JHEP} {\bfseries 11}
  (2022) 143} [\href{https://arxiv.org/abs/2208.02259}{{\ttfamily
  2208.02259}}].

\bibitem{Brown:1986nw}
J.D.~Brown and M.~Henneaux, \emph{{Central charges in the canonical realization
  of asymptotic symmetries: An example from three-dimensional gravity}},
  \href{https://doi.org/10.1007/BF01211590}{\emph{Commun. Math. Phys.}
  {\bfseries 104} (1986) 207}.

\bibitem{Dolan1996}
L.~Dolan, P.~Goddard and P.~Montague, \emph{{Conformal field theories,
  representations and lattice constructions}},
  \href{https://doi.org/cmp/1104286871}{\emph{Commun. Math. Phys.} {\bfseries
  179} (1996) 61}.

\bibitem{Collier:2016cls}
S.~Collier, Y.-H.~Lin and X.~Yin, \emph{{Modular bootstrap revisited}},
  \href{https://doi.org/10.1007/JHEP09(2018)061}{\emph{JHEP} {\bfseries 09}
  (2018) 061} [\href{https://arxiv.org/abs/1608.06241}{{\ttfamily
  1608.06241}}].

\bibitem{Siegel1935}
C.L.~Siegel, \emph{{\"Uber die analytische Theorie der quadratischen Formen}},
  \href{https://doi.org/10.2307/1968644}{\emph{Ann. Math.} {\bfseries 36}
  (1935) 527}.

\bibitem{Conway1982}
J.~Conway and N.~Sloane, \emph{{On the enumeration of lattices of determinant
  one}}, \href{https://doi.org/10.1016/0022-314X(82)90084-1}{\emph{J. Number
  Theory} {\bfseries 15} (1982) 83}.

\bibitem{Niemeier1973}
H.-V.~Niemeier, \emph{{Definite quadratische Formen der Dimension 24 und
  Diskriminante 1}},
  \href{https://doi.org/doi.org/10.1016/0022-314X(73)90068-1}{\emph{J. Number
  Theory} {\bfseries 5} (1973) 142}.

\bibitem{Schellekens:2016hhf}
A.N.~Schellekens, \emph{{Big numbers in String Theory}},
  \href{https://arxiv.org/abs/1601.02462}{{\ttfamily 1601.02462}}.

\bibitem{Bilous2002}
R.T.~Bilous and G.H.J.~van Rees, \emph{{An enumeration of binary self-dual
  codes of length 32}},
  \href{https://doi.org/10.1023/A:1016544907275}{\emph{Designs, Codes and
  Cryptography} {\bfseries 26} (2002) 61}.

\bibitem{Bilous2006}
R.T.~Bilous and G.H.J.~van Rees, \emph{{Enumeration of the binary self-dual
  codes of length 34}}, {\emph{J. Combin. Math. Combin. Comput.} {\bfseries 59}
  (2006) 173}.

\bibitem{Harada2012}
M.~Harada and A.~Munemasa, \emph{{Classification of self-dual codes of length
  36}}, \href{https://doi.org/10.3934/amc.2012.6.229}{\emph{Advances in
  Mathematics of Communications} {\bfseries 6} (2012) 229}
  [\href{https://arxiv.org/abs/1012.5464}{{\ttfamily 1012.5464}}].

\bibitem{Guruswami2010}
V.~Guruswami, ``Introduction to coding theory (course notes).''
  \url{https://www.cs.cmu.edu/~venkatg/teaching/codingtheory/}, 2010.

\bibitem{McEliece1977}
R.~McEliece, E.~Rodemich, H.~Rumsey and L.~Welch, \emph{{New upper bounds on
  the rate of a code via the Delsarte-MacWilliams inequalities}},
  \href{https://doi.org/10.1109/TIT.1977.1055688}{\emph{IEEE transactions on
  Information Theory} {\bfseries 23} (1977) 157}.

\bibitem{Polyakov:1981rd}
A.M.~Polyakov, \emph{{Quantum geometry of bosonic strings}},
  \href{https://doi.org/10.1016/0370-2693(81)90743-7}{\emph{Phys. Lett. B}
  {\bfseries 103} (1981) 207}.

\bibitem{Belavin:1986cy}
A.A.~Belavin and V.G.~Knizhnik, \emph{{Algebraic geometry and the geometry of
  quantum strings}},
  \href{https://doi.org/10.1016/0370-2693(86)90963-9}{\emph{Phys. Lett. B}
  {\bfseries 168} (1986) 201}.

\bibitem{DHoker:1986eaw}
E.~D'Hoker and D.H.~Phong, \emph{{On determinants of Laplacians on Riemann
  surfaces}}, \href{https://doi.org/10.1007/BF01211063}{\emph{Commun. Math.
  Phys.} {\bfseries 104} (1986) 537}.

\bibitem{Voros:1986vw}
A.~Voros, \emph{{Spectral functions, special functions and Selberg Zeta
  function}}, \href{https://doi.org/10.1007/BF01212422}{\emph{Commun. Math.
  Phys.} {\bfseries 110} (1987) 439}.

\bibitem{Sarnak1987}
P.~Sarnak, \emph{{Determinants of Laplacians}},
  \href{https://doi.org/10.1007/BF01209019}{\emph{Commun. Math. Phys.}
  {\bfseries 110} (1987) 113}.

\bibitem{Krasnov:2000zq}
K.~Krasnov, \emph{{Holography and Riemann surfaces}},
  \href{https://doi.org/10.4310/ATMP.2000.v4.n4.a5}{\emph{Adv. Theor. Math.
  Phys.} {\bfseries 4} (2000) 929}
  [\href{https://arxiv.org/abs/hep-th/0005106}{{\ttfamily hep-th/0005106}}].

\bibitem{Gaberdiel:2010jf}
M.R.~Gaberdiel, C.A.~Keller and R.~Volpato, \emph{{Genus two partition
  functions of chiral conformal field theories}},
  \href{https://doi.org/10.4310/CNTP.2010.v4.n2.a2}{\emph{Commun. Num. Theor.
  Phys.} {\bfseries 4} (2010) 295}
  [\href{https://arxiv.org/abs/1002.3371}{{\ttfamily 1002.3371}}].

\bibitem{Tan:2014mba}
H.S.~Tan, \emph{{Closed string partition functions in toroidal
  compactifications of doubled geometries}},
  \href{https://doi.org/10.1007/JHEP05(2014)133}{\emph{JHEP} {\bfseries 05}
  (2014) 133} [\href{https://arxiv.org/abs/1403.4683}{{\ttfamily 1403.4683}}].

\bibitem{Zograf1987}
P.G.~Zograf and L.A.~Takhtadzhyan, \emph{{A local index theorem for families of
  $\bar\partial$-operators on Riemann surfaces}},
  \href{https://doi.org/10.1070/RM1987v042n06ABEH001501}{\emph{Russian Math.
  Surveys} {\bfseries 42} (1987) 169}.

\bibitem{Zograf1987b}
P.G.~Zograf and L.A.~Takhtadzhyan, \emph{{On the uniformization of Riemann
  surfaces and on the Weil-Petersson metric on the Teichm\"uller and Schottky
  spaces}}, \href{https://doi.org/10.1070/SM1988v060n02ABEH003170}{\emph{Math.
  USSR Sbornik} {\bfseries 60} (1987) 297}.

\bibitem{Zograf1989}
P.G.~Zograf, \emph{{Liouville action on moduli spaces and uniformization of
  degenerate Riemann surfaces}}, {\emph{Leningrad Math. J.} {\bfseries 1}
  (1990) 941}. [{\em Algebra i Analiz} {\bf 1} (1989) 136].

\bibitem{McIntyre2002}
A.~McIntyre, \emph{{Analytic torsion and Faddeev-Popov ghosts}}, Ph.D. thesis,
  Stony Brook, 2002.

\bibitem{McIntyre:2004xs}
A.~McIntyre and L.A.~Takhtajan, \emph{{Holomorphic factorization of
  determinants of Laplacians on Riemann surfaces and a higher genus
  generalization of Kronecker's first limit formula}},
  \href{https://doi.org/10.1007/s00039-006-0582-7}{\emph{Geom. Funct. Anal.}
  {\bfseries 16} (2006) 1291}
  [\href{https://arxiv.org/abs/math/0410294}{{\ttfamily math/0410294}}].

\bibitem{Alessandrini:1971dd}
V.~Alessandrini and D.~Amati, \emph{{Properties of dual multiloop amplitudes}},
  \href{https://doi.org/10.1007/BF02731520}{\emph{Nuovo Cim. A} {\bfseries 4}
  (1971) 793}.

\bibitem{Montonen:1974jj}
C.~Montonen, \emph{{Multiloop amplitudes in additive dual-resonance models}},
  \href{https://doi.org/10.1007/BF02785444}{\emph{Nuovo Cim. A} {\bfseries 19}
  (1974) 69}.

\bibitem{DiVecchia:1987uf}
P.~Di~Vecchia, M.~Frau, A.~Lerda and S.~Sciuto, \emph{{A simple expression for
  the multiloop amplitude in the bosonic string}},
  \href{https://doi.org/10.1016/0370-2693(87)91462-6}{\emph{Phys. Lett. B}
  {\bfseries 199} (1987) 49}.

\end{thebibliography}\endgroup

\bibliographystyle{JHEP.bst}

\end{document}